%% file: main.tex
\documentclass[oneside,12pt]{report}
\usepackage[paper=a4paper]{geometry}
\usepackage[utf8]{inputenc}
\usepackage{graphicx}

\usepackage{xpatch}
\usepackage{svg}
\usepackage{algorithm}
\usepackage{algpseudocode}
\usepackage{amsmath}
\usepackage{hyperref}
\hypersetup{
    colorlinks=false,
    breaklinks=true
}
\usepackage{caption}
\usepackage{adjustbox}
\usepackage{longtable}
\usepackage{booktabs}
\usepackage{pdflscape} 
\usepackage{xcolor} % to access the named colour LightGray
\definecolor{LightGray}{gray}{0.9}

\usepackage{listings}
\usepackage{xcolor}

\renewcommand\lstlistlistingname{List of Listings}

\usepackage{tocloft}
\newlistof{lstlistoflistings}{lol}{\lstlistlistingname}
\newcommand{\listoflistings}{\lstlistoflistings}

\usepackage{chngcntr}

\usepackage{multirow}
\usepackage{array}
\usepackage{amssymb} % For checkmark
\usepackage{pifont}  % For crossmark
\usepackage{booktabs, caption, makecell}

\usepackage{threeparttable}

\newcommand{\cmark}{\textcolor{green}{\checkmark}} % checkmark
\newcommand{\xmark}{\textcolor{red}{\ding{55}}}  % crossmark

\usepackage[style=numeric,backend=biber,url=true, doi=true, sorting=nyt, natbib=true]{biblatex}
\usepackage{bibentry}
\usepackage[title, titletoc]{appendix}

\usepackage{datetime}
\newdateformat{monthyeardate}{%
  \monthname[\THEMONTH], \THEYEAR}

\newcommand\thesistitle{Software Unclonable Functions for IoT Devices Identification and
Security}
\newcommand\authorname{Saeed Alshehhi}

\newcommand\supervisor{Professor Nicholas Race}

\newgeometry{left=25mm, right=25mm, top=25mm, bottom=25mm, includeheadfoot}

\usepackage{fancyhdr}
\fancypagestyle{preamble}{ %
  \fancyhf{} % remove everything
  \fancyfoot[C]{\thepage}

}
\fancypagestyle{main}{
  \fancyhf{}
  \fancyfoot[C]{\thepage}
  \fancyhead[L]{\textit{ \nouppercase{\leftmark}} }
  \fancyhead[R]{\textit{ \nouppercase{\rightmark}} }

}

\fancypagestyle{plain}{ %
  \fancyhf{}
  \fancyfoot[C]{\thepage}

}

\usepackage{lipsum} % filler text

\begin{document}
\pagenumbering{roman}

\clearpage

\input{abstract}
\clearpage

%\input{acknowledgements}
%\clearpage

\tableofcontents
\clearpage

\listoffigures
\clearpage

\listoftables
\clearpage

\listoflistings
\clearpage

\pagestyle{main}
\pagenumbering{arabic}

\chapter{Introduction}
\input{chapters/introduction}

\chapter{Background}
This chapter provides an overview of the key concepts related to IoT device security and the existing techniques that would be employed to achieve our research target. First, common threats targeting embedded systems, such as firmware attacks, botnets, and denial-of-service attacks, are introduced. Further, the chapter looks into the contemporary methods used to protect these systems, such as Traditional-based and AI-based approaches. This would be followed by introducing the hardware performance counters (HPCs) and the potential to improve IoT security. This would lead to software unclonable functions (SUFs) being utilised and defined in this research. Then, sequence classification models such as LSTM and GRU are covered, highlighting their role in detecting patterns of the functions’ activities and characterisation.

%Next, the chapter explores the current methods used to protect these systems, including traditional and AI-based approaches. This will be followed by introducing the hardware performance counters (HPCs) and their potential to enhance IoT security. This will lead to the concept of software unclonable functions (SUFs) being employed and defined in this research. Finally, sequence classification models like LSTM and GRU are covered, highlighting their role in detecting patterns of functions' activities and characterization.\\ 

%This chapter sets the stage for examining the security landscape of IoT devices, the limitations of existing solutions, and the potential of hardware-based approaches like HPCs. This foundational knowledge is critical for appreciating the research's objectives and the innovative solutions it seeks to introduce.
\input{chapters/background}

\chapter{Literature Review}
\input{chapters/Rel}

\chapter{Design}
\label{Chapter:Design}
%Change  This chapter is reserved for the research design stage 
This chapter dives into the research design stage, where the theoretical concepts and ideas are translated into usable solutions. It identifies the common malicious functions that pose significant threats to embedded systems, such as port scanning, brute-forcing, command and control (C\&C) communication, and persistence mechanisms. It clarifies the process of designing the given functions in the controlled environment, thus allowing for the specific characterisation of the behaviour using the HPCs. This chapter further explores the methodology for extracting the HPC data from the embedded systems, mainly focusing on Reduced Instruction Set Computing (RISC) architecture, which is mostly found in IoT devices. By linking the specific HPC events to the malicious functions, this chapter sets the stage for the various implementation phases. Design principles appointed here are important for developing a security framework that is efficient in detecting threats and efficient in operation, making it highly suitable for deployment in actual IoT environments.
\input{chapters/Design}

\chapter{Implementation}
\label{chapter:Implemt}
This chapter describes the implementation at the proof-of-concept level, translating the ideas defined in Chapter \ref{Chapter:Design} to a running prototype. We will delve into the development process of the malicious function prototypes and the setup of the data collection systems that facilitate the acquisition of HPC data for the different scenarios. We will also describe the configuration of HPCs for event monitoring and the tools used to collect the performance data, for example, the Linux perf utility.

The chapter also outlines the preprocessing of this collected data, including normalisation, feature extraction, and creation of training datasets for a machine-learning model. Technical implementation challenges are also elaborated on, such as cross-compiling for different architectures and ensuring the accuracy and reliability of the data. All these experiments were conducted at the Technology Innovation Institute (TII) \cite{TII}, where their team fixed and provided the experimental setups. This chapter provides the most comprehensive description of how the theoretical framework was turned into a real-world security solution.

\input{chapters/Implementation}

\chapter{Evaluation}
This chapter evaluates the formulated security framework, detailing the experimental results obtained from the selected IoT devices from Chapter \ref{chapter:Implemt}. It begins by outlining the metrics used to assess the performance of the AI models and the methods used to determine whether the HPC-based SUFs can detect the identified security threats. These metrics involve standard ways to examine the trained AI models, such as the validation accuracy, precision, recall, and F1 score, as well as the use of confusion matrices to analyse the results.

Next, the obtained metrics will be analysed to evaluate the different experiments and enhance the model initially used to achieve a more accurate classification. By the end of the chapter, we will summarise the overall performance evaluation and the capabilities achieved by the developed model.

\input{chapters/Experements}

\chapter{Conclusion and Future work}
\input{chapters/Disscusion}

\clearpage % Required to make sure the references table of contents page number is correct.

\printbibliography[heading=bibintoc,title=Reference]

\end{document}

%% file: abstract.tex
\begin{center}
\textbf{\thesistitle}

\authorname \\
Saeed.alshehhi@tii.ae \\
Supervised by  \supervisor \\
n.race@lancaster.ac.uk

School of Computing and Communications, Lancaster University

A dissertation submitted for the degree of \textit{Master of Science} in Cyber Security.
\newline September, 2024
\end{center}

\section*{\centering Abstract} \vspace{8pt}

\textbf{In the evolving landscape of IoT ecosystem, distinguishing between legitimate and compromised devices is a critical challenge. This research investigates the effectiveness of hardware performance counter (HPC)-derived signatures' uniqueness under the umbrella of a concept that we introduced as software unclonable functions (SUFs). Our experiments demonstrate that HPC-derived SUF signatures can accurately distinguish between legitimate and compromised behaviors in IoT devices, achieving accuracy rates between 83.77\% and 96.47\% with high precision. The presented work delves into the technical details, the various experiments that we conducted to collect an appropriate dataset using a custom implementation, and the design choices that have been initiated to evaluate the collected data. Furthermore, we propose an enhanced machine learning model based on Gated Recurrent Units (GRUs) tailored to solve this specific problem. To further fortify the deployment framework for SUFs, we explored the integration of Machine Learning techniques with HPC data. Our findings indicate that using constant compilation options across different devices results in stable signatures that can be reliably identified. Additionally, we show that employing statistical tools and techniques such as scaling contributed to the robustness of our detection approach, demonstrating good performance across various experimental setups.
We also addressed the integration of HPC-based SUFs into current IoT security paradigms, ensuring minimal overhead on operational efficiency. The proposed modular architecture features independent data collection and classification pipelines, allowing seamless incorporation into existing IoT devices without compromising functionality. This lightweight and flexible integration allows for seamless incorporation into existing IoT devices without compromising their functionality. Our research confirms that HPC patterns for specific functions can serve as a reliable signature for identifying malicious threats targeting IoT devices or firmware. The proposed model achieved high accuracy in both binary and multiclass classification scenarios, underscoring its potential as a powerful tool for IoT security.}

%% file: chapters/introduction.tex
\section{Objective}

%The Mirai botnet and other similar cyber attacks have highlighted the critical vulnerabilities in IoT devices, exposing them to exploitation and control by malicious actors. These attacks leverage weak security measures, resulting in massive disruptions and breaches that can cripple services and compromise sensitive data.

%To effectively counter these threats, traditional security measures are no longer sufficient. We need innovative solutions that can offer more robust and resilient protection. One such promising approach is the use of software unclonable functions (SUFs).

%What are Software Unclonable Functions (SUFs)?

%SUFs harness the unique physical characteristics of hardware to generate a digital "fingerprint or signatures" for each function. This fingerprint is derived from minute variations in the manufacturing process, making it virtually impossible to duplicate or clone. Unlike traditional security methods that rely solely on software, SUFs provide a hardware-based layer of security that is inherently more secure against tampering and replication.

The ever-evolving landscape of cybersecurity threats demands innovative and robust solutions to safeguard the vast array of Internet of Things (IoT) devices permeating our daily lives. Today's IoT spread is expanding beyond simple applications and automation towards more critical infrastructure. This expansion comes at the cost of exposing such infrastructures to more serious incidents, such as infrastructure blackouts and national security. According to \cite{caburntelecom_iot} IoT devices are one of the foundational pillars of smart cities, smart homes, supply chain management, agriculture, enterprise solutions, healthcare, traffic, and smart grids. Recently, during the Russian-Ukranian war, the Five Eyes Intelligence reported \cite{ucsc_ukrain_2024} that the Industroyer II operation was used as a war tactical intimidation by the adversary. The operation led to a complete blackout of the smart grid of the Kyiv city. The Five Eyes report highlights that such an operation is a glance at modern warfare that combines physical and cyber acts as a tactical strategy. The Mirai botnet and similar cyber-attacks highlighted the critical vulnerabilities inherent in IoT devices \cite{cloudflare_mirai}, exposing them to exploitation and control by malicious actors. These attacks leverage weak security measures, resulting in massive disruptions and breaches that can incapacitate services and compromise sensitive data. An example of these attacks is when cybercriminals attacked Synnovis, NHS England \cite{nhs_synnovis_cyber_attack} and stole sensitive data.   \\

%23 August 2024 > I need to change the we named. to we propose 
According to \cite{Mazhar2023IoTSecurity}, Traditional security measures are proving inadequate when facing such sophisticated threats. Hence, there is an urgent need for more resilient and adaptive security mechanisms. One promising approach to address this challenge is using our proposed concept, Software Unclonable Functions (SUFs), which will be detailed in this work. \\

% Here, you are defining PUF. The objective is to explain SUF PUF could be used for analogies only
\textbf{What are Software Unclonable Functions (SUFs)?} The basic concept is similar to Physically Unclonable Functions (PUFs) \cite{pufs}, which use inherent manufacturing variations in hardware components to create a unique, unclonable identifier for each device. These identifiers, derived from unpredictable physical properties, are highly secure and cannot be replicated, even by the manufacturer. SUFs that are modelled after the PUFs, are designed to develop distinct identifiers for software operations that make them difficult to duplicate in other situations or systems. As a kind of fingerprint, these SUFs produce distinct patterns per the inherent characteristics of the software and the environment in which it is executed. We think this strategy works especially well in embedded and Internet of Things contexts, where typical security techniques are less practical due to resource limitations and the specialised nature of firmware. SUFs are all distinctive because of the present heterogeneity in software execution across various hardware platforms, especially in devices with ARM and RISC architectures. Processing efficiency and simpler pipelining are made possible by RISC architecture \cite{arm_risc}, which is known for its excellent performance per watt and streamlined instruction set.

Due to these features, SUFs are mainly effective in preventing cloning and guaranteeing the integrity and authenticity of Hardware Performance Counters (HPCs), which are counters provided directly by the CPU, in Internet of Things devices. SUFs can also reliably generate unique identifiers tied to the specific execution environment. They're employed in debugging as well. This study investigates how developing and using SUFs might improve IoT security using HPCs. This project seeks to provide a new security approach by creating reliable signatures (SUFs) using HPC patterns to differentiate between compromised and benign IoT devices. This would be accomplished by extracting and evaluating various capabilities from firmware using HPCs.

%thIS DOWN Is what I want to write but idk < 23 August 2024
%This research explores the utility of Hardware Performance Counters (HPCs), which are counters directly provided by the CPU used for debugging, for enhancing IoT security through the development and application of SUFs. By extracting and analyzing various functionalities that exist in firmware via HPCs, this study aims to pioneer a novel security methodology by defining reliable signatures (SUFs) through HPC patterns to distinguish between benign and compromised IoT devices.

The primary objective of this dissertation is to investigate how HPC-derived properties can be used to characterise a software function that has been designed on a particular firmware. Through a comprehensive study and experimentation, we aim to:

\begin{itemize}
    \item Determine the accuracy with which deep learning models, like RNN, could be used to distinguish between benign and malicious behaviour on a range of IoT devices when using HPC-based characterisation. 
    
    \item When combined with HPC metrics, determine software-based approaches and pre-processing strategies that majorly increase the robustness of SUF identification. 
    
    \item Examine ways to enhance firmware integrity checks and device authenticity verification with HPC-based SUFs integrated into current IoT security frameworks, all while minimising the impact on device performance.
\end{itemize}

The significance of this research lies beneath its innovative approach to addressing the critical security gaps in current IoT devices. It offers a novel solution that combines hardware and software-based security measures to create a more secure and resilient IoT ecosystem with reduced overhead.

% to be revised later    
\section{Goals and Motivation} 

%The primary goal of this research is to advance the security of Internet of Things (IoT) devices by leveraging Hardware Performance Counters (HPCs) to identify and mitigate various threats within embedded systems. This research explores whether specific HPC patterns can be reliable signatures for detecting malicious activities targeting IoT devices or their firmware by identifying a set of patterns that we could qualify as SUFs. Following the first study outcomes, the identified patterns would be employed to detect various security-related concerns, such as malicious activities.

The main goal of the research is to enhance the security of Internet of Things (IoT) devices by leveraging HPCs within embedded systems. The research identifies whether specific HPC patterns could be reliable signatures for detecting malicious activities targeting IoT devices or firmware by identifying the set of patterns that could qualify as SUFs. Using the first study outcomes, the identified patterns will be employed to detect different security-related concerns, like malicious activities.

The major motivation behind the research is also grounded in the rising complexity of IoT environments, where the devices are mostly used in large numbers across diverse and hostile atmospheres. The main issue with IoT devices comes from the fact that little to no standardisation efforts are implemented in this sector. Furthermore, each manufacturer's firmware could be uniquely implemented compared to its competitors, making the application of traditional security approaches very challenging. Traditional security solutions, which rely heavily on software-based mechanisms, often fall short in these contexts due to the constrained resources of IoT devices. Hardware-level security measures, such as those provided by employing the HPCs, offer a promising alternative that could significantly bolster the security of these devices.

\subsection{Research Question}
This research aims to address the security benefits of introducing a hardware-aided approach to identify various threats in the context of embedded systems. The main research question that would be clarified by the end of this document is: 
\textbf{Can HPC patterns for specific functions serve as a reliable signature for identifying malicious threats that target IoT devices or firmware? }
\hfill\\ To clarify our research objective, the research question will be addressed through the following sub-questions: \\

\textbf{Question 1:} How effectively can HPC-derived SUF signatures distinguish between legitimate and compromised devices, thereby enhancing the nuanced approach required for robust IoT security? \\

%\textbf{Objective:} To ascertain the precision with which HPC-based SUFs can differentiate between benign and malicious behavior across a variety of IoT devices, considering the inherent diversity in their hardware and software configurations.

\textbf{Question 2:} Which software-based techniques, integrated with HPC data, significantly fortify the deployment framework for SUFs, ensuring a resilient identification? \\
%\textbf{Objective:} To identify software-based methods that, when integrated with HPC metrics, substantially improve the robustness of SUFs in detecting unauthorized firmware modifications or accurately identifying a series of targeted behaviors.

\textbf{Question 3:} In what ways can HPC-based SUFs be seamlessly incorporated into current IoT security paradigms without compromising operational efficiency? \\

%\textbf{Objective:} To explore avenues for integrating HPC-based SUFs into existing IoT security frameworks to strengthen firmware integrity checks and device authenticity verification while minimizing the impact on device performance.

\section{Document Structure}
This document is structured as follows:
\begin{itemize}
    \item Chapter 2 - Background: This chapter introduces the key elements that serve as a foundation for this research, including embedded systems threats, mitigation strategies, hardware performance counters (HPCs), and the implementation of AI models.
    \item Chapter 3 - Literature Review: This chapter reviews existing work related to this topic, identifying gaps in the current research and linking them to this study's objectives.
    \item Chapter 4 - Design: This chapter examines the design of malicious functions, the data collection approach, and the AI model design.
    \item Chapter 5 - Implementation: This chapter details the implementation process, including data collection, cross-compiling, data processing and analysis, SUFs implementation, and AI model implementation.  
    \item Chapter 6 - Evaluation: This chapter discusses the evaluation metrics and experimental setup for two devices, followed by an in-depth analysis of the experimental results.
    \item Chapter 7 - Conclusion and Future work: This chapter provides an overview of the conducted research, discusses the research significance, and evaluates the initial research objective through the question answering. The chapter concludes by highlighting the future work and the potential impact of this work.
\end{itemize}

\clearpage

%% file: chapters/background.tex
\section{Embedded Systems Threats}
\label{section:threats}
    % What threats are faced in Embedded systems; cyber-attacks in general. 

Embedded systems play an essential part in modern technology, making them a potential target for many modern cyber-attacks. Embedded systems range from white goods to complex factory machinery, making their security an absolute concern. This section will examine the current threats facing such systems, including firmware attacks, botnets, ransomware, denial of service (DoS) attacks, and malware.

%This section will examine the current threats facing such systems: firmware attacks, botnets, ransomware, Denial of Service (DoS) attacks, and malware. 

%, with a focus on their impact and the importance of strong security mechanisms. 
%\begin{itemize}
    %\item Mirai Botnet: The Mirai botnet exploited default credentials and unsecured devices to create a network of compromised devices, which were then used to launch DDoS attacks. \\
    
    %By implementing SUFs, each device's unique fingerprint would make it significantly more difficult for attackers to clone devices or gain unauthorized access, thereby reducing the botnet's ability to proliferate.
    
    %\item Other Malicious Attacks: Beyond Mirai, various other attacks target IoT devices, including data breaches, ransomware, and control hijacking. \\
    %SUFs provide a robust defense mechanism by ensuring that only legitimate, authenticated devices can operate, making it much harder for attackers to infiltrate and compromise the network.
    
%\end{itemize}

%Add statistics about how often such system is exposed to each of the threats 
\subsection{Firmware Attacks}

Attackers are able to exploit the vulnerabilities in firmware by using firmware attacks that happen while updating the firmware of the device. A few of the threats are installing unauthorised firmware, transforming the firmware, and reverse engineering.
One study \cite{firmwarecase} demonstrated, for example, how malicious code introduced into the firmware of the smart plug might grant access and remote control, thereby opening a back channel to the attacker's computer. According to a Microsoft study \citep{microsoft_firmware_study}, firmware attacks on IoT devices are becoming more prevalent, with 80\% of enterprises experiencing at least one firmware attack in the past two years. 

%For instance, a study by \citep{firmwarecase} demonstrated how bad code inserted into a smart plug’s firmware allowed remote control and access, creating a back channel to the attacker's computer. 

\subsection{Botnet}
Botnets are a major threat to embedded systems, as they harness multiple devices to coordinate malicious operations such as distributed denial-of-service (DDoS) attacks, data theft, and malware propagation. Embedded systems, generally, lack sufficient security measures, which makes them an accessible target for recruiting into botnets. Once hackers gain access to these systems, they can commandeer them remotely and direct them towards starting harmful activities without users’ knowledge. To thwart this threat, it is crucial to have robust security measures with regular updates and use advanced methods like AI-based anomalies and classification techniques to detect botnets early.

Data from \citeauthor{socradar2024} \citep{socradar2024} indicates that the largest botnet of Q1 2024 contained 51,400 devices compared with 16,000 devices in the previous quarter, but lower than the same period a year earlier, with botnets of up to 131,628 devices. A Mirai-variant botnet initiated the largest DDoS attack of 2024, peaking at two terabits per second.

\subsection{Ransomware}
In ransomware attacks, the attacker encrypts the victim’s data so that the organisation or individual can’t access their data unless they pay a ransom; if the victim chooses to pay the ransom, they will be able to re-access their data, which could cripple organisations and individuals by preventing them from accessing their data until they do. As of 2024, 59\% of organisations reported ransomware attacks, according to \citeauthor{stateofransomware2024} \cite{stateofransomware2024} They stated: ‘The overall cost of attacks in 2023 exceeded \textdollar1 billion of ransomware payments'.

\subsection{DoS \& DDoS}
A Denial of Service (DoS) attack floods a target, such as a computer server, with internet traffic to render it inaccessible to everyone trying to use it, whereas a Distributed Denial of Service (DDoS) attack enlists multiple compromised systems to target a network resource. Both attack strategies aim to make a device or resource unavailable and, if successful, can degrade service accessibility or result in a system outage for a significant period of time. Cloudflare \citep{cloudflare2024ddosq2} reported a 20\% year-over-year increase in recorded Distributed Denial of Service (DDoS). According to a recent report in \citep{zayo2023}, unprotected organisations pay \textdollar 6,000 per minute, regardless of frequency, duration, or volume of attack.

\subsection{Network Scanner}
%Network scanners are tools used by attackers to identify vulnerable devices on a network. By scanning for open ports and services, attackers can find entry points to exploit and compromise devices, making network scanners a common tool in cyber-attacks.

A network scanner is a tool that attackers use to find open ports and services for devices on the same network. For example, an attacker uses a tool like nmap \citep{nmap} that scans the network and returns any open ports or services of all hosts on the network. This allows them to identify any host with an improperly configured port or service, increasing its vulnerability to hacking. Ultimately, a network scanner helps a hacker identify the entry points for an attack. Therefore, a network scanner is a common component of cyberattacks. Botnet operators are constantly trying to identify potential targets and grow their network of infected machines. They look for vulnerable machines on unprotected ports for botnet addition and hacking.

\section{Mitigation Strategy of Embedded Systems}

This section discusses various mitigation strategies for embedded systems, focusing on Signature-based and AI-based approaches. The advantages and limitations of each method are explored to provide a comprehensive understanding.

\subsection{Signature Based}
%Signature unclonable functions (Malicious behaviour)
Signature-based approaches usually rely on identifying unique patterns for each attack variant without any capabilities to generalise or mitigate emerging threats. This approach is usually implemented through a set of Indicators of Compromise (IoC) that describe a specific threat. These IoC could be extracted from many sources, such as network packets, system logs, and hardware logs. Several studies have employed signature-based methods for IoT security, including those by \citep{ioulianou2018signature} \citep{otoum2021ids} \citep{nawaal2024signature} and \cite{thankappan2024signature}.

\subsubsection{Network Based}

Signature-based network packet analysis looks for known malicious patterns embedded in the data packets sent over the network to stop potentially harmful traffic. This is how intrusion detection systems (IDS) often work, detecting and blocking malicious traffic through traditional signature-matching attack detection. The strong point of this method is that it works well for known attacks, catching and blocking threats, such as worms, that allow hackers to attack a system by comparing network traffic to a list of signatures. However, because this method cannot detect new or previously unknown threats (a zero-day attack), this type of attack detection needs to be constantly updated with new signatures and requires ongoing operational monitoring and maintenance, which can get expensive.

%https://gcelt.org/signature-based-ids-advantages-and-disadvantages-in-cybersecurity/
%reference this to the limitation of this approach 

\subsubsection{System Logs}
System logs record detailed information about the system activities and events. These logs are analysed to detect patterns (an Indicator of Compromise) of known malicious behaviour, aiding in incident response and forensic investigations. The advantage of this method is that it provides a comprehensive record of system activities, useful for tracing attacks and understanding their scope. On the other hand, this method will generate large volumes of data, making real-time analysis challenging. Attackers may also \textbf{alter or erase logs to hide their activities}. 

%https://www.comptia.org/blog/log-data-key-to-identifying-cybersecurity-threats 
%in the above link there is IoT logs. 

%https://www.tributech.io/blog/log-data-integrity
%this shows the important of alert or erase logs to hide thier activities. 

\subsubsection{Hardware Logs}
    %- what is the advantage and what is the limitation 
Hardware logs, generated by components such as Trusted Platform Modules (TPMs), provide a tamper-resistant record of system activities in embedded systems, capturing events at the hardware level. It is more resistant to tampering compared to software logs, enhancing the integrity and trustworthiness of the logging system, but its limited storage capacity and granularity of data may prevent capturing all relevant security events, especially those at higher software layers.

\subsection{AI Based}
AI-based approaches in IoT security represent a significant advancement over traditional methods, offering dynamic and adaptive protection against the unique and evolving threats IoT devices face. Several studies have employed AI to defend their IoT networks \cite{altulaihan2024anomaly} \cite{bacha2024anomaly} \cite{alshahrani2023optimising} and \cite{behiry2024cyberattack}.

\subsubsection{Network Based}
%network based (packets)
 %   - what is the advantage and what is the limitation 
Processing and analysing network traffic as it occurs, AI-based network analysis offers real-time threat detection and quickly identifies malicious patterns and behaviours. This method is good at detecting attacks immediately and assisting in decreasing response times. However, its success rests largely on the quality of the data and the labelling used to train machine learning models. When data is poorly or incorrectly labelled, the models produced are erroneous. It is thus incumbent on those arranging these data sets to ensure that AI can effectively detect malicious network activity.

\subsubsection{System Logs}
AI techniques applied to analysing system logs like \cite{systemlog_ai} efficiently search large logs, and their collecting power is second only to that of humanoid computers. However, the advantage of using this method for detection lies in its ability to automatically carry out case analysis and thus better monitor potential target risks beyond detection thresholds. Operation in practice clearly indicates that detecting blogs is an extremely difficult business. The bottleneck currently lies in the large volume of log data and how to ensure that AI models are trained on properly labelled datasets containing comprehensive examples. This will reduce both false positives and negatives.
%https://www.mdpi.com/1424-8220/20/9/2451

%system logs
%    - what is the advantage and what is the limitation 
\subsubsection{Hardware Logs}
%hardware logs
%    - what is the advantage and what is the limitation 
Using AI-based analysis of hardware logs allows for taking advantage of the detailed, low-level insights that hardware components already produce. This includes the detection of real or potential threats from software that may have evaded your defenses and is currently operating in the background, rather than being halted by an alert system on your computer screen, which only catches a partial sentence. However, the limited storage capacity and granularity of computer logs will determine how much data is available to analyse. The complexity of interpreting these logs also means that you need powerful AI models trained on vast and representative records.

\subsection{Anomaly Based}
A survey conducted by \citeauthor{anomalydetection_ai} \cite{anomalydetection_ai} explores the anomaly detection methods for IoT and its applications. Anomaly detection can start an analysis process to detect possible anomalies by comparing the data with normal behaviour. This process starts by collecting data of many types, such as network traffic, system logs, and hardware event logs. Machine-learning algorithms can compare this data with past learning to identify typical patterns and behaviours. They build these models using techniques like clustering and neural networks. They compare present data streams against designed earlier profiles to detect anomalies. When the system detects something different, it will flag the potential threat. The effectiveness of this method relies on adequate training data with well-labelled and high-quality tagging to avoid mistakes and ensure accurate detection.
%anomaly based:
 %   - how AI approach detect anomaly : e.g. by learning user behavior or device. 
 %   - We don't care about the fucking user we care about the device itself. 
\subsection{Classification Based}

%classification based . 
%    - learn  the both malicious and normal and approximate into on of these class. 
%    - this is what we will use in this work. 
%    - AI and Hardware based approaches ( I need to focus on this. why because it will show me the importance of my work. or what I'm trying to achieve.)

AI methods based on classification, as can be seen in \cite{classification_ai}, are algorithms that learn how to differentiate normal from abnormal. This deterministic method usually involves training on datasets with labelled data, where we feed in examples of both benign and malicious behaviour to make the decision boundary clear and fine using supervised learning algorithms such as decision trees, support vector machines, or neural networks to establish decision boundaries that distinguish between normal and malicious activities. When new data comes in, these models approximate it into one of the pre-defined classes according to patterns they have learned, resulting in unified error rates even when we have never seen specific information before.

In this work, classification-based methods are employed due to their effectiveness in detecting known patterns with a focus on both AI and hardware-based techniques to indicate how important the combination of these two may be for improving malware detection and security at system levels. This makes the best of both AI and computer resources. It stresses the value of hardware-based approaches in showing up the need for deep low level data analysis to detect highly abstract types of attack that would otherwise not be detected by any software method alone. 
\section{Hardware Performance Counters}
\label{background:HPC}
%HPCs provide insights into how specific parts of the code perform, making them invaluable for profiling and debugging complex software. It also helps monitor the system's health and performance in real-time, which is crucial for system administrators. 
According to the \citep{cambridge2021hwpmc}, "Hardware performance counters are a processor facility that gathers statistics about architectural and micro-architectural performance properties of code execution and data access".  HPCs are crucial for profiling and debugging complex software as they shed light on the behaviour of individual code segments. For system administrators, it facilitates real-time system health as well as performance monitoring. Cybersecurity methodologies employ Hardware Performance Counters (HPCs) to enhance defences against cyberattacks \citep{hpc_cybersecurity}. These counters provide comprehensive insights into hardware-level performance metrics, assisting in identifying and mitigating security breaches. HPCs are important for ensuring the system and data integrity utilising statistical and computational analysis methods. Through the successful handling of the complex nature of cyber threats, they serve as a complement to traditional security systems such as firewalls and IDSs. Hus, HPCs are an important component of cybersecurity plans as they offer a more thorough insight into the system's performance, leading to increased security.

Before the use of HPCs effectively, it's important to understand the distinctions between the Complex Instruction Set Computer (CISC) used in x86 architecture design and the Reduced Instruction Set Computer (RISC) that will be discussed in Section \ref{designrisc} and is used in ARM architecture design. The type of architecture highly influences the behaviour of programs in terms of HPC metrics. Since the RISC architectures have simple, predictable instruction sets, embedded systems often select them as they make it easier to read HPC data and identify malicious activity based on the measured activities.

HPCs are considered low overhead for several reasons:
\begin{itemize}
    \item \textbf{Integration into Processor Architecture}: The architecture of CPU includes HPCs directly. They do not require a lot of extra computational power or sophisticated software layers in order to function because they are a component of the hardware. The performance of the system as a whole is protected from being mainly impacted by performance monitoring due to such integration.
    
    \item \textbf{Minimal Resource Utilisation}: Dedicated registers are being used by HPCs to count particular occurrences like cache misses, branch predictions, and instruction executions. These records function independently of the principal processing activities so as to prevent competition for CPU resources between them and the primary workload.
    
    \item \textbf{Passive Monitoring}: HPCs monitor the events in a passive manner, which means they don't impede the software's flow of execution. By using passive monitoring, it is ensured that performance counters can collect data without affecting the behaviour or efficiency of processor-running apps.

    \item \textbf{Transparent Measurements}: The HPCs’ design enables them to collect the data with minimal intervention. They can monitor and record events without the need for frequent context switches or additional instructions, which would otherwise consume CPU cycles and slow down the system.
    
    \item \textbf{Hardware-Level Implementation}: Because HPCs are implemented at the hardware level, they bypass the need for software-based instrumentation, which can introduce significant overhead. Software-based monitoring tools often require additional code to be inserted into the application, leading to increased execution time and resource consumption.
\end{itemize}

Using the HPC peripheral typically involves several steps to ensure accurate and efficient monitoring of hardware events, as shown in Figure \ref{fig:statemachine}. As shown in the figure, HPC is a memory-mapped device that connects to the AXI/AHB interconnect through the ABP adapter. The HPC peripheral can monitor various events within the CPU on the other peripherals. Based on the article from \citep{hpcflow}, The following steps are required:

\begin{figure}[!h]
    \centering
    \includegraphics[scale=0.3]{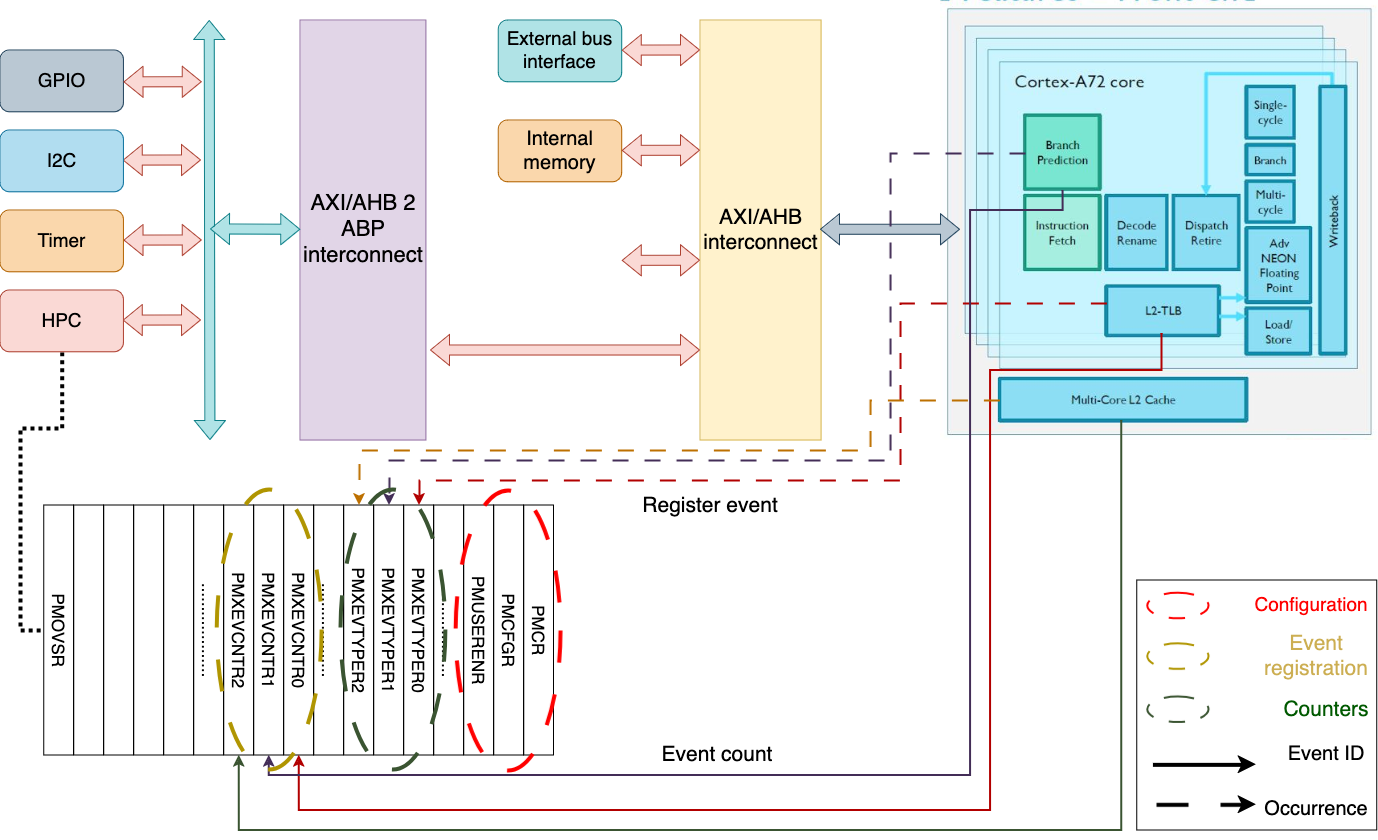}
    \caption{HPC overview}
    \label{fig:statemachine}
\end{figure}

\begin{enumerate}
    \item \textbf{Initialisation}:
    \begin{itemize}
        \item The PMU must initialise, configure the necessary registers, and enable the User Enable Register (USEREN).
        \item Set the event selection registers to indicate which hardware events should be monitored.
        \item The application activates only the necessary counters, preparing the system for future tasks.
    \end{itemize}
    
    \item \textbf{Monitoring}:
    \begin{itemize}
        \item Zero all counters to ensure an accurate starting point for valid data.
        \item Enable event counting by setting appropriate controls (PMXEVTYPER).
        \item The PMU collects specified hardware events, gradually accumulating recognisable data from a known start point.
    \end{itemize}
    
    \item \textbf{Reading Counters}:
    \begin{itemize}
        \item Sample counters (PMXEVCNTR) at regular intervals (or in response to specific triggers, such as interrupts) to ensure continuous data collection.
        \item Handle counter overflows (PMOVSR) effectively by incrementing a higher-order counter or setting a flag when a counter exceeds its maximum value.
    \end{itemize}
    
    \item \textbf{Stop Monitoring}:
    \begin{itemize}
        \item Transition to stop monitoring state when the required monitoring period ends or the temporary stop is required.
        
        \item Counting Halt event by setting appropriate control registers, preventing extra data collection.
    \end{itemize}
    
    \item \textbf{Final Reading and Data Processing}:
    \begin{itemize}
        \item Take the final reading of counter values after stopping monitoring to account for all the events during the review period.

        \item Process collected data to yield meaningful metrics about monitored events.
        
        \item  The phase could also involve recalibration and resetting of the counters to change the raw data into actionable insights for the optimisation of hardware performance.
    \end{itemize}
    
    \item \textbf{Cleanup}:
    \begin{itemize}
        \item Shut down PMU if it is not required, releasing the resources and power.
        \item Proper cleanup makes sure that the system remains in a good state for the next step, thus maintaining efficiency and smoothness of operation.
    \end{itemize}
\end{enumerate} 

The transparent nature and the low level of interference with the target system of HPCs' designs are justified. Due to the minimal overhead, HPCs are a good fit for IoT devices with limited resources, as they make standard software-based anomaly detection techniques less practical. IoT device restrictions are well-suited for low-cost hardware-based monitoring solutions like perf. RISC architectures are widely used in embedded systems because of their simple and predictable instruction sets that facilitate the reading of HPC data and the detection of malicious activities without adversely impacting system performance. The way HPCs are designed ensures they work as a discrete yet powerful tool for keeping an eye on and safeguarding IoT devices.

Table \ref{table:hpcevents} gives an overview of common hardware performance counter (HPC) events that include event numbers and descriptions for each. In the context of cybersecurity approaches that make use of HPCs to strengthen defences against cyberattacks, these events are mainly important for understanding and analysing the behaviour and performance of hardware during code execution and data access. 

\begin{table}[!h]
    \centering
    \begin{tabular}{|c|l|}
     \hline
     Event Num & Description \\ [0.3ex] 
     \hline\hline
        0x01 & Instruction fetch causing a cache refill at the lowest level. \\
        0x02 & Instruction fetch causing a TLB refill at the lowest level. \\
        0x03 & Data read/write causing a cache refill at the lowest level. \\
        0x04 & Data read/write causing a cache access at the lowest level. \\
        0x10 &  Counts every pipeline flush due to misprediction. \\
        %Branch mispredicted/not predicted
        \hline
    \end{tabular}
    \caption{Common feature numbers from \citep{arm2024}}
    \label{table:hpcevents}
\end{table}

The events are taken from \citep{arm2024} and are important sources of information for security monitoring and performance tweaking. It offers important information for identifying possible security breaches and enhancing hardware performance. By examining these events, scientists and engineers can obtain a deeper comprehending of how systems behave, which can result in hardware optimisations and cybersecurity strategies that are more successful.

\section{Sequence Classification Model}
Academia has studied various ways of handling challenges in sequence classification related to Natural language processing (NLP). The common technique is to use classical models with feature extraction techniques for complicated sequence identification. Features can include word frequencies, part-of-speech tags, n-grams, TF-IDF, etc. %These feature vectors can then be applied to elements in traditional machine learning algorithms like support vector machines (SVM) or logistic regression for classification.

However, the feature-based method often loses out when dealing with all the nested dependencies and long-range relationships among sequences of data points. To overcome shortcomings like these, researchers turned their eyes back onto deep learning techniques like RNNs and their offshoots - LSTMs (long short-term memory networks) and GRUs (gated recurrent units). These models are designed specifically for sequential data and can learn dependencies that stretch over long enough runs \citep{bengio1994learning, hochreiter1997long, chung2014empirical, lipton2015critical}.

Recalling that RNNs operate one element at a time and keep a state with a specific size that holds the information about what was seen before in the sequence to build up an impression of the semantic relationships of complex inputs. LSTMs and GRUs enhanced the RNNs capabilities by adding mechanisms for controlling the flow of information and dodging problems that come up in ordinary RNNs, such as vanishing and exploding gradients \citep{bengio1994learning, hochreiter1997long}.

As stated by \citep{Phi2018}, Long Short-Term Memory (LSTM) and Gated Recurrent Units (GRUs) are advanced types of recurrent neural networks (RNNs) specially designed to capture long-range dependencies. These models have proven particularly effective at handling sequential data; they are well-suited to tasks in natural language processing, time series prediction and other situations involving ordered data. This dissertation will explore these models in depth in the following two subsections \ref{LSTM} \& \ref{GRU}.

\begin{figure}[!h]
    \centering
    \includegraphics[scale=0.5]{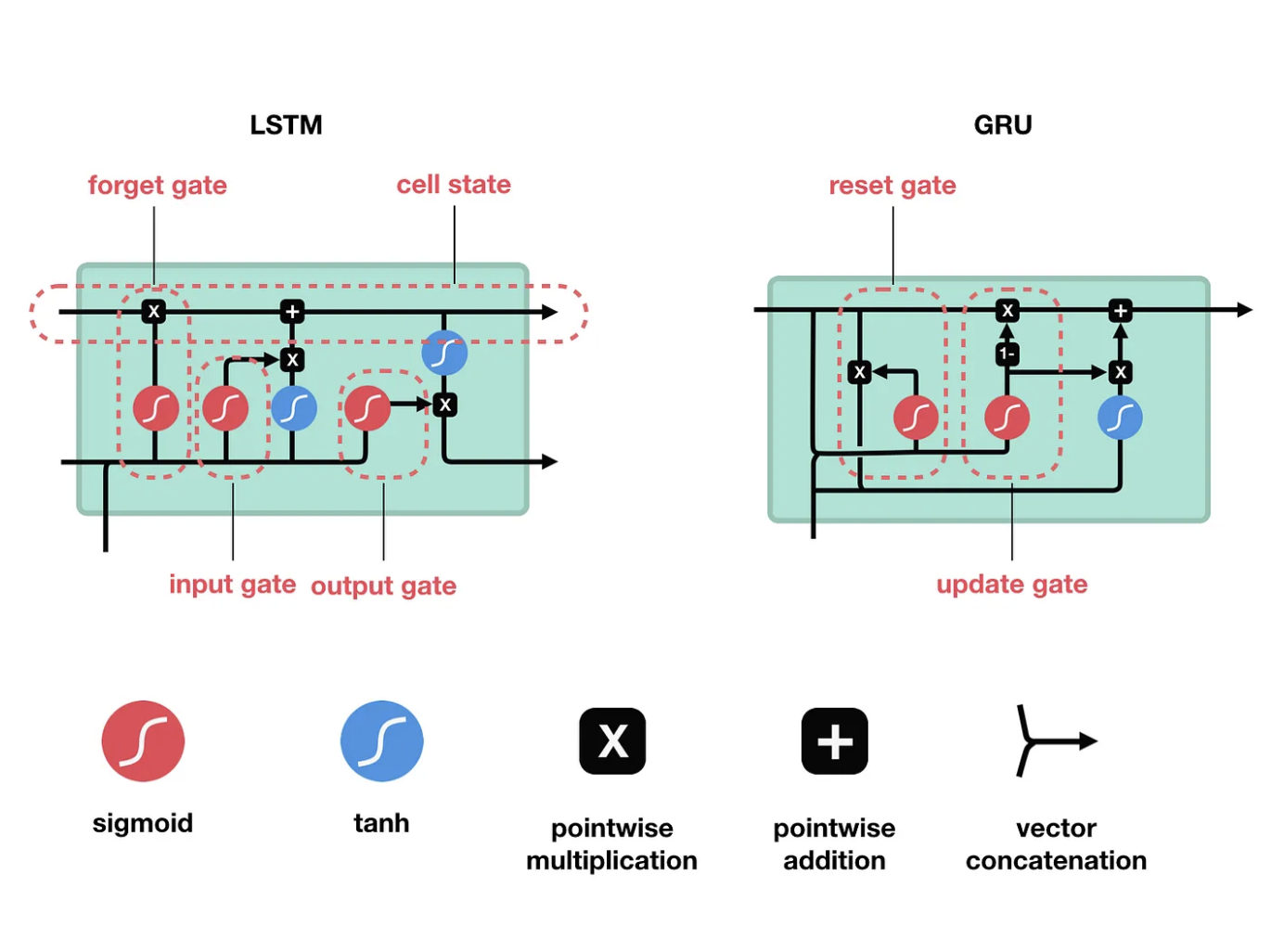}
    \caption{LSTM \& GRU  \citep{Phi2018}}
    \label{fig:LSTMANDGRU}
\end{figure}

\subsection{LSTM}
\label{LSTM}
Unlike traditional RNNs that become unable to retain information over long sequences, Long-Short Term Memory (LSTM) models were devised to rectify this problem, as stated by \citep{Phi2018}. The introduction of the memory cell, allowing information to be stored over long periods, constitutes their main innovation. These memory cells are governed by three different types of gate: the input gate, the forget gate, and the output gate.

\hfill\\ Figure \ref{fig:LSTMANDGRU} explains the LSTM:
\begin{enumerate}
    \item Input Gate: How much new information from the current input should be added to the memory cell is governed by the input gate. It chooses which values to update, which to nearly leave unchanged from before, and which to disregard on the input and previous hidden states.
    
    \item Output Gate: The output gate tells future hidden states and ultimately the output what information from the memory cell to pass. It does this by designing to make sure that only important bits of information are utilised in predictions or transferred up to a higher layer.
    
\end{enumerate}
All of these gates work together to permit LSTM networks to remember and update over extended sequences. In this way, they can satisfactorily retain long-range dependencies without suffering from vanishing or exploding gradients.

\subsection{GRU}
\label{GRU}
According to \citep{Phi2018}, GRU are a streamlined version of LSTMs that also seek to solve the problems raised by ordinary RNNs. This simplified architecture, however, combines functions found in the input and forget gates of an LSTM network into one update gate. Meanwhile, GRU also gets rid of an output gate. This turn of events has resulted in fewer parameters and faster training times, as well as maintaining the network's ability to capture long-range dependency.

\hfill\\ Figure \ref{fig:LSTMANDGRU} explains the GRU:
\begin{enumerate}
    \item Update Gate: The update gate decides how much past information needs to be passed from the memory. It serves a role that is similar to and replaces the combined work done by input and forget gates in LSTMs, so it makes this whole process more efficient by needing only one gate instead of two.

    \item The reset gate is responsible for deciding how much information from the past should be discarded. When the reset gate is approximately equal to zero, it encourages the model to discard the previous state entirely. It allows to selectively clear away irrelevant sections of the sequence.
    
\end{enumerate}
GRUs can simplify the architecture by using these two gates while maintaining the ability to manage long-range dependencies. Therefore, GRUs could perform as well as LSTMs in many applications but with less computational complexity due to both reset and update gates. \\

LSTM and the GRU models are well-suited for identifying the complex, time-dependent patterns typical of IoT-based cyber risks due to their capacity to capture long-term dependencies in sequential data. The research attempts to improve the precision and dependability of SUF-based security measures by using the GRU model to analyse HPC data, thus offering more sophisticated and potent protection against IoT security breaches.

%I need to edit this. so I can be more clear. and add more 
\section{Summary}
This chapter elaborates on the security landscape of IoT devices, highlighting the major threats such as botnets, ransomware, firmware attacks, and DDoS attacks. It further discusses the challenges of traditional signature-based and AI-based security measures. This chapter also introduces Hardware Performance Counters (HPCs) as an effective tool for enhancing IoT security, laying the base for Software Unclonable Functions (SUFs). The utilisation of HPCs, along with sequence classification models such as LSTM and GRU, is a main method for detecting time-dependent patterns, subtle in software behaviour, that can show security threats. \\

Having established the need for advanced security measures and introduced HPCs and SUFs, the literature review in the following chapter will explore existing research and methodologies related to these concepts. This review will identify gaps in current knowledge and guide the dissertation's focus towards innovative solutions for IoT security.

%% file: chapters/Rel.tex
The literature on the use of HPCs for behavioural malware detection is large and diverse. Here, we present a general overview of the different methodologies, algorithms, and applications where HPCs are used to improve or harden the security of the devices. In order to make our review more comprehensive and informative, we broadly classify the studies into five subsections – \ref{relatedwork:recognition} Function Recognition and Reverse Engineering, \ref{relatedwork:anomaly} Anomaly Detection Using HPCs, \ref{relatedwork:firmwareintegrity} Firmware Integrity Monitoring, \ref{relatedwork:timeseries} Time Series-Based Classifiers Using HPC, and \ref{relatedwork:machinelearning} Machine Learning Classifiers for Malware Detection Using HPCs. By the end of this chapter, we will conclude the literature review by pointing out the identified gaps. \\ 

\section{Function Recognition and Reverse Engineering} 
\label{relatedwork:recognition}
The study by \citep{Shepherd2022} explores the use of  HPCs to identify some function's signatures in a black-box fashion. The proposed approach aims at easing reverse engineering, cryptographic function identification, and discovering security bugs in compiled libraries and trusted execution environments (TEEs). Using these counters, the authors developed a supervised machine-learning approach for classifying hardware events. They were able to apply it to a variety of CPU architectures, including Intel/X86-64, ARM, and RISC-V, using the MiBench benchmarking suite, as well as cryptographic libraries such as WolfSSL, Intel’s Tinycrypt, Monocypher, and LibTomCrypt. Measurements were taken before and after the invocation of each function to train a classifier (e.g., naïve Bayes, logistic regression, decision trees), simulating a compiler optimisation that the attacker does not know.

%They conducted experiments in Dell Latitude 7410 with Intel (R) i5-10310U CPU @ 1.70GHz (X86-64), Raspberry Pi 3B with Broadcom BCM2837 SoC (ARM Cortex-A53 CPU), and another recently released board, SiFive HiFive Rev. B with FE310-G002 SoC (RISC-V). The HPC measurements were taken using the PAPI library for both X86-64 and ARM devices. For the RISC-V measurements, they used the assembly instructions.

For the experimental setting, they collected HPC events for 64 functions (including cryptographic and non-cryptographic functions) from the three devices and divided the dataset into training 80\% and testing 20\% sets. Features were first normalised using a Z-score, and then classifiers (such as naïve Bayes, logistic regression, decision trees, and random forests) were trained in environments that simulate situations where all the compiler optimisations are known (under O0, Os and O3).

The results showed high classification accuracies, particularly with privileged mode counters: 97.33\%–99.70\% on X86-64, 90.68\%–96.81\% on ARM, and 86.22\%–93.34\% on RISC-V. Mixed compilation parameters caused some accuracy degradation but still maintained reasonable effectiveness. Tree-based models and ensembles, especially random forests, performed best.  This study is particularly relevant to this work, as the study's scope was identifying specific functions, a key aspect of functions and libraries identification was introduced. Their work was limited to particular cryptographic functions and did not explore the broader generalisation of the approach to other functions. However, this work aims to extend this by focusing on specific malicious functions and exploring the potential of SUFs to make a signature for detecting malicious behaviour in IoT devices. We consider introducing such an approach to identify some libraries as an initial entry to the SUFs concept that we are targeting. \\ 

%Despite these promising results, the study's scope was limited to specific cryptographic functions, and it did not explore the broader generalization of the approach to other functions or device behaviours. This limitation is significant to our research, which aims to extend this approach by focusing on specific malicious functions and exploring the potential of Software Unclonable Functions (SUFs) to create signatures for detecting malicious activity in IoT devices.

%This study is particularly relevant to our research, as it demonstrates the potential of HPCs in identifying specific functions, which is a key aspect of Software Unclonable Functions (SUFs) in IoT security. However, this work aims to extend this by focusing on specific malicious functions and exploring the potential of SUFs to make a signature for them to detect anomalies within the same device behaviour. 

% Critical Analysis 

\citeauthor{basu2020theoretical} \cite{basu2020theoretical} offers a new analytical framework for evaluating the security efficacy of HPC-based malware detection methods and identifies the technical limitations of such methods. The paper uses control flow graphs (CFGs) to model how programs behave and mathematical modelling to calculate the probability that malware will be detected by taking readings of HPC during the program runtime at different time points. The paper finds that different types of malware could produce the same HPC counts, making them harder to distinguish.

The experimental method is to see if the theoretical model holds for a particular test platform by running CHStone benchmarks (a set of benchmarking programs that capture the characteristics of typical data-intensive programs) on it. The experiments were conducted in two steps. Firstly, they derived CFGs for each of the benchmarks. Secondly, they enumerated the number of paths for the CFGs created in the first step and measured the probability of matching HPC readings between two different paths, given a fixed number of cycles. They have also tested the impacts of varying HPC sampling intervals and different HPC parameters.

The results of the experiments reflected the theoretic predictions. The probability of matching the HPC readings between various paths was reflected to decrease significantly, with more HPCs evaluated concurrently for the data-intensive programs. It was also verified experimentally that by raising the number of HPCs and reducing the sampling interval, the probability of the malware escaping detection became smaller. The efficiency of HPC-based malware detection was effectively improved by the proposed framework effectively

%This study's emphasis on the limitations of HPC-based detection methods provides a crucial foundation for understanding the inherent challenges in using HPCs for security. It guides the direction of our research toward overcoming these limitations by focusing on SUFs.\\ 

\section{Anomaly Detection Using HPCs} 
\label{relatedwork:anomaly}

\citeauthor{Garcia-Serrano2015} \cite{Garcia-Serrano2015} focuses on utilising HPCs to identify anomalies in program execution that may indicate the presence of malware.
The experiments involved a controlled environment using a modified version of a simple web server called nweb. This web server was chosen for its simplicity and was specifically modified to be vulnerable to common attack techniques like stack overflow and Return Oriented Programming (ROP). Three potential methods were evaluated to measure HPCs during these attacks: instrumenting the source code with libraries like PAPI, developing a kernel module for periodic measurements, and using the Linux perf utility to measure HPCs during these attacks. The perf utility was selected due to its ease of use and effectiveness.

In the experimental setup, the researchers collected HPC data at different intervals (1ms, 10ms, and 100ms) to determine the optimal frequency for anomaly detection. They simulated normal web browsing activity followed by an attack to capture the relevant HPC data. The study identified the key HPC events, iTLB-load-misses, dTLB-loads, bus cycles, LLC-store-misses, LLC-loads, and LLC-load-misses as effective malware indicators. These events were selected based on their ability to show significant deviations during an attack compared to normal execution.

The Local Outlier Factor (LOF) method was employed to detect anomalies, which assesses the local density deviation of data points. The results showed that a 100ms interval for measuring HPC events was most effective for real-time analysis, as higher frequencies did not improve detection and sometimes even reduced accuracy. The study successfully demonstrated that by calculating an attack factor from the average LOF value of the selected HPCs, ongoing malware attacks could be detected with a low false positive rate. 

As per the findings, measuring HPC events at a 100 ms interval worked best for real-time analysis, with higher frequencies not helping with identification and occasionally even decreasing accuracy. The research effectively showed that a low false positive rate could be achieved in detecting malware attacks by computing an attack factor based on the average LOF value of the chosen HPCs. According to these results, these techniques can successfully detect threats without the requirement for substantial previous training or further statistical adjustments, opening the door for cybersecurity solutions that are more flexible as well as durable. The authors gave evidence of the approach's efficacy, but they provided no information regarding the online training procedure that ought to be customised for every device prior to creating anomaly detection.\\

%critical analysis 

%The experiments were conducted in a highly controlled environment, which may not accurately reflect the complexities and variability of real-world systems

The study by \citeauthor{krishnamurthy2020anomaly} \citep{krishnamurthy2020anomaly} explores how HPCs might be used to enhance embedded cybersecurity with a focus on programmable logic controllers (PLCs). By using HPCs to profile routine processes and looking for variances that could be signs of anomalies or assaults, the authors develop a baseline performance pattern. The process uses the black-box strategy to gather HPC data without changing the original code. The one-class Support Vector Machine (SVM) is trained using these data as a time series to differentiate between normal and aberrant conduct. Because of this, the method is easy to implement on a range of devices and reasonably priced. The major platform was the Wago PLC, and extra validation was supplied by a Raspberry Pi running the OpenPLC system. The approach achieved high accuracy in detecting different types of code alterations and parameter changes. When applied in a deterministic environment like an RTOS (real-time operating system), the experimental results indicate a potential future for HPC in anomaly detection. Even though the authors were able to get this in their experimental setup, real-world situations might present more crucial challenges, mainly in settings with limited resources. It is essential to address the performance concerns in order to provide realistic practicals. Furthermore, the author did not provide any study to confirm the seamless integration of the approach with the right priorities into a preemptively scheduled environment. \\

%ott2018hardware
%Title: Hardware Performance Counters for Embedded Software Anomaly Detection

Another work on anomaly detection by \citep{ott2018hardware} presents two methods utilising HPCs to identify anomalies in embedded software, primarily focusing on runtime integrity validation and safeguarding against malicious intrusions. The authors developed models based on Hidden Markov Models (HMMs) and Long Short-Term Memory (LSTM) neural networks.

The experiments were conducted using the CoremarkPro benchmark suite on an Intel Core processor running Linux. The data collection setup ensured minimal performance overhead and accurate HPC data. The methodology employs HPCs to collect performance data of software execution without modifying the source code, forming a time series representation of the program's behaviour. This data is then used to train HMMs and LSTMs. The HMMs use a probabilistic approach, calculating the probabilities of observation sequences to detect anomalies, while the LSTMs predict the next value in a sequence, flagging deviations as anomalies. This black-box approach utilises existing HPCs in modern processors, ensuring that software protection does not require any changes at the source or binary level.

The classification achieved 100\% accuracy for both LSTM and HMMs in offline anomaly detection and 95\% and 98\% accuracy for online detection for HMMs and LSTM, respectively. While these proposed methods show high accuracy and low false positive rates, the study is centred on particular Intel processors and HPC events. Additionally, the unpredictable nature of HPC measurements, influenced by factors such as OS behaviour and multi-threading, could potentially affect detection accuracy. Although the experimental results are optimistic, the testbed uses an Intel CPU i7 for the experiments, and the data collection was performed on an architecture that is not very common in an embedded software context. \\

%fix this 
%The outliers are extracted compared to the majority behaviour.
The work in \citep{Bourdon2020},  describes a lightweight anomaly detection approach for massively deployed smart industrial devices based on HPCs. The approach monitors low-level processor events that are collected through HPCs and then compares the HPC values across devices to detect abnormalities. Specifically, the method monitors the behaviour of many devices without the need to model the legitimate behaviour of the applications running on the devices. Since HPC values are collected at a low level, the behaviour of any application would display some sort of abnormality whenever malicious activity is present. The proposed method performs anomaly detection based on the presence of outliers in the HPC measurements. The authors designed the proposed approach to be computationally lightweighted, enabling massively deployed scenarios. 

Their setup consisted of 100 Raspberry Pi 3B devices to simulate an industrial environment with different software profiles and attack scenarios. %By monitored the activities of the target processor using six software counters: four L1 cache access counters measuring different components of the cache, one instruction execution counter, and one branch prediction counter. The results from these counters (primarily the cache access counters) were normalised and fed as input to multiple outlier detection algorithms. Using the results, the researchers extrapolated the data to a set of 10,000 devices to simulate a large-scale deployment.
The proposed method achieved extremely high true positive rates and very low false positive rates while detecting outliers. It was compared with seven other outlier detection algorithms, and although some of them performed better for specific scenarios, the overall result proved the robustness of the approach for a wide range of software profiles and attack modalities. The true positive rate remained high even for small fractions of infected devices (as low as 1\%), while the false positive rate was very low. The proposed method was also very scalable, as it could handle large numbers of devices in good time. While the authors demonstrated the effectiveness of their approach to some extent, the authors did not consider any of the odd scenarios where one or multiple devices are in sleep mode or performing some software updates which could debilitate the whole algorithm functioning.

The paper by \citeauthor{ElBouazzati2023} \citep{ElBouazzati2023} introduces a host-based intrusion detection system (HIDS) against IoT memory corruption attacks using HPCs for real-time detection. Their approach begins with a threaded model where the victim is an IoT SoC that includes a CPU for user applications and a network processor for wireless connectivity, and the attacker is remote and capable of performing attacks such as DoS, MITM, and RCE. They generate datasets using a simulation testbed to train and build machine learning models, reducing the cost and complexity of obtaining real-world data. They monitored specific hardware events during network packet processing using the HPMtracker, which collected cumulative HPC values at the end of processing each network packet. These collected HPC data are stored locally every 5 seconds and then sent to the detection module every 30 minutes for analysis, balancing the need for timely data analysis with resource constraints.

The HIDS has implemented Field Programmable Gate Arrays (FPGA), enabling real-time detection of stack and heap overflow attacks with a detection accuracy of 99.98\%. The detection process involved using a decision tree machine learning classifier trained on the collected HPC data to differentiate between legitimate and malicious activities. The dataset used for training consisted of 3,000,000 samples, covering scenarios like stack and heap overflow. The system was tested on a RISC-V ISA, specifically utilising the CV32E41P soft-core as the network processor for wireless connectivity.  \\

\section{Firmware Integrity Monitoring}
\label{relatedwork:firmwareintegrity}
Confirm suggested by \citeauthor{Confirm} \citep{Confirm}, A novel, inexpensive method to identify malicious changes in embedded systems' firmware by quantifying the occurrence of low-level hardware events during operation. The study shows that this approach can be applied on ARM and PowerPC platforms with minimal performance overhead. and demonstrate this with case studies on ARM-based as well as PowerPC-based platforms. The method successfully detects such modifications in both cases, sometimes even with minimal deviations.  
Confirm takes advantage of HPCs present in many embedded processors and counts low-level hardware events like instructions loads, branches, and memory-related operations during firmware execution. The paper states that monitoring these events can indeed be used to identify a firmware's fingerprint, which allows the detection of deviations that stem from tampering.
%ConFirm aims to collect information on the hardware events generated by the device in two phases: an offline profiling phase and an online checking phase. In the offline profiling phase, the clean firmware is executed on the device, and the hardware events generated are collected and stored to build the corresponding HPC signature. This signature is then stored in the database. In the online checking phase, the transient hardware events are measured and monitored in real-time during the firmware’s execution, and the collected HPC data will be compared with the reference signatures to determine the deviations. 
Experimental results show that ConFirm detects all tested modifications with minimal overhead performance on ARM and PowerPC based processors. For instance, in the case of the wireless access point firmware, ConFirm detects denial-of-service attacks with a minimal deviation of 8.7\%, proving the approach's efficiency and practicality. Furthermore, the performance overhead of ConFirm is low enough to make the approach feasible in an embedded environment where computational resources are constrained. The researchers plan to investigate whether machine learning techniques can decrease the noise threshold, thus strengthening the detection capability of ConFirm while maintaining a low memory and performance footprint. 

Building upon these findings, they addressed Machine learning techniques in \cite{wang2016malicious} as planned on \cite{Confirm} to automatically mine relations among different hardware events and thus significantly reduce the number of pre-stored valid HPC signatures without sacrificing the detection accuracy. ConFirm uses One-Class Support Vector Machine (OC-SVM) classifier. This unsupervised learning technique benefits anomaly-based detection because it is trained solely with HPC measurements from clean executions. The OC-SVM classifier creates a model of the normal behaviour based on the measurements and then classifies test data as similar to or different from the training set. The classifier uses a non-linear Radial Basis Function (RBF) kernel to map data points to a higher-dimensional space where they can be linearly separated, thus enhancing detection precision. \\

\section{Time Series-Based Classifiers using HPC}
\label{relatedwork:timeseries}

A paper written by \citep{TimeSeries} explores the use of time series-based classifiers (TSCs) in combination with Hardware Performance Counters (HPCs) for malware detection. It enhances accuracy and lowers false positives, making it particularly helpful for devices with constrained profiling capabilities. The sequential order of data is frequently ignored by traditional HPC-based malware detection techniques, which results in a high false-positive rate and the misclassification of benign applications as malware. 
In order to reduce profiling overhead, the authors devised a Sequential Time Series-based Detection (SEQ-TSD) architecture that makes use of the single HPC. Using a majority voting mechanism, they classified the data after training numerous classifiers, each on the different time series. Every classifier had a limit assigned to it, and the classifiers' majority vote determined the main categorisation.

The authors trained various forest classifiers and Time Series Forest (TSF) models to assess performance enhancements utilising various time series lengths and different HPCs. The experimental results showed that the time series-based classification mainly improved accuracy and reduced false positives compared to traditional HPC-based methods. The experimental results demonstrated that time series-based classification significantly improved accuracy and reduced false positives compared to traditional HPC-based methods. The SEQ-TSD framework enhanced the average accuracy to 90.41\% (ARM) and 92.5\% (Intel) with the single HPC, reflecting an extremely low false positive rate. Also, combining the multiple HPCs within the SEQ-TSD method enhanced accuracy to 97.36\% (ARM) and 97.5\% (Intel), majorly outperforming traditional techniques, thereby providing a reliable solution for detecting malware on the restricted devices and improving the system security. \\

%sayadi2020stealthminer
%Stealthminer: Specialized time series machine learning for run-time stealthy malware detection based on microarchitectural features},
\citeauthor{sayadi2020stealthminer} \citep{sayadi2020stealthminer} proposes a novel time-series machine learning-based methodology for detecting stealthy malware based on microarchitectural features. Common Hardware-assisted malware Detection (HMD) methods fail to detect embedded malware that hides within benign applications. StealthMiner resolves such an issue by applying a new time series ML approach that mines only one HPC feature, namely branch instructions. This feature detects malware embedded in HPC traces with an average detection accuracy of 94\%, 42\% better than state-of-the-art methods. The document visualises the challenge of detecting stealthy malware using the t-SNE algorithm, demonstrating how traditional methods struggle to differentiate between malware and benign data when malware is embedded.

It was implemented on a machine with an Intel Xeon X5550 processor, on which they executed more than 3,500 benign and malware samples of backdoor, rootkit, and Trojan types. They used Linux Containers (LXC) to avoid any contamination of the results and have clean HPC data. They selected the important features to train the ML classifier: branch instructions, cache references, branch misses, etc.

StealthMiner processes time series data from HPC systems using the lightweight Fully Convolutional Neural Network (FCN) model. The architecture consists of a fully connected neural network, a global average pooling layer, and two 1-D convolution layers. At runtime, the architecture recognises polluted intervals in the HPC-based time series, enabling it to automatically detect the incorporated malware.

StealthMiner's performance was evaluated using F-score, detection accuracy, precision, and recall. The findings showed strong detection capability with an average Area Under the Curve (AUC) of 0.94, demonstrating good detection performance across various implanted malware variants. Compared to more conventional machine learning techniques, StealthMiner outperformed them, successfully tackling the issue of identifying covert malware nested inside legitimate apps. StealthMiner is a potential solution for actual applications as it greatly increases detection performance by concentrating on essential HPC features and using a time-series machine learning approach. \\

\section{ML Classifiers for Malware Detection}
\label{relatedwork:machinelearning}

%Abraham 
%kuruvila2020analyzing
%Analyzing the Efficiency of Machine Learning Classifiers in Hardware-Based Malware Detectors
\citeauthor{kuruvila2020analyzing} \citep{kuruvila2020analyzing} Examines the efficiency of various machine learning classifiers in detecting malware, including Neural Networks, Support Vector Machines (SVM), K-Nearest Neighbors (KNN), Logistic Regression, Naive Bayes, Decision Trees, and Random Forests. The method they utilised involved collecting the HPC data during the execution of both benign and malicious programs. In order to assess the performance of the classifiers the parameters were adjusted to optimise the detection metrics. They conducted their experiments on a dataset collected from a Raspberry Pi 3 Model B, involving 300 benign and 300 malware programs. The Random Forest classifier outperforms other classifiers, achieving a detection accuracy of 83.04\% at a max depth of 15, allowing the model to capture more detailed patterns in the data, increasing its complexity and potentially improving accuracy. It's important to remember that all classifiers must have their parameters carefully adjusted to achieve the best possible performance. The parameters hypertuning in some cases could be related to the used dataset which has not been highlighted effectively in this work.

Building upon this foundation, \citeauthor{kuruvila2020defending} \cite{kuruvila2020defending} has also addressed the vulnerability of these machine learning-based malware detectors to adversarial attacks. Adversarial attacks introduce perturbations in the HPC traces to misclassify programs. A Moving Target Defense (MTD) mechanism has been proposed in response. The MTD enhances security by dynamically altering the set of HPCs used and the specific classifiers employed, making it extremely difficult for attackers to successfully reverse-engineer the system and manipulate the detection process. 
Their experimental setup trains machine learning classifiers – Decision Trees (DT) and Neural Networks (NN) – on HPC data from a Raspberry Pi 3 Model B. The authors applied adversarial attacks while monitoring the HPC traces to manipulate the HMD. They also test the impact of the MTD as a mitigation technique. The main finding of this work points out that while adversarial attacks reduced the precision and accuracy of the classifiers to 50\%, the MTD significantly recovered these metrics to over 70\%. The MTD robustly maintained malware detection performance under adversarial conditions. The work also pointed out that using two classifiers provided the best balance between defence effectiveness and computational overhead. This approach improves the classification accuracy of adversarially robust HPC classifiers by up to 31.5\%, restoring nearly perfect accuracy (99.4\%) of the original detection model. \\

%%%%%%%%%%%%%%%%%%%%%%%%%%%%%%%%%%%%%%%%%%%
% Left to revise
\citeauthor{Sayadi2018} \cite{Sayadi2018} describes how ensemble learning techniques, including boosting and bagging, can be employed to circumvent the limitations arising due to the availability of a restricted number of HPCs in modern processors such as the Intel Xeon X5550 used in this study, and improve the effectiveness of hardware-based malware detection using HPCs. Methodology: HPC data from numerous benign and malware applications collected on an Intel Xeon X5550 processor are parsed and stored in CSV files. The most informative feature combinations are found via correlation analysis. Eight general machine learning classifiers are trained, including J48, BayesNet, IBk, ZeroR, RandomForest, Logistic, SMO and REPTree. Bagging and boosting ensemble learning techniques were used to combine the results of all the classifiers and make them more accurate and robust. The empirical study showed that ensemble techniques significantly improved general machine learning classifiers' detection accuracy and robustness, even when the number of HPCs is reduced. Specifically, the Boosted REPTree classifier achieved the best detection accuracy with 88\%. While bagging and boosting are generally used to cope with the unbalanced dataset, their impact on classification is not well justified through this work. % and 2 HPCs selected using the feature selection approach, which is comparable to models that typically require 16 HPCs. \\ 
 
%Fine-Grained Malware Detection using HPC
%Kadiyala2020

\citeauthor{Kadiyala2020}  \citep{Kadiyala2020} works present a novel technique for identifying malware using HPC. Their method leverages the dynamic behaviour of programs, potentially outperforming conventional static analysis techniques that clever coding methods can circumvent. By focusing on micro-architectural events, the proposed method can detect malware that employs dead code insertion, register reassignment, and code transportation techniques. Static analysis often misses these techniques, but the dynamic nature of HPCs captures them. Their experiments use an ALTER SoCit with a 650 MHz dual-core Cortex-A9 processor. In their experiments, they collected data on a Linux-based embedded OS, by extracting HPC data for each system call of an unknown program. They handled the large volume of HPC data using fine-grained extractions, an alternative approach to extracting HPCs for system calls during runtime that provides a more detailed view of program execution. They perform dimensionality reduction with Principal Component Analysis (PCA) to reduce the dimensions of each column of the feature matrix to a smaller dimension, which helped in identifying the components that have the maximum variance and reduced the data to a manageable size without losing significant information. The test set included a malware sample from OpenMalware and 293 benign programs from public databases such as MiBench, SPEC CPU 2017, and EEMBC. This work's primary machine learning model was the Random Forest classifier. The model was selected due to its ability to generalise effectively and perform well in new, unknown situations. This classifier achieved a 98.4\% detection rate with a 3.1\% false positive rate, demonstrating its high effectiveness. This methodology could be vulnerable to adversarial attacks that manipulate hardware counters, such as those leveraging excessive page faults to bring HPC counts beyond defined thresholds. Also, the advanced features, such as system calls, are susceptible to manipulation.\\

%early detection of ransomware 
%Anand2023

\citeauthor{Anand2023} \cite{Anand2023}focused on detecting ransomware early using HPCs. They performed tests using an Intel Core i7-9700 CPU, which can handle up to ten HPC registers. They utilised the Linux tool 'Perf' to collect HPC data. The Boruta algorithm was used to analyse the data and identify the most important HPC registers for ransomware detection. The tests were carried out in a Windows 7 sandbox environment to securely run ransomware and benign samples. HPC statistics were recorded at different time intervals (100ms, 500ms, 3s, 5s) to find the best timeframe for early detection.

The data collection involved executing 183 ransomware samples from 27 different families and an equal number of benign samples obtained from software informed. Then they used machine learning models, including Support Vector Machines (SVM), Random Forest (RF), Gradient Boosting Machine (GBM), and AdaBoost, were used to analyse the HPC data. Adaboost got the best result and had an accuracy above 90\% while capturing the HPC data at 100ms intervals for a total duration of around 3s. The study found shorter, more frequent data captures yielded better results than longer, less frequent captures. \\

%Deep Neural Network and Transfer Learning for Accurate Hardware-Based Zero-Day Malware Detection
%He2022 

\citeauthor{He2022} \cite{He2022} They suggested the use of the Deep-HMD framework to improve the detection of previously unknown malware by leveraging HPCs with deep neural networks (DNN) and transfer learning. They used Mutual Information to identify important HPC features and transformed them into 2D images to be fed into a ResNet18-based DNN, which was adjusted using pre-existing ImageNet weights. They gathered data from more than 5,000 benign and malware applications, obtaining benign applications from real-world software and malware samples from VirusShare and VirusTotal. The profiling was conducted on an Intel Xeon X5550 machine. The Deep-HMD framework operates in two main stages. The first stage involves encoding each row of four HPC tabular data points into an image. All data rows are initially normalised using Scikit Learn’s standard scaler, which standardises the data by removing the mean and scaling to unit variance. Then, using the OpenCV library, these four numeric data points are projected into a 256 x 256 x 3 resolution image, ensuring equal spacing and no overlapping. In the second stage, the generated image data is used to train a robust DNN model for zero-day malware detection. 
The proposed model demonstrated that this new method is much more effective than traditional machine learning classifiers. It achieved an impressive 97\% detection rate with high accuracy, precision, recall, and a low false positive rate. \\ 

\citeauthor{He2023Image} \cite{He2023Image} The paper proposes a new hybrid and adaptive framework called DRL-HMD, which leverages HPC data and advanced deep learning and deep reinforcement learning algorithms to detect zero-day malware on IoMT devices. The proposed method converts the HPCs data to images and trains the Deep Neural Network (DNN) models using these images. An Advantage Actor-Critic (A2C)-based deep reinforcement learning agent is used to dynamically select the best DNN model at runtime, depending on the training data, which is frequently changing. The selection is based on the F1-score and AUC to optimise the detection performance.

The main goal of this research was to improve the detection of zero-day malware on IoMT devices. Traditional software-based detection methods monitor the suspicious behaviours of software programs, but they often suffer from a high false-positive rate. Moreover, those methods cannot easily detect zero-day malware, which essentially are unknown malware variants, as they are not present in the blacklist databases. Their experiment used data from Intel Xeon processors, simulating an IoMT environment with benign and malicious software data streams. The framework tested both known and zero-day malware. The model proved to be able to improve adaptiveness.

The experiment's results reflect a 99\% detection rate with extremely low false positive (0.01\%) and false negative (1\%) rates compared with traditional machine learning models and standalone DNNs. This demonstrates the potential of the proposed DRL-HMD framework to provide a very high real-time detection rate of malware, which is critical for securing health systems from zero-day attacks.\\

\section{Summary}
The literature review presented in Table \ref{table:summryrelatedwork} lists the significant HPCs use cases related to detecting malicious behaviours in cybersecurity research. Various methodologies, algorithms, and applications demonstrate the versatility and effectiveness of HPCs in identifying functions of black-box programs, detecting anomalies, ensuring firmware integrity, and improving malware detection using machine learning classifiers.

Function recognition and reverse engineering, as explored by \cite{Shepherd2022}, show high classification accuracies across multiple CPU architectures, enabling practical applications in vulnerability detection and cryptographic functions. Anomaly detection studies, such as those by Garcia-\cite{Garcia-Serrano2015} and \cite{krishnamurthy2020anomaly}, illustrate the effectiveness of HPCs in identifying deviations from normal behaviour with minimal false positives. \cite{ott2018hardware} and \cite{Bourdon2020} further reinforce these findings by employing advanced models like HMMs and LSTM neural networks, despite challenges related to hardware diversity and environmental noise.

In the domain of firmware integrity monitoring, the  \cite{Confirm} approach, enhanced by \cite{wang2016malicious}, showcases the practical applications of HPCs on ARM and PowerPC platforms with minimal performance overhead. The review also highlights the use of machine learning classifiers for malware detection, as demonstrated by \cite{kuruvila2020analyzing}, \cite{kuruvila2020defending}, \cite{Anand2023}, and \cite{He2022}, where techniques like Random Forest and deep neural networks achieve high detection rates and low false positives. 

The review concludes with studies on time series-based classifiers using HPCs, such as those by \cite{sayadi2020stealthminer}, which achieve high detection accuracy by focusing on key HPC features and utilising a time-series ML approach. This study allowed us to identify various topics that were not covered by the state of the art. 
Our work will address the following research gaps:
\begin{itemize}

    \item One of the significant limitations observed was the absence of a standardised dataset for evaluating HPC-based security approaches. 
    
    \item Existing studies often employ simulated or controlled environments that do not fully capture the complexities and constraints of real-world IoT devices, limiting the practical applicability of the results. 
    
    \item While sequence classification models like LSTMs and GRUs have been proposed, there is limited research evaluating their effectiveness specifically for function recognition within IoT devices. 
    
    \item Current research predominantly focuses on threats that significantly alter the control flow, leaving a gap in addressing subtler malicious intents that operate within the normal flow of functions. 

\end{itemize}

To address these gaps, this dissertation has been designed with the following considerations: 

\begin{itemize}
    \item Building a custom Dataset: Due to not having an available data set public, this study will build a dataset of HPC performance data acquired from IoT devices to be run under various operation conditions.
    
    \item Realistic IoT Environment: The study applies classification methods aided by HPC on a realistic IoT environment, utilising real devices like Raspberry Pi and Turris Omnia Router systems, to ensure that the results can be directly transferable to a real-world scenario.
    
    \item Evaluating Sequence-Classification Models: This model will be evaluated across the generated custom dataset, particularly the performance of sequence classification models such as the GRU in recognising and classifying functions in IoT devices.
    
    \item Addressing Subtle Malicious Intents: The research adapts the new approach of HPC data to detect much more subtle threats that keep the control flow graph unchanged.
\end{itemize}

Having established these gaps, the next chapter will discuss the design of this dissertation, detailing how the study is structured to address these identified gaps and contribute to the advancement of IoT security research.

%Overall, %I need to add something related to my topic and how I can integrate this to my project. 

\input{chapters/RelatedWorkTable}

%% file: chapters/RelatedWorkTable.tex
\begin{table}[!ht]
    \centering
    \tiny
    \begin{tabular}{|p{3cm}|p{1cm}|p{4cm}|p{5cm}|}
    \hline
          \textbf{Category} & \textbf{Study} & \textbf{Focus Area} & \textbf{Key Findings} \\  
          \hline
        Function Recognition and Reverse Engineering & \cite{Shepherd2022} & Use of HPCs to identify functions in black-box programs for reverse engineering, cryptographic function guessing, and vulnerability discovery. & High classification accuracies across various CPU architectures; practical applications in detecting vulnerabilities and cryptographic functions. \\ \hline

        Anomaly Detection Using HPCs & \cite{Garcia-Serrano2015} & Utilisation of HPCs to detect anomalies indicating malware. & Evaluates methods for measuring HPCs and employs the LOF method to detect attacks with low false positive rates. \\ \hline

        ~ & \cite{krishnamurthy2020anomaly} & Use of HPCs for anomaly detection in programmable logic controllers (PLCs). & Utilises a one-class SVM to identify deviations from normal behaviour. \\ \hline

        ~ & \cite{basu2020theoretical} & Proposes a mathematical framework to assess the security provided by HPC-based malware detection techniques. & Theoretical analysis and experiments using CHStone benchmarks demonstrate that monitoring multiple HPC parameters significantly reduces the likelihood of undetected malware. The study confirmed that increasing the number of monitored HPCs and reducing the sampling intervals enhances detection accuracy, even in data-intensive programs. \\ \hline

        ~ & \cite{ott2018hardware} & Models based on HMMs and LSTM neural networks to identify software anomalies using HPCs. & Achieved 100\% accuracy for offline detection and high accuracy for online detection centred on Intel processors.\\ \hline

        ~ & \cite{Bourdon2020} & Outlier identification techniques with HPCs to detect compromised devices. & High detection efficiency but highlights challenges related to hardware diversity and environmental noise.\\ \hline

        ~ & \cite{ElBouazzati2023} & Host-based intrusion detection system (HIDS) for IoT memory corruption attacks using HPCs. & Demonstrates high detection accuracy with minimal resource usage. \\ \hline

        Firmware Integrity Monitoring & \cite{Confirm} & Inexpensive method to detect malicious changes in embedded systems' firmware by monitoring low-level hardware events with HPCs. & Practical applications on ARM and PowerPC platforms; minimal performance overhead. \\ \hline

        ~ & \cite{wang2016malicious} & Enhances ConFirm by integrating machine learning techniques to reduce the number of HPC signatures needed without sacrificing accuracy. & Utilises OC-SVM classifier for anomaly-based detection, enhancing detection precision.  \\ \hline

        Machine Learning Classifiers for Malware Detection using HPCs &  \cite{kuruvila2020analyzing} & Analyses the efficiency of various machine learning classifiers in hardware-based malware detection using HPCs. The study focuses on optimising classifier parameters to achieve the best performance metrics. & Random Forest classifier achieved the highest accuracy (83.04\%) with precision (82.28\%) and recall (83.55\%). Decision Trees were noted for their balance of accuracy and computational efficiency, while Naive Bayes and Logistic Regression performed poorly due to sensitivity to correlated data. \\ \hline

        ~ & \cite{kuruvila2020defending} & Introduces a Moving Target Defense (MTD) mechanism to defend against adversarial attacks on hardware-based malware detectors. The MTD dynamically changes the set of HPCs and classifiers used, making it challenging for attackers to predict and manipulate the detection process. & Experiments on a Raspberry Pi 3 Model B showed that the MTD could restore up to 99.4\% of the original detection accuracy after adversarial attacks, significantly improving the system's resilience to such threats. \\ \hline

        ~ & \cite{Sayadi2018} & Proposes an ensemble learning framework to improve the accuracy of hardware-based malware detectors using fewer HPCs. Boosting and Bagging techniques are applied to various classifiers to achieve high detection accuracy with only 2-4 HPCs, matching the performance of models using 8-16 HPCs. & Experiments using an Intel Xeon X5550 processor showed that the Boosted REPTree classifier achieved 88\% accuracy with only 2 HPCs, significantly improving detection efficiency while reducing the need for a large number of HPCs. \\ \hline

        ~ & \cite{Anand2023} & Focus on early ransomware detection using HPCs. & Demonstrates that frequent short-interval data captures yield better detection results and high accuracy with the AdaBoost classifier. \\ \hline

        ~ & \cite{He2022}& Deep-HMD framework leveraging DNN and transfer learning to improve zero-day malware detection with HPCs. & Achieved a high detection rate with a low false positive rate using the ResNet18-based DNN. \\ \hline
        
        ~ & \cite{He2023Image} & Proposes a hybrid AI-enabled framework (DRL-HMD) for zero-day malware detection in IoMT devices, combining deep learning with deep reinforcement learning. & Achieves a 99\% detection rate (F1-score and AUC) with only 0.01\% false positive rate and 1\% false negative rate. The method dynamically selects the optimal DNN model during run-time, significantly improving detection accuracy for unknown malware variants. \\ \hline
        
        ~ & \cite{Kadiyala2020}& Novel technique for detecting malware using HPCs, focusing on dynamic program behaviour to outperform conventional static analysis techniques. & High detection rate with Random Forest classifier; recognises potential vulnerabilities to adversarial attacks and suggests combining features for improved detection. \\ \hline

        Time Series-Based Classifiers using HPC & \cite{TimeSeries} & Use of time series-based classifiers (TSCs) with HPCs to detect malware. & Significantly improves accuracy and reduces false positives; reliable solution for detecting malware on constrained devices. \\ \hline

        ~ & \cite{sayadi2020stealthminer}& Time series machine learning-based methodology for detecting stealthy malware using HPCs. & Achieved high detection accuracy (94\%) by focusing on key HPC features and using a time-series ML approach, outperforming state-of-the-art methods by 42\%. \\ \hline

    \end{tabular}
    \caption{Literature Review Summary Table}
    \label{table:summryrelatedwork}
\end{table}

%% file: chapters/Design.tex
%3.1 Research Design - 1,500 words
%3.2 Data Collection Methods - 1,000 words
%3.3 Data Analysis Techniques - 1,000 words
%3.4 Ethical Considerations - 500 words
%3.5 Limitations - 500 words
%4.1

%chapter introduction on what im going to explain in this chapter. 

\section{Malicious Function Identification and Design}

We developed distinct binaries so that we could examine and assess the behaviour of different harmful functions. We were able to evaluate and investigate each function separately as well as in conjunction with various other regular functions using this method, which helped us comprehend their importance. We made sure the code was portable and functional across a given range of hardware architectures and operating systems by developing it to run on multiple different kinds of platforms and utilising just standard libraries. We ensured that the mitigation measures based on our discoveries could be deployed successfully to a variety of embedded systems, including IoT devices, by cross-compiling the code onto different platforms. By doing away with platform-specific or proprietary libraries, the adoption of standard libraries also made code easier to use and change. This gives a stable and documented base for the implementation.
%In order to study and analyse the behavior of various malicious functions, we implemented them as discrete binaries. This approach allowed us to test and examine each function in isolation, as well as in combination with normal functions, to understand their impact. By writing the code to run on many different types of systems, and using only standard libraries, we ensured that the code would be portable and functional regardless of the operating system and hardware architecture. Cross-compiling the code onto various platforms enabled us to ensure that mitigation strategies derived from our findings could be effectively applied across different embedded systems, including IoT devices. Using standard libraries also made the code easy to use and modify, eliminating the need for platform-specific or proprietary libraries. This provides a stable and documented base for implementation.

\subsection{Common Malicious Functions in Embedded Systems}

According to the given study in \ref{section:threats}; major harmful behaviours in embedded systems include persistence mechanisms, port scanning, brute-force assaults, command and control (C\&C) communication, and using the services like Telnet. In order to demonstrate several common malicious functions that target the IoT context, we will utilise the Mirai malware as an example in the following. Sources with in-depth information about Mirai's functionality and complexities underlying the operations include \cite{sensorstechforum}, \cite{stratosphereips}, and \cite{threatdown}. 
%Port scanning involves searching for open ports on devices within a network, identifying potential entry points for malicious activities. Brute-force attacks are used to gain access by systematically trying a large number of possible passwords or keys. Once access is gained, C\&C communication enables the malware to receive instructions from remote servers, allowing attackers to control compromised devices. Persistence mechanisms ensure that the malware remains active on the system, often by modifying startup processes or hiding its presence. Exploiting services like Telnet, which is commonly used for remote communication, allows malware to leverage weak or default credentials to gain access.
%Mirai targets IoT devices by scanning for open Telnet ports, then using brute-force attacks to gain access with default or weak credentials. Once compromised, Mirai communicates with command and control servers to receive instructions and execute attacks. Although it doesn't focus heavily on long-term persistence, it can re-infect devices if necessary. Mirai primarily uses these compromised devices to launch large-scale Distributed Denial of Service (DDoS) attacks, severely impacting network performance and device functionality. In the following subsections I will explain each. 

\subsubsection{Port Scanning}
%defind port scanning 
%how miri use this function 
The method for finding the open ports and services on a networked device is called port scanning. To find out which ports are open and mainly running services on them entails sending a series of messages to every port on a device and evaluating the responses. Potential vulnerabilities that could be utilised in order to gain unauthorised access can be found using this information. For instance, port scanning is one of Mirai's main features. Using TCP SYN packets, Mirai attempts to connect to the devices with open Telnet ports (23 and 2323) by methodically searching the network for the devices. Once the open port is found, it will try to do brute-forcing.

\subsubsection{Brute-Forcing }
%defind brute-force ! 
%how miri use this function
Brute-forcing is described as an unauthorised access approach that involves repeatedly trying a huge number of passwords or keys until the right one is found. When a device owner doesn't change their password, this technique frequently takes advantage of weak or default passwords. After locating devices through port scanning, Mirai employs brute force to breach them. Following the discovery of the open Telnet port, Mirai attempts to authenticate using the list of known default credentials. \textbf{Admin/admin} and \textbf{root/root} are two major combinations on this list that are frequently used by default in many IoT devices. If Mirai successfully executes the brute-force attack, the device could be added to the botnet and used for other malicious purposes.

\subsubsection{C\&C Communication}
%defind c2 communication . 
Command and Control (C\&C) communication is the method that is used by the malware in order to maintain communication with the central server. The server issues the instructions and also receives reportsfrom the compromised devices, allowing attackers to manage and coordinate their malicious activities.

The C\&C server is essential to the operation of Mirai. By employing a hardcoded IP address and port, the virus connects to the C\&C server. When Mirai is connected, it waits for the server to give commands. These commands may be to start a DDoS assault, halt the scanning process, or change the list of IP addresses that are targets. In order to keep the botnet operating in unison across infected devices, the malware also sends periodic reports back to the C\&C server with details about successful brute-force efforts and the state of the network.

\subsubsection{Persistence Mechanisms}
%defind Persistence Mechanisms
%defind the use case like mirai how it use this 
Malware makes use of persistence mechanisms to make sure that, even after reboots or attempts to remove it, it stays alive and functions on the infected systems. These defences frequently involve disabling system tools that could interfere with the malware's functioning, disguising its existence, and preventing detection. Mirai uses a number of persistence techniques to make sure it stays active on compromised systems. It prevents the virus from being removed from the memory by disabling the watchdog timer, a system utility that would mainly reset the device if problems were discovered. It does this by renaming the process to look like a legitimate process, like as \texttt{sshd} or \texttt{dropbear}. By checking and terminating any competing instances via a control port, Mirai makes sure that only one instance of itself is operating on the device.
This is enforced through functions like assure single \textbf{instance()} and \textbf{killer\_init()}, terminate other processes that might potentially interfere with Mirai's functionality, like those that are using popular ports like Telnet (23), HTTP (80), and so on. Mirai is a serious threat to IoT devices because of its persistence tactics, efficient communication with the command and control servers, and capacity to search, brute-force, and initiate DDoS attacks. Developing strong security measures to ward off such malware requires an understanding of these functions.

\subsection{Malicious Functions Design}
\label{maliciousfunctionsdesign}
% Bilal Notes :  sequance diagram . for each attack 
% 
% Explain briefly why we did not use mirai without doing any design
One key consideration in our decision not to use the Mirai botnet codebase directly was the potential impact of the included dependencies on our measurement methodology. 

The Mirai source code incorporates several external libraries and modules that, while necessary for the malware's full functionality, could introduce confounding factors into our analysis.

Additionally, Mirai's architectural design presented another challenge. The malware implements a dedicated command parsing stack responsible for interpreting and executing the instructions received from the command-and-control (C2) infrastructure. While essential for the malware's functionality, this abstraction layer could potentially alter the behaviour of targeted malicious functions to complicate the identification and characterisation process. We aim to focus on the core malicious behaviours rather than the peripheral mechanisms Mirai employs for its command delivery and execution.
% 3- mirai includes a full c2 communication stack
The comprehensive C2 communication stack included in the Mirai codebase represented an unnecessary layer of complexity for the purposes of our study. While this functionality is critical for the botnet's real-world operation, it would have introduced additional variables and considerations that could have detracted from our primary research objectives. By avoiding the inclusion of the full Mirai architecture, we could streamline the test environment and concentrate our efforts on the specific malicious behaviours we aimed to investigate.
\subsubsection{DoS Function Design}
% for each diagram the text should be an explanation of the diagram AKA design explanation.
\label{dosdesign}
\begin{figure}[hbt!]
    \centering
    \includegraphics[scale=0.5]{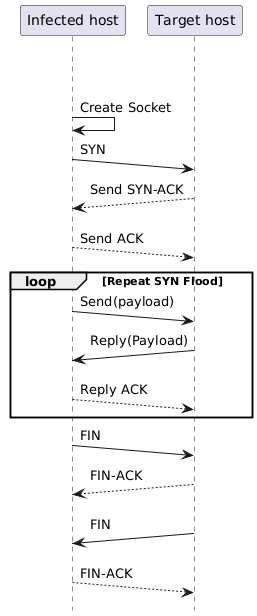}
    \caption{DoS Attack}
    \label{fig:dos_attack}
\end{figure}
A Denial of Service (DoS) attack aims to make a target service or network resource unavailable to its intended users by overwhelming it with traffic. The attacker sends high requests or data packets to the target, which struggles to process the excessive load. The target server becomes slow or unresponsive, disrupting normal operations. As shown in Figure \ref{fig:dos_attack}, we chose a design based on creating a socket and sending a fixed payload within a loop. After creating the appropriate sockets, a TCP connection is established with the victim host. Once the connection is established, the infected host sends a fixed-length payload to the target host within an infinite loop. Once we interrupt the loop (at the end of the experiment). The socket is automatically closed.

\subsubsection{Telnet Function Design}

\begin{figure}[!ht]
    \centering
    \includegraphics[scale=0.5]{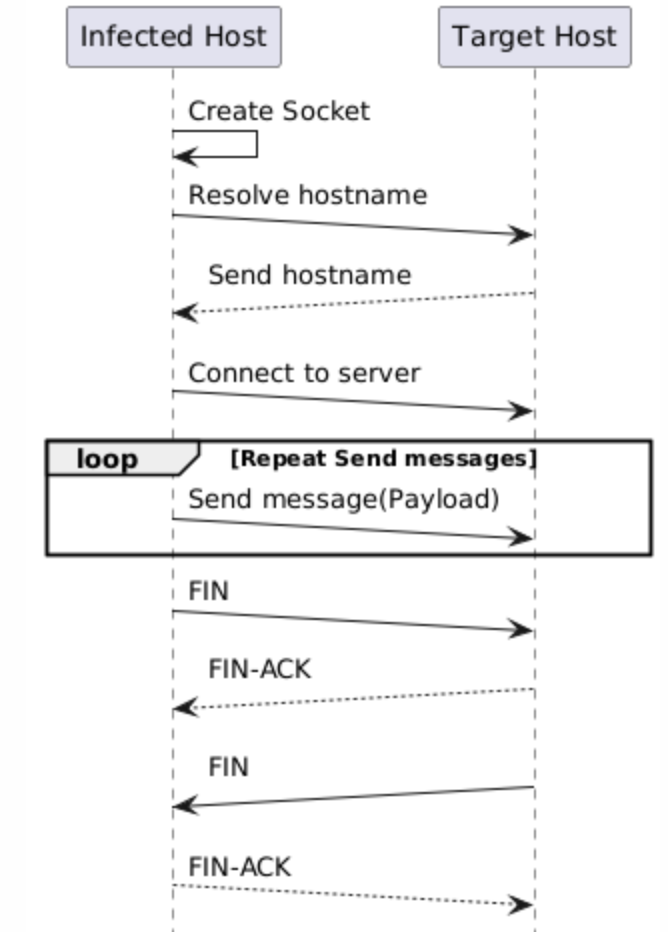}
    \caption{Telnet Attack}
    \label{fig:telnet_attack}
\end{figure}

Telnet is a protocol-use connection that lets users log into and interact with a remote machine on a TCP/IP network. Figure \ref{fig:telnet_attack} A telnet-based attack where an infected host (attacker) targets a victim (target host). Initially, the attacker creates a socket and resolves the target's hostname. The target responds with the hostname, allowing the attacker to connect to the server. The attacker then enters a loop where it repeatedly sends messages containing payloads to the victim. Finally, the attacker initiates a connection termination by sending a FIN packet. The victim responds with a FIN-ACK, and then the victim sends its own FIN packet. The attacker completes the termination process by sending a final FIN-ACK. This sequence represents a persistent attack where the attacker floods the target with messages before properly closing the connection.

\subsubsection{PortScan Function Design}

\begin{figure}[!h]
    \centering
    \includegraphics[scale=0.5]{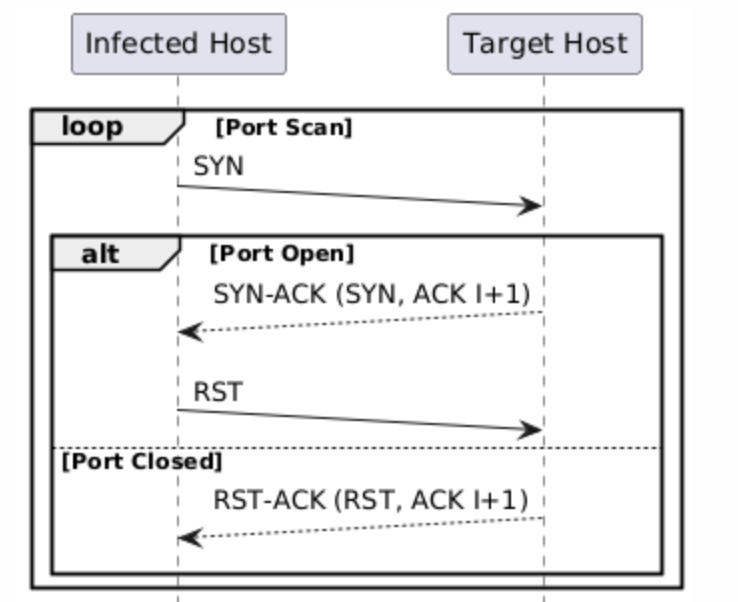}
    \caption{Port Scaning}
    \label{fig:Port_Scan}
\end{figure}

A Port Scan is a technique to find a network's open ports and weak points. In a port scan attack, the attacker uses a scanning tool to probe a target device for open ports. The tool systematically sends SYN packets to a range of ports and records the responses. Open ports respond with SYN-ACK packets, while closed ports respond with RST packets. By identifying which ports are open, the attacker can gather information about potential vulnerabilities on the target device. As shown in Figure \ref{fig:Port_Scan}, if the port is open, The receiver will receive an SYN J, ACK I+1; if it is closed, the receiver will receive RST, ACK I+1. Miri uses port scanning to identify accessible services on a specified host. Also, Miri can pinpoint potential entry points for more detailed security testing. Additionally, port scanning is valuable for monitoring and managing network resources, allowing administrators to keep track of active services and maintain network health effectively.

\subsubsection{ICMP function desgin}
\begin{figure}[hbt!]
    \centering
    \includegraphics[scale=0.4]{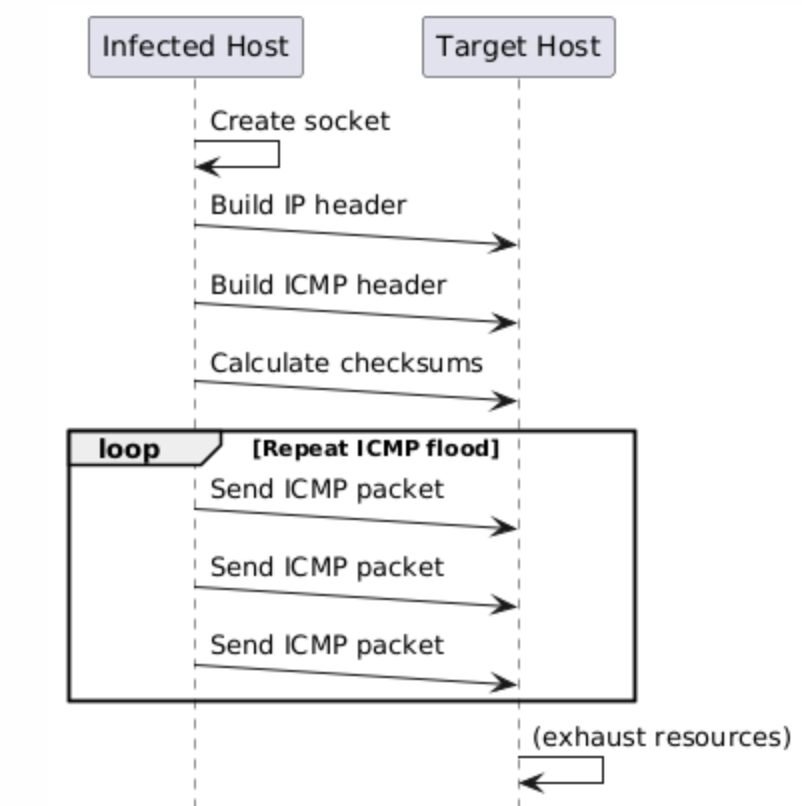}
    \caption{ICMP Attack}
    \label{fig:ICMP_Attack}
\end{figure}

ICMP (Internet Control Message Protocol) is primarily used to send operational information in network environments. The most common use of ICMP is in the ping command, which checks the connectivity between two network nodes. Figure \ref{fig:ICMP_Attack} is an ICMP ping flood attack design where an infected host creates a raw socket and continuously sends ICMP Echo Request packets to a target host. The target host responds with ICMP Echo Replies. This process is repeated in a loop, flooding the target with requests. The slanted arrows in the diagram represent the delays between sending and receiving packets, emphasising the asynchronous nature of the attack. Finally, the infected host closes the connection, and the target host acknowledges the closure. This attack can overwhelm the target host, resulting in resource exhaustion and service denial.

\subsubsection{TCP\_SYN flood function desgin}

\begin{figure}[hbt!]
    \centering
    \includegraphics[scale=0.5]{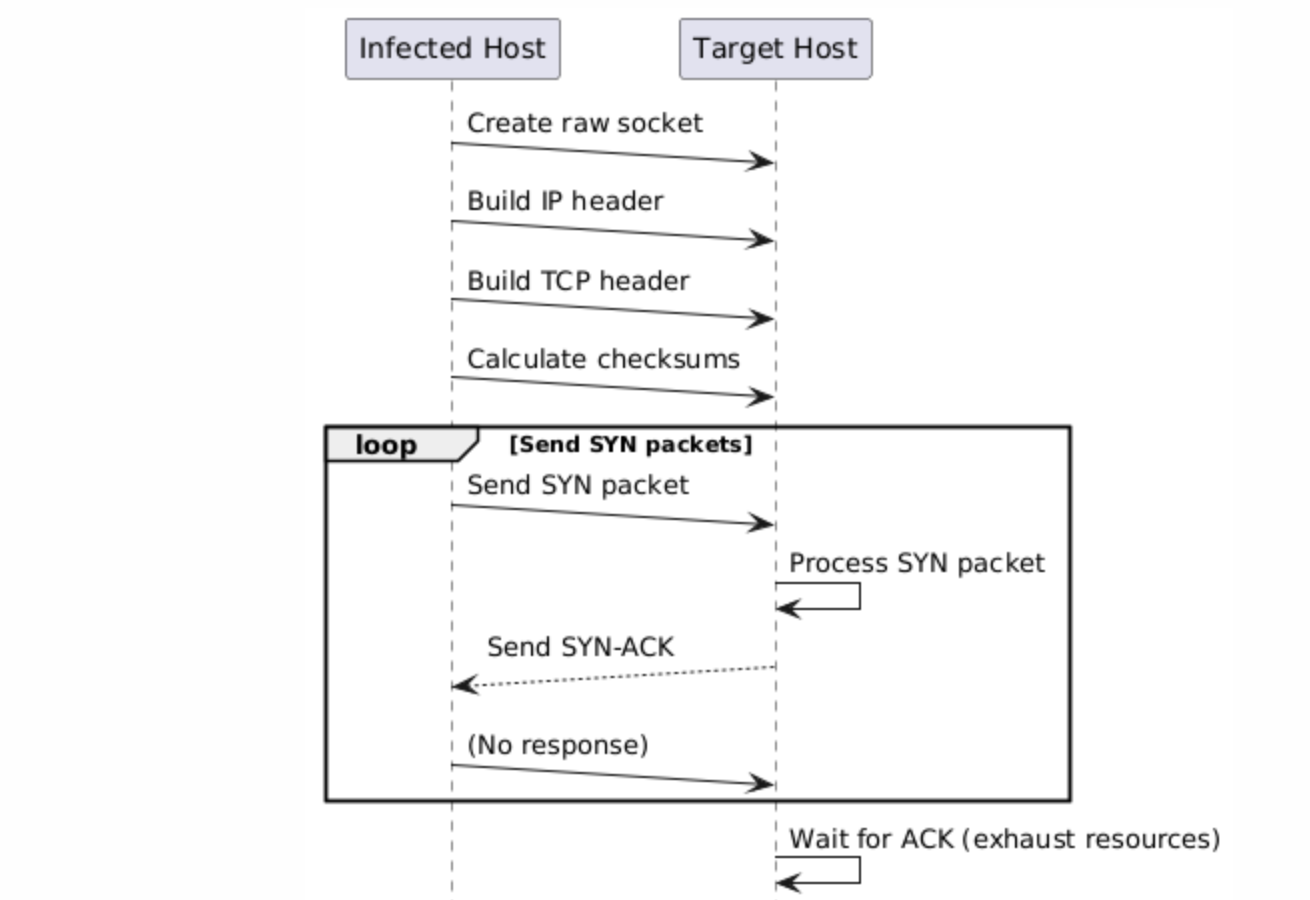}
    \caption{TCPSYN Flood Attack}
    \label{fig:tcpsyn_flood}
\end{figure}

TCP\_SYN flood is a denial-of-service (DoS) attack that relies on the TCP three-way handshake to hammer a target server with requests to make a connection. Figure \ref{fig:tcpsyn_flood} shows a TCP\_SYN flood attack where an infected host (attacker) sends SYN packets to a target host. The attacker creates a raw socket, builds IP and TCP headers, and calculates checksums. It then enters a loop, continuously sending SYN packets to the target host, which processes these packets and responds with SYN-ACKs. However, the attacker never completes the handshake, causing the target host to wait for ACK responses indefinitely, leading to resource exhaustion.

%The \texttt{tcp\_syn\_flood.c} file contains code to perform a TCP SYN flood attack. This attack works by repeatedly sending TCP SYN packets to a specified target address and port, with the goal of overwhelming the target with connection requests. Such an attack can lead to a denial of service, as the target system becomes inundated with fake connection requests and is unable to handle legitimate traffic.\\ 

 %The code was implemented as a separate binary where we could define the different parameters for these malicious functions. It was designed to be portable on any platform through cross-compiling. The code was based solely on standard libraries, ensuring portability and ease of use. It was written or modified from publicly available resources. We chose to design this to be suitable for ARM architecture. In order to do that, we needed to understand the RISC design architecture, which will be explained in this chapter.

     %Malware types: 
     %Mirai < function explains how it 
     %port scan, and do %ddos %telnet, %brute-force,  

        %- we implemented these in a saparate binary were we can defined the different parameters for this malicious functions. 
        %- the code was implemented to be portable on any platform throw cross-compiling. 
        %-  the code was based on standard libraries only.
        %- the code was written or modified from publicly available resources.
        %we choose to design this to be sutable for arm arch. in order to do that we need to understand the risc design arch. that will be explained in this chapter. 

\section{Malicious Functions Characterisation } 

The journey from source code (e.g., '.c' files) to an executable binary involves a critical transformation. The code is compiled into a set of instructions that the CPU can understand and execute. This instruction flow dictates how the hardware components of the systems, particularly the CPU, will perform during the binary's execution. 

By analysing the HPC values, we can gain insights into how the function interacts with the hardware. This analysis is important as deviation from the expected HPC patterns can indicate potential issues, such as bugs or malicious behaviour. For example, the sudden spike in branch mispredictions or the cache misses could mean that the binary is executing unexpected or unusual instructions, mainly pointing toward a malicious function.

For instance, to show how deviations in HPC values could indicate malicious behaviour, consider Spectre attack that is known for exploiting the speculative execution in the modern CPUs, including the ARM processors.

During a Spectre attack done by \cite{li2022detecting}, the HPC metrics showed abnormal patterns. For instance, there was an increased rate of branch mispredictions as the branch predictor was manipulated. Also, there were higher cache miss rates as the attack continually flushed and reloaded the cache to exploit the speculative execution side channel. They have reflected that analysing the HPC features makes it possible to know about the Spectre attacks with a high level of accuracy. 

Malicious functions may not always have immediate effects but could instead aim to alter the system's performance over time; the use of HPC is our key to characterising malicious functions. Understanding how each function impacts the CPU and other hardware components allows us to establish expected patterns of HPC value for normal, benign behaviour. By studying how HPC values evolve, we can build a signature for each function while executing it in the device and then analyse this signature with the baseline that we have. Also, during prolonged execution, we can identify patterns that indicate long-term malicious activity, such as resource exhaustion or persistent backdoor operations. By doing so, we can detect malicious activities, helping to detect if the device is malicious or not

ARM's RISC architecture is designed to execute simpler and more predictable instructions, making it easier to analyse and interpret the resulting HPC values. In contrast, x86's CISC architecture involves more complex and variable instructions, which can complicate the process of correlating instruction flow with hardware behaviour.  Focusing on ARM architectures, we can leverage the simplicity and predictability of the RISC design as will be explained in section \ref{designrisc}.

%They may have impact on the long term. example optimization and implementation variants. 

%Arch between x86 or arm etc but our fouces will be on arm arch only. 

%Understanding the impact of instruction flow on hardware is critical for identifying malicious functions. When the .c code is first compiled into a binary, it generates an instruction flow that directs the hardware's behaviour. By analysing this hardware behaviour, we can further characterise the binary. Once executed on the hardware, the instruction flow changes the hardware performance counter (HPC) values. While this work focused on HPC values, it is crucial to understand the impact of instruction flow on hardware for characterising binaries, especially for detecting anomalies or malicious activities.

%Given the focus on arm architectures, we must first distinguish between them and understand the significance of HPC values for us. 

\subsection{RISC}
\label{designrisc}
%According to \citeauthor{stanford_risc} \cite{stanford_risc} 

The RISC (Reduced Instruction Set Computing) architecture is the foundation for our design approach in characterising malicious functions on ARM processors. RISC's simplified instruction set and the architecture's inherent properties provide several advantages when analysing and interpreting hardware behaviour, particularly through HPCs

RISC architectures, like ARM \cite{arm_risc}, are characterised by their streamlined and efficient instruction sets. These instructions are designed to execute in a single clock cycle, making the CPU's operation more predictable and easier to analyse. ARM’s RISC architecture includes conditional execution as a core feature, which is not as prevalent in other RISC designs. Conditional instructions allow the CPU to execute or skip certain instructions based on the evaluation of condition flags without requiring explicit branch instructions. This property reduces the need for branching, minimising the risk of pipeline stalls and making instruction flow more efficient and predictable.

\subsubsection{HPC Extraction from RISC}
RISC’s predictability and simplicity allow for more straightforward extraction and interpretation of HPC data, as explained in section \ref{background:HPC}. The uniform instruction length and the focus on register-based operations make it easier to correlate HPC events with specific parts of the code. Since each RISC instruction is designed to be completed in a single cycle and has a direct impact on specific hardware components, the resulting HPC metrics can be more easily mapped to the executed instructions. This straightforward mapping simplifies the detection of anomalies in hardware behaviour that might suggest malicious activity.

The design leverages predictability by setting up the baselines for normal HPC behaviour for each device for various types of instructions depending on every device’s HPC events. Deviations from these baselines during program execution could signal potential anomalies that warrant further investigation. 

\subsubsection{Branch Predication in ARM}
Branch prediction units in ARM processors make an effort to forecast, before the actual condition is resolved, the result of the conditional branches (i.e., whether a branch will be taken or not). The pipeline flush, which could be expensive in terms of performance, will happen if the prediction is incorrect; otherwise, the pipeline will proceed without incident. 

ARM processors use various branch predictor types  \cite{arm_branch_prediction}, each with a unique technique for predicting a branch's conclusion. Simple predictors, like static prediction, and more complex ones, such as dynamic two-level predictors, are among them. The branch predictor selection impacts the CPU's overall performance as well as prediction accuracy.

The branch predictors from the ARM mainly use algorithms that use past branch outcome data to enhance their accuracy. Pipeline execution runs smoothly and efficiently when the predictions come true and the CPU could carry out instructions without stopping. In general, the efficient execution maintains the HPC metrics—like CPU cycles and the branch mispredictions within the predicted bounds.

%Malicious code can exploit branch prediction mechanisms to cause mispredictions deliberately, which might be a strategy for certain side-channel attacks (e.g., Spectre). By monitoring unusual branch prediction behaviour via HPCs, our design can identify and flag potential security threats.

\subsubsection{Cache-Memory in ARM} 
ARM processors mainly have a multi-level cache hierarchy to optimise access to frequently used data. The RISC architecture’s load/store model shows that all memory operations need to go through explicit load/store instructions, which makes cache interaction highly transparent and easier to control.

\subsubsection{Linking Program Execution to HPCs} 
The main justification for focusing on the ARM’s RISC architecture is the direct relationship between program execution, CPU instruction processing, and the resulting HPC values. The relationship allows us to develop a coherent model that links all the aspects together to effectively detect malicious functions. 

In RISC architectures, the simplicity and predictability of instructions mean that each operation has a clear and direct impact on specific hardware components. HPCs capture this direct correlation, allowing us to comprehensively understand how a program behaves at a hardware level.

%features on risc 
%depth on risc instruction. and have proparte that have conditional instruction that is specific for arm. aka risc. 
%why we said that its easy hpc extracted from 

%how branch prediction work in arm. 
%the instructions of arm 
%   1. (see confitional instruction )
%   2. cache-momery and its relation with       instrcution. in arm aka risc 
%   3.branch prediction work in arm.

%justification is that there is a relation with program and instruction executed from cpu to hpc. that will link all togther. 
%

\section{Data Collection Approach Design}

\begin{figure}[!ht]
    \centering
    \includegraphics[scale=0.2]{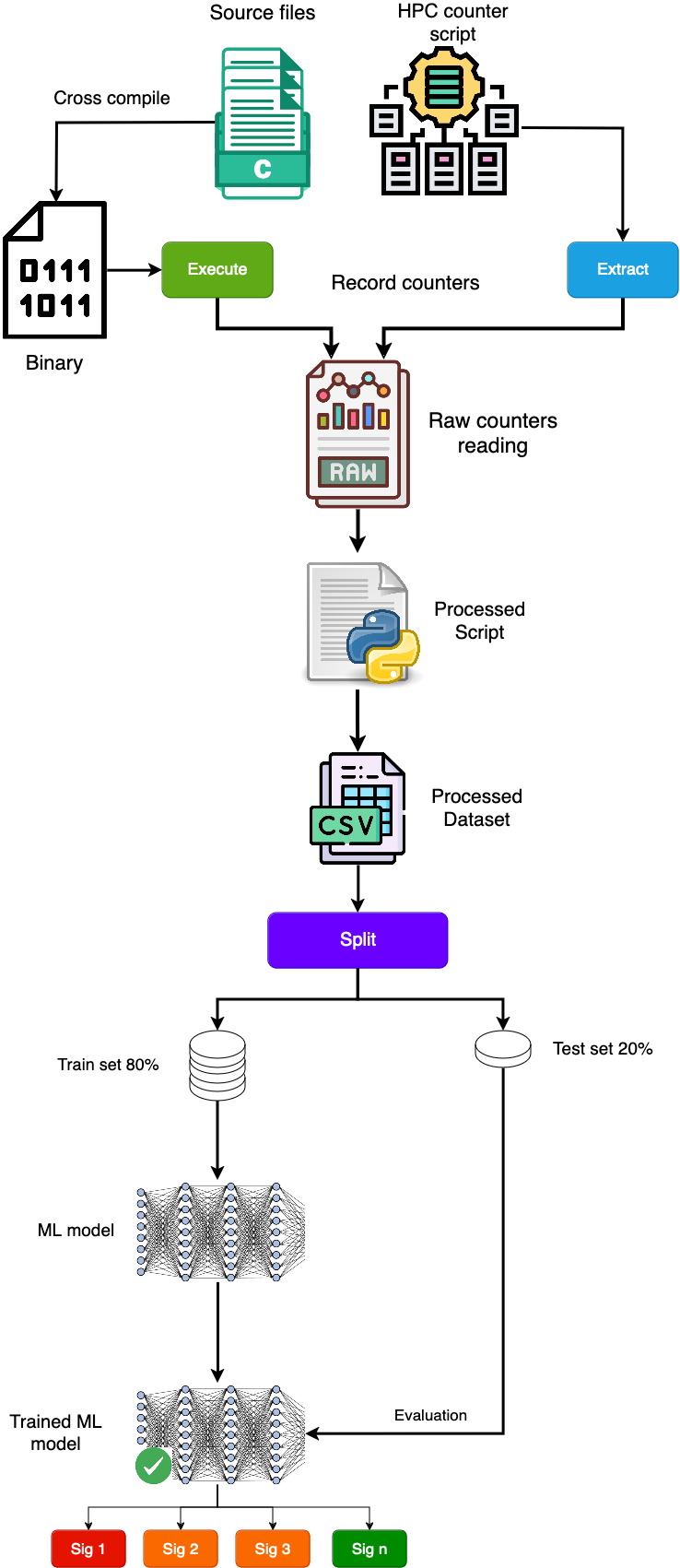}
    \caption{Overall data collection approach}
    \label{fig:hpchighview}
\end{figure}

As explained in the previous section, many steps are involved in using the HPC counters. To extract the low-level details of the HPC data collection process, we designed the overall architecture shown in Figure \ref{fig:hpchighview}. The first step is to pick the right sources corresponding to the function under evaluation and configure the HPC through the configuration module. At this step, the source code is compiled to the right target architecture, and the HPC counters are initialised, zeroed, and configured with the right set of events. Once both the HPC and the function under evaluation are ready. The next two blocks, named extract and execute, start cycling through executing the under-evaluation function in a process and the HPC measurements in a separate process. Since the HPC is hardware-based, the Extraction module does not introduce any significant overhead to our setup. We chose to record the HPC raw counter values to avoid any extra processing overhead during the data collection process. Once the initial process in the design is complete, the raw data is taken by a second module that preprocesses the raw data to extract significant and labelled datasets. The processed data is separated into multiple tabular datasets corresponding to different scenarios. By the end of this process, the dataset can be loaded and split into training and testing subsets to experiment with the ML designs in order to get a significant identification of the evaluated functions.
%    - here what concept I use. 
%    - high level of how to collect data. 
%   - I have malicious functions > which is found in any binary. 
%   - we isolate malicious function that have potanital signature. (DoS or any)
%   -  running these malicious in the target devices. 
%   - read and record the counters valuse during defined period. 
   
\label{section:datacollection}

\section{Model Design}
\begin{figure}[!htb]
\centering
\minipage{0.4\textwidth}
\includegraphics[width=\linewidth]{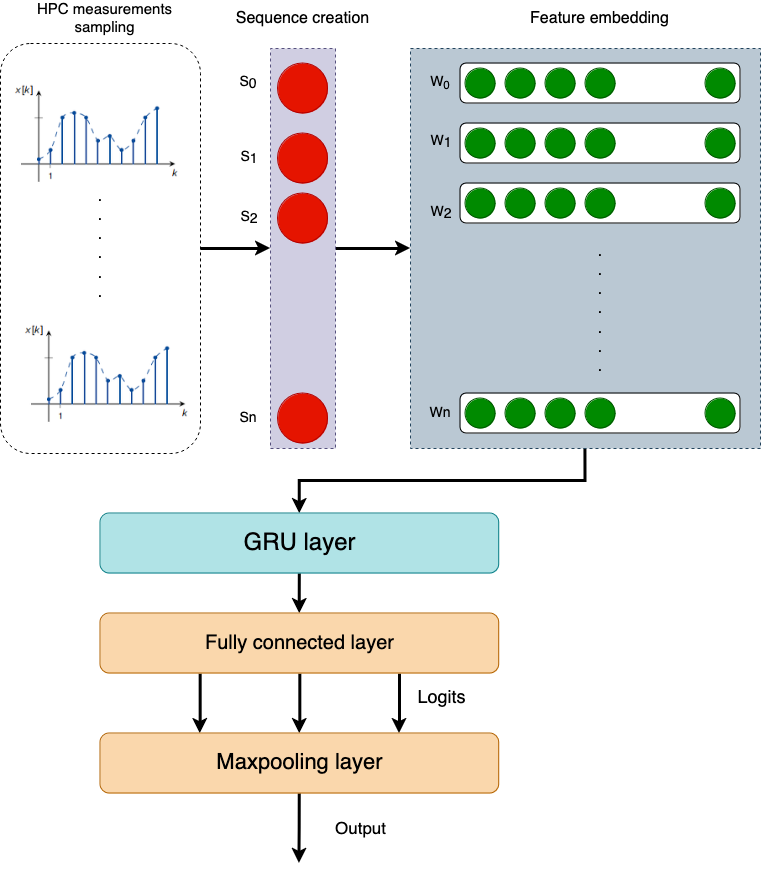}
 \caption{AI Model design}
    \label{fig:AIDesign}
\endminipage
\end{figure}

As shown in Figure \ref{fig:AIDesign}, the initial design of the model includes various layers to adapt the sampled data to the model's input. After sampling the data from the HPC peripheral, the measurements are gathered together in the form of a sequence. Each sequence is formed by concatenating the various measurements together according to the sampling frequency. Each sequence segment represents a measurement taken at the instant t from the HPC counters readings. These readings represent various event counts, such as the instruction cache misses, the data cache misses, the branch misprediction, etc. Each sequence is passed by an embedding layer that transforms the sequence into a matrix with vectors representing each input sequence segment. The Embedding Layer provides a mapping function from input sequences to a higher-dimensional space to provide a more fine-grained input representing each sequence. The embedding vectors that are extracted from the sequenced input are fed to a single GRU layer that plays the role of a feature extractor. The advantage of using the GRU layer is that this model can memorise the successive sequences, which makes it more adapted to pattern identification on successive input embedding. We designed the model to be trainable on more than a single GRU layer to cope with the complexity of our problem. Two parameters can be adapted in such design during the training process: the number of the GRU layer and the embedding size. Such flexibility is needed to experiment with the efficiency of our model in classifying the HPC extracted sequences.\\
Following the feature extraction layer comes the classification layer components. The classification layer is implemented as a fully connected layer that matches the GRU layer output size in terms of the input size. In our case, we chose to use a linear layer. The linear layer applies an affine linear transformation to the incoming data:
\[ y = xA^T + b\]
Given that A and b represent the learnable parameters of our model, this layer learns to classify the extracted features into an appropriate class. The output of the linear layer is adapted to the unique label count. At the output of this layer, we obtain the set of logits that represents the probability of the input belonging to one of the defined label classes. The logits are then fed to a max pooling layer to extract the input sequence's appropriate classification class. 

%\begin{figure}[!ht]
%    \centering
%    \includegraphics[scale=0.35*0.3]{figures/Design_Diagram/gru_design.drawio.png}
%    \caption{AI Model design}
%    \label{fig:AIDesign}
%\end{figure}

%% file: chapters/Implementation.tex
\label{implementation}

\section{Malicious function implementation}
\label{maliciousfunctionimplementation}
As explained earlier in Section \ref{maliciousfunctionsdesign}. Implementing five specific Software Unclonable Functions (SUFs) is designed to perform different types of functions used in cyberattacks. The SUFs discussed are Denial of Service (DoS), TCP SYN Flood, Port Scan, Telnet Flood, and ICMP Flood. This aims to observe the hardware counters for the device running these functions.

After implementing the C files for each function, we did compile it to the desired architecture.  Then we installed Per

%I need to add something more here. 

\subsection{DoS Flood}
\vspace{1cm}
\label{dosFloodFunction}

\input{chapters/Code/DoSCode}

The httpFlood function, as shown in Listing \ref{code:DoSAttack}, performs a Denial of Service (DoS) attack by repeatedly sending HTTP GET requests to a specified server address and port as a parameter. It establishes a TCP connection to the target server using a \textbf{'fd'} in line 3 and continuously writes the HTTP GET request in line 4 to the server in an infinite loop. \\
%The goal of this function is to observe the hardware counters. \\ 

\subsection{TCPSYN}
\label{httpflood}
\input{chapters/Code/HTTP}

The provided Listing \ref{code:TCPSYNAttack} code snippet is a loop that performs a TCP SYN flood attack. After setting up the necessary headers and initialising the socket, the loop begins. The \textbf{sendto} function is used to send the prepared datagram to the target address continuously. This loop runs indefinitely, aiming to overwhelm the target system with TCP SYN packets. The process continues without pause, repeatedly sending packets to the target to perform the flood attack until the program is manually stopped or the system running it runs out of resources.

\subsection{Port Scaning Flood} 
\label{scanHostFunction}

The \textbf{scanHost} function, as shown in Listing \ref{code:PortAttack}, determines the open or closed status of a specific port on a designated computer based on its IP address. 

\input{chapters/Code/port_scan}

It begins by announcing the port it is about to scan. The function then proceeds to create a communication endpoint, known as a socket, necessary for sending and receiving data over the network. Using this socket, the function attempts to connect to the specified port on the target computer. The function reports an error and terminates if it cannot open the socket. If the socket successfully opens but the connection fails, it prints a message indicating the port's closure. Conversely, if the connection is successful, it confirms that the port is open. The function closes the socket after port checking and returns the port number.

\subsection{Telnet Flood}
\label{telnetFunction}

\input{chapters/Code/telnet}

The provided snippet code, as shown in \ref{code:TelnetAttack}, is a client-side implementation that establishes a connection to a server, sends a message, and reads the server's response. a connection to a server using a socket \textbf{socktfd} and a server address \textbf{serv\_addr}. When the users enter a message, it holds the user input and reads it using \textbf{fgets()}, then the \textbf{write()} function sends the message to the server. The goal of this function is to send messages to the attacker simultaneously.    

%The provided code snippet is a client-side implementation that establishes a connection to a server, sends a message, and reads the server's response. Initially, the \textbf{connect} function attempts to establish a connection to the server using the socket file descriptor \textbf{sockfd} and the server address \textbf{serv addr}. If the connection fails, an error message is printed, and the program exits. Once connected, the client prompts the user to enter a message, which is read into the \textbf{buffer} using \textbf{fgets}. This message is then sent to the server via the \textbf{write} function. If writing to the socket fails, an error message is printed, and the program exits. The client then waits for a response from the server, reading it into the \textbf{buffer} using the \textbf{read} function. If reading from the socket fails, an error message is printed, and the program exits. Finally, the server's response is printed to the standard output. This sequence facilitates basic client-server communication where the client sends a message to the server and processes the server's reply.

\subsection{ICMP Flood}
\label{icmpflood} 
The provided Listing \ref{code:ICMPAttack} a code snippet is a loop that performs an ICMP ping flood attack. After setting up the necessary headers and initialising the socket, the loop begins by filling the ICMP payload with random data using \textbf{memset}. It then recalculates the ICMP header checksum to account for the new payload data. The \textbf{sendto} function is used to send the packet to the target address, and if sending fails, an error message is printed using \textbf{perror}, and the loop breaks. If sending is successful, the number of sent packets is incremented and printed using \textbf{printf}. The loop then pauses for a short interval using \textbf{usleep} to control the flood rate. This process aims to continue indefinitely, repeatedly sending ICMP packets to the target to perform the flood attack.

\input{chapters/Code/ICMP}

\section{Data Collection Setup}
\begin{figure}[!ht]
    \centering
    \includegraphics[width=8cm,angle=270]{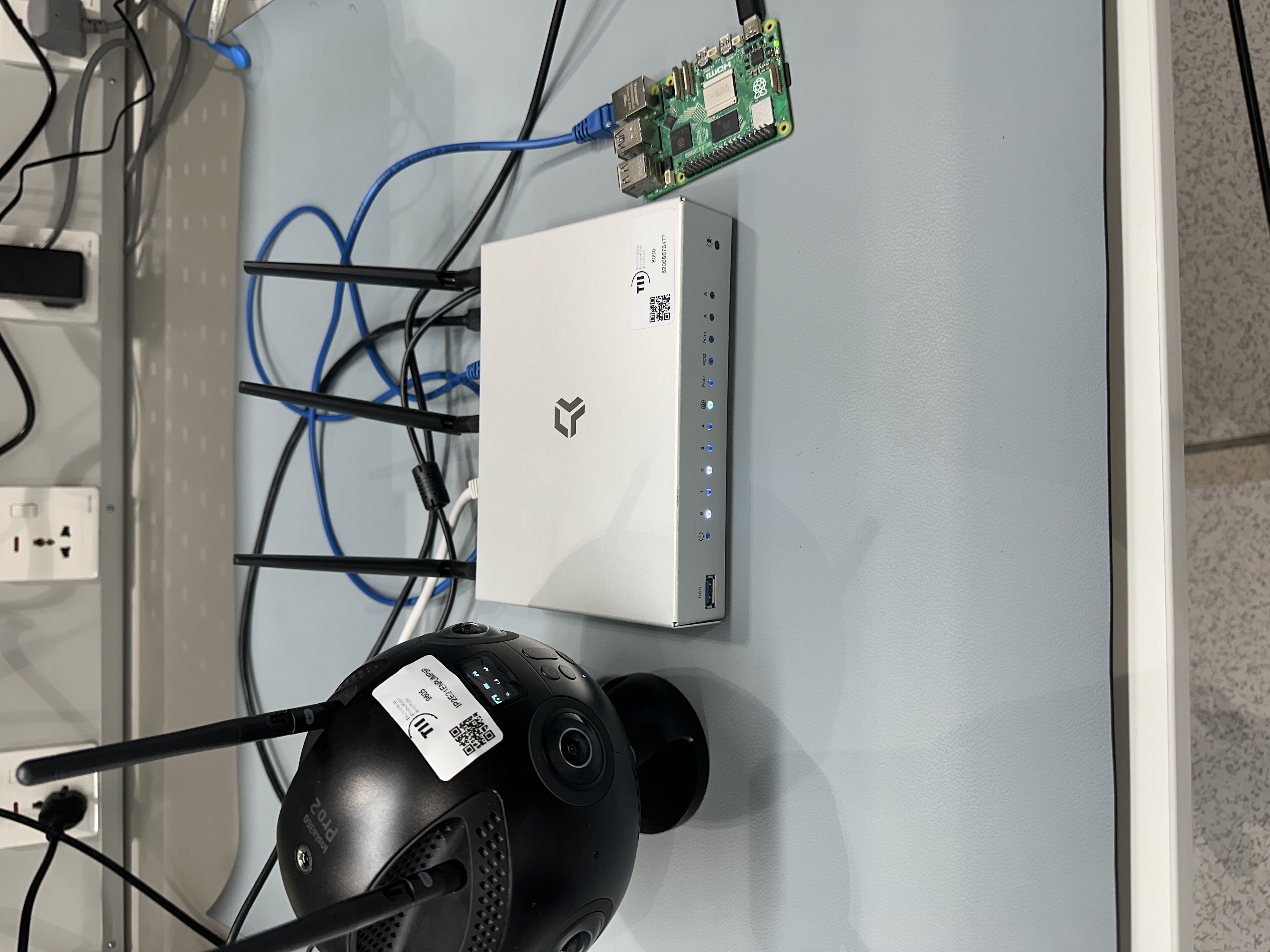}
    \caption{Data collection flow}
    \label{fig:setup}
\end{figure}

%I need to edit this and revise it and make it more clear to explain the full section later ! 

Figure \ref{fig:setup} shows the devices used to implement the experiments, as explained in Section \ref{section:datacollection}. Figure \ref{fig:hpchighview} shows a high overview of the whole process. It demonstrates all steps by using hardware performance counter data to identify software anomalies. There are several crucial steps and utilities in this figure. They converge to ensure the effective and efficient detection of anomalies.
%Check this with Bilal on Sunday 
% The process starts with executing a bash script, starting the malicious binary and 'perf' tool collection, as shown in the code snippet. 

\subsection{Perf Tool}

% Re do this perf definition and examples. 23 Aug 2024 
On Linux systems, Perf \citep{perftutorial} is an indispensable performance analysis tool. This utility allows you to monitor and analyse the performance of system software applications by pulling together hardware performance counter (HPC) data. 
The Linux hardware performance counters (HPCs) sample the processor's simple hardware events while generating high-level and software events. These counters are special registers in the CPU that count various low-level activities, such as executed instructions. These counters track cache hits and misses and cycles on the CPU (refer to Section \ref{background:HPC}). Applications can provide a comprehensive picture of all aspects of procedure behaviour by utilising performance counters.

We cannot emphasise enough the difficulty of modern software and systems. Numerous factors influence performance, often presenting subtle challenges. While the performance tool provides critical insights into how well one's system and applications are performing, it also enables developers and system administrators to pinpoint such performance bottlenecks exactly, understand what their software is doing from underneath the hood, and make optimisations based on accurate, data-driven decisions.

The perf tool can collect profiles on CPU, Cache, and Memory Events, allowing users to backtrack through the performance data for review and analysis. This added historical viewpoint is extremely valuable in identifying malicious behaviour or long-term performance patterns. We used the command below to see the available HPC events.

\begin{verbatim}
    Using this command: "sudo perf list"    List of pre-defined events
\end{verbatim}

\hfill\\ HPC Events:  All of these events were extracted from the device we have using the perf tool to extract them. The description for each of these events can be found in Table \ref{Table:PerfEvents}. These Events are not available in all of these devices Table \ref{table:HPC-events} is the available events for each device.

%HPC EVENTS 
\input{chapters/HPCeventsName} 
\input{tables/Design/HPCEventsdevices}

\subsection{Process of Cross-Compiling}

Cross-compiling is the process of creating software on one system (host) so that it can run on a different type of system (target). This technique is particularly important because, in some cases, the target system has limited resources or things terrible with hardware architecture, making direct software development and compilation difficult, if not impossible.

\begin{figure}[!ht]
    \centering
    \includegraphics[scale=0.8]{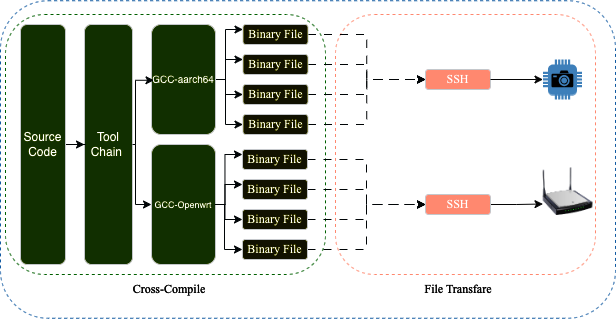}
    \caption{Cross-Compilation Process}
    \label{fig:crosscompiler}
\end{figure}

Figure \ref{fig:crosscompiler} illustrates the process of cross-compiling software for different target architectures. It begins with source code files on the left as explained in section \ref{maliciousfunctionimplementation}, which need to be compiled using a toolchain. The toolchain installation includes specific compilers like \textbf{gcc-aarch64-linux-gnu} for ARM 64-bit architecture and \textbf{gcc-openwrt-linux-gnu} for OpenWrt. These compilers convert the source code into executable binaries suitable for the target CPUs on the right after getting the \textbf{Binary File}. It is then sent through \textbf{SSH} to transfer these compiled binaries to the target systems, allowing them to run the newly compiled software. \\

\textbf{Command} example for the cross-compiling cmd  

\begin{verbatim}

arm-openwrt-linux-gcc -o icmp-attack-router icmp-attack.c 

\end{verbatim}

The diagram depicts the result of the cross-compilation process: an executable malicious binary specifically designed to run on an ARM-based OpenWrt device.

\subsection{Bash } 
\label{section:bash}

The data collection process is managed by two primary scripts, collection.sh and startcollection.sh. The collection.sh script is designed to run a specified binary while simultaneously using the perf tool to monitor various hardware events at 200-millisecond intervals according to (arm link ). For each 200 millisecond, the HPC is counted. These events are described in Table \ref{Table:PerfEvents}. The performance data is captured in a file named after the binary which can be seen in Listing \ref{code:collection} \textbf{\textdollar 2} is the binary name where is \textbf{\textdollar 3}, \textbf{\textdollar 4} and \textbf{\textdollar 5} depends on each binary parameters requirements while it runs the binary it will lunch as will the perf command as can be seen in line \textbf{8} where the \textbf{EventsName} is explained in Table \ref{Table:PerfEvents} and this depends on the device available events as can be seen in \ref{table:HPC-events}.

\input{chapters/CodeBash/Collection}

The startcollection.sh script orchestrates the collection process for multiple binaries. It sequentially runs the collection.sh script for each binary, specifying a collection duration of one hour as shown in Listing \ref{code:startcollecting}

\input{chapters/CodeBash/StartCollection}

For each binary, after the one-hour collection period, the script kills the running processes and ensures the data is stored in a file named appropriately for that binary. This process is repeated for each binary listed in the startcollection.sh script as shown in Listing \ref{code:startcollecting} example where \textdollar 1 you need to insert the time in our case \textbf{3600} seconds. After which, the script terminates the running processes and stores the performance data in \textbf{.txt} files that will be later used in section \ref{dataprocessing}.

\section{Data Processing }
\label{dataprocessing}
Once the performance tools record the required measurements, they are stored as unstructured data in text files on the disk. The text files which has been generated in \ref{section:bash} are loaded on a Python script (. PY) to structure the data and transform it into a usable object (data frame). 

\input{chapters/CodeData/Data_Parsing}

As shown in Listing \ref{code:DataPreprocessingParsing}, The first loop loads the unstructured data lines and stores them on a variable. The following loops clean the content of the lines and clean them from a couple of characters. By the end of the final loop, we get an array of objects with different measurement values.

% Part to describe the formating step
As shown in Listing \ref{code:DataPreprocessing}, The data that have been extracted is further cleaned and formatted in a dictionary called dataexp. The dictionary key for each element is the timestamp of the measurement. Each key holds as a value a list of tuples. The tuples contain the name of the extracted counter value and the counter value. After Populating the whole dictionary, the last code section takes the dictionary, iterates through it, and creates a CSV file that holds the different HPC measures as separate columns. By the end of the process, we obtain a structured CVS file for each of the experiments.
\input{chapters/CodeData/Data_Prepocessing}

\section{Data Analysis}
\label{DataAnalysis}
After obtaining the data in CSV format from Section \ref{dataprocessing}, we proceeded with a thorough statistical analysis. The objective was to derive meaningful insights and understand the underlying distribution and variability within our dataset. The following statistical measures were computed:

\begin{itemize}
    \item \textbf{Mean:} The average value of each feature gives an indication of the central tendency.
    
    \item \textbf{Standard Deviation (std):} The measure of dispersion or variability around mean, showing how spread out the given values are. 
    \item \textbf{Minimum (min):} Smallest observed value for each of the features, identifying the lower bound of data.
    \item \textbf{Maximum (max):} Largest observed value for each of the features, identifying the upper bound of data.
    \item \textbf{Variance (var):} Squared standard deviation provides another measure of the spread and variability within data.
\end{itemize}

The statistical measures offer an in-depth overview of the dataset, facilitating analysis as well as interpretation. They are important for understanding the different characteristics of each feature, identifying the potential outliers, and ensuring the effectiveness of subsequent analyses. 

Through examining the metrics, one gains insights into the behaviour and distribution of various features under normal conditions and different attack scenarios. The foundational understanding is important for developing efficient anomaly detection and mitigation strategies.

\subsection{Pi Dataset Statistics}
Table \ref{static_comparison_pi_normal_dos} shows the statistical analysis of different performance metrics that are collected from PI datasets. The analysis of collected performance data shows multiple critical observations.

\begin{table}[!ht]
    \centering
    \begin{adjustbox}{max width=\textwidth}
    \begin{tabular}{|l|l|l|l|l|l|l|l|l|}
\hline
\textbf{Feature} & \textbf{Min (Normal)} & \textbf{Min (DoS)} & \textbf{Max (Normal)} & \textbf{Max (DoS)} & \textbf{Std (Normal)} & \textbf{Std (DoS)} & \textbf{Mean (Normal)} & \textbf{Mean (DoS)} \\ \hline
L1-dcache-load-misses & 3,694,179 & 1,960,085 & 9,931,251 & 3,346,366 & 197,495.79 & 89,843.13 & 4,162,510.61 & 2,259,924 \\ \hline
L1-dcache-loads & 144,430,577 & 55,283,572 & 564,125,847 & 115,208,522 & 29,900,102.93 & 3,828,157 & 166,898,311.2 & 60,546,280 \\ \hline
L1-icache-load-misses & 4,334,868 & 11,988,798 & 14,935,291 & 23,845,377 & 476,794.17 & 542,301.83 & 7,383,203.66 & 21,747,695.11 \\ \hline
L1-icache-loads & 153,729,689 & 119,887,798 & 550,247,352 & 238,453,777 & 27,168,996.52 & 542,301.83 & 181,891,032.1 & 217,476,95.11 \\ \hline
LLC-load-misses & 2,591,956 & 1,935,723 & 9,448,360 & 3,393,875 & 314,625.14 & 95,823.33 & 2,940,353.55 & 2,259,892 \\ \hline
LLC-loads & 8,452,497 & 5,521,460 & 15,392,321 & 11,217,462 & 362,198.20 & 4,010,832 & 9,327,951.78 & 6,052,755 \\ \hline
branch-load-misses & 2,308,776 & 4,807,422 & 5,209,952 & 9,389,385 & 183,896.31 & 206,308.00 & 2,986,880.53 & 8,347,052 \\ \hline
branch-loads & 89,497,978 & 52,146,008 & 344,407,371 & 110,217,462 & 21,289,043.41 & 4,010,832 & 105,450,715.6 & 60,527,580 \\ \hline
branch-misses & 2,507,594 & 1,935,723 & 5,096,456 & 3,393,875 & 183,181.98 & 95,823.33 & 2,986,123.26 & 2,259,892 \\ \hline
bus-cycles & 416,516,962 & 520,860,422 & 1,077,054,794 & 648,209,507 & 50,029,090.58 & 10,988,368.99 & 474,301,365.2 & 557,045,053.5 \\ \hline
cache-misses & 3,753,468 & 1,935,723 & 10,038,657 & 3,393,875 & 203,649.07 & 95,823.33 & 4,161,924.25 & 2,259,892 \\ \hline
cache-references & 143,309,657 & 52,146,008 & 522,535,456 & 110,217,462 & 30,083,358.81 & 4,010,832 & 166,899,408.4 & 60,527,580 \\ \hline
cpu-cycles & 422,464,162 & 520,860,422 & 1,067,247,088 & 648,209,507 & 49,608,950.36 & 10,988,368.99 & 474,358,448.8 & 557,045,053.5 \\ \hline
dTLB-load-misses & 1,134,244 & 1,619,787 & 5,434,899 & 6,504,337 & 114,798.71 & 634,449.75 & 1,334,402.997 & 2,259,046 \\ \hline
dTLB-loads & 143,718,436 & 52,146,008 & 518,730,956 & 110,217,462 & 29,981,673.15 & 4,010,832 & 166,962,703.5 & 60,527,580 \\ \hline
iTLB-load-misses & 129,621 & 1,935,723 & 3,424,760 & 3,393,875 & 99,503.11 & 95,823.33 & 164,253.34 & 2,259,892 \\ \hline
iTLB-loads & 142,829,458 & 52,146,008 & 449,225,254 & 110,217,462 & 24,923,036.38 & 4,010,832 & 165,321,664.4 & 60,527,580 \\ \hline
instructions & 423,369,339 & 168,377,524 & 1,581,925,852 & 350,625,667 & 89,918,209.17 & 5,984,553 & 492,356,210 & 191,641,573.7 \\ \hline
stalled-cycles-backend & 130,643,312 & 372,718,288 & 504,411,919 & 511,850,597 & 13,976,568.16 & 7,885,738 & 145,687,285.9 & 426,475,802.2 \\ \hline
stalled-cycles-frontend & 128,058,304 & 88,138,223 & 340,998,483 & 143,933,966 & 13,496,242.76 & 4,544,204 & 143,269,667.3 & 115,176,038.7 \\ \hline
    \end{tabular}
    \end{adjustbox}
    \caption{Comparison of Statistics between Pi Normal and Pi DoS}
    \label{static_comparison_pi_normal_dos}
\end{table}

The observation comparing the Pi system’s performance metrics under normal conditions and during the DoS attack shows many notable changes. During the DoS attack, mean CPU cycles rise from 474.36 million to 557.05 million, showing a significant amount of increase in the computational load, while standard deviation also grows from about 49.6 million to 10.99 million, reflecting more variability in the processing demands. Cache performance is mainly affected, with mean L1-dcache loads decreasing sharply from 166.90 million to 60.55 million, and L1-dcache load misses falling from 4.16 million to 2.26 million, meaning, meaning reduced cache efficiency during the attack. Standard deviation for L1-dcache load misses increases, thus, pointing towards more erratic performance of the cache.

The execution of the instructions has a dramatic fall in mean number from 492.36 million to 191.64 million, showing that the attack mainly impacts the system’s processing efficiency. Also, the maximum number of instructions executed fell from 1.58 billion to 350.62 million, further showing the strain on the given system. Branch operations also become less predictable, with the standard deviation of branch misses increasing from 183,182 to 206,308, reflecting more variability and potential inefficiencies in control flow.

Memory access patterns show a similar trend, with dTLB-load misses rising from 1.13 million to 2.26 million and iTLB-load misses increasing as well, indicating more frequent and erratic memory management issues during the attack. These changes demonstrate that the DoS attack imposes significant strain on the Pi system, leading to increased computational load, reduced cache efficiency, and greater variability in performance metrics, ultimately compromising the system's stability and reliability.

\subsection{Pi Data Box Plot Analysis}
\begin{figure}[htbp]
    \centering
    \begin{minipage}[b]{0.95\textwidth}
        \centering
        \includegraphics[width=\textwidth]{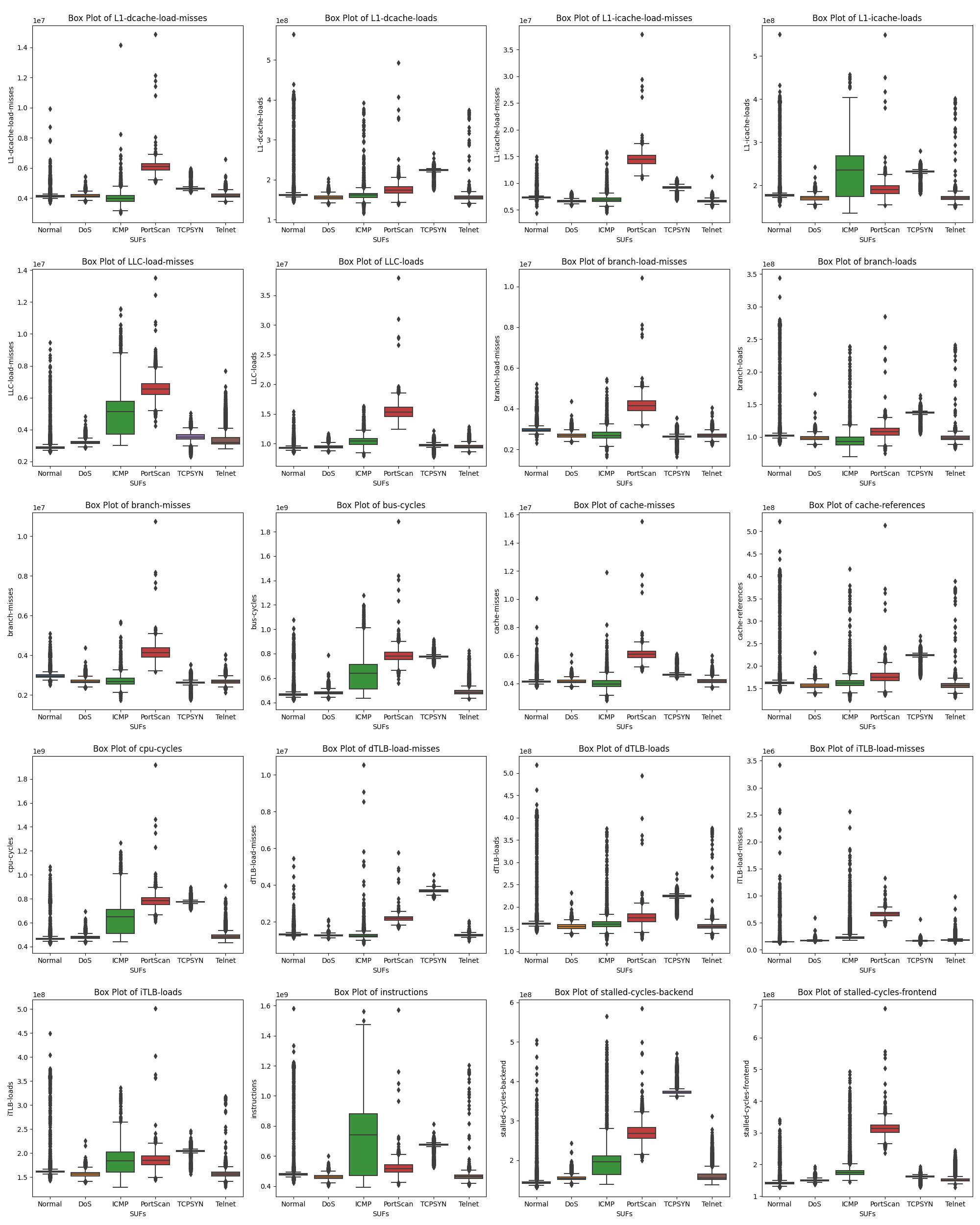}
        \caption{Box Plot of all features Pi}
        \label{fig:allpifeatures}
    \end{minipage}
\end{figure}

A BoxPlot visualises data distribution and outliers for comparative analysis in Figure \ref{fig:allpifeatures} below the overall observation.

\subsubsection{ICMP Flood}
\textbf{Significant Impact:} The ICMP flood attacks cause a noticeable increase across several key performance metrics, particularly in \texttt{Branch-Loads}, \texttt{Cache-References}, \texttt{CPU-Cycles}, and \texttt{Stalled-Cycles}. The box plots demonstrate that during an ICMP flood, the system experiences a significant rise in CPU utilisation and memory access, reflecting the high disruptive potential of this type of attack. This suggests that ICMP floods impose heavy demands on both the CPU and memory systems, leading to considerable inefficiencies.

\subsubsection{PortScan}
\textbf{High Branch-Related Metrics:} PortScan attacks result in a substantial increase in \texttt{Branch-Load-Misses}, \texttt{Branch-Misses}, and \texttt{L1-dcache-Load-Misses}. This indicates that PortScan activities induce complex branching operations, leading to more frequent branch mispredictions and data cache misses, which in turn increases the CPU load and reduces overall system efficiency.

\textbf{Moderate Impact:} In addition to the branch-related inefficiencies, PortScan moderately impacts other metrics such as \texttt{Bus-Cycles}, \texttt{dTLB-Loads}, and \texttt{Stalled-Cycles}. This broader impact suggests that PortScan attacks place a significant, though not overwhelming, demand on system resources, leading to increased processing and memory access overhead.

\subsubsection{DoS Attack}
\textbf{Minimal Impact:} Interestingly, the DoS attacks show performance metrics that are quite similar to those observed under normal traffic conditions for most indicators. This suggests that, despite the potential to overwhelm systems through high traffic volume, the per-request resource consumption of DoS attacks is not significantly different from normal traffic. This may indicate a more uniform distribution of resource demands, where the system is taxed primarily by the volume of requests rather than their complexity.

\subsubsection{TCPSYN Flood}
\textbf{Moderate Increase in Specific Metrics:} TCPSYN floods show moderate increases in \texttt{dTLB-Loads}, \texttt{Branch-Loads}, and \texttt{Branch-Misses}. This suggests that while the TCPSYN flood does not heavily tax the system across the board, it introduces notable overhead in specific areas, particularly in memory and branching operations, where the system's control flow and data retrieval are disrupted.

\subsubsection{Telnet Attack}
\textbf{Low to Moderate Impact:} The Telnet attacks generally result in low to moderate increases across most metrics, with some notable increases in \texttt{dTLB-Load-Misses} and \texttt{Branch-Load-Misses}. This indicates that while Telnet attacks do add some overhead to the system, they are less disruptive compared to ICMP floods or PortScans. The impact is more confined to specific operations, particularly in memory access and branch prediction.

\subsubsection{General Observation}
\textbf{Branch Prediction and Caching:} Metrics related to branch prediction such as \texttt{Branch-Load-Misses}, \texttt{Branch-Misses} and caching including \texttt{Cache-Misses}, \texttt{L1-dcache-Load-Misses} are significantly impacted by attack traffic, particularly by ICMP floods and PortScans. These attacks introduce inefficiencies in control flow and data retrieval operations, resulting in increased overhead and reduced performance.

\textbf{Stalled Cycles:} Both \texttt{Stalled-Cycles-Backend} and \texttt{Stalled-Cycles-Frontend} see significant increases during ICMP floods and PortScan attacks, indicating that these types of attacks cause substantial pipeline stalls. This delays instruction execution, leading to overall inefficiencies in processing.

\subsubsection{Efficient Attacks}
\textbf{DoS and TCPSYN:} DoS and TCPSYN floods appear to be more "efficient" in terms of resource usage per request, with many metrics remaining close to normal traffic levels. These attacks seem to rely more on the volume of requests rather than the complexity of each request to achieve their disruptive effects, making them less taxing on the system per individual operation.

%This input below is just an example and its not real !, need to be modifed for real impact 
%\input{chapters/AnalysisDevicePi}

  % Switch to landscape mode
\input{chapters/AnalysisDeviceLOWHIGH}

\subsection{Router Dataset Statistics}

%Table \ref{static_comparison_router_normal_dos} presents the statistical analysis of various performance metrics collected from Router datasets. 

\begin{table}[!ht]
    \centering
    \begin{adjustbox}{max width=\textwidth}
    \begin{tabular}{|l|l|l|l|l|l|l|l|l|}
\hline
\textbf{Feature} & \textbf{Min (Normal)} & \textbf{Min (DoS)} & \textbf{Max (Normal)} & \textbf{Max (DoS)} & \textbf{Std (Normal)} & \textbf{Std (DoS)} & \textbf{Mean (Normal)} & \textbf{Mean (DoS)} \\ \hline
branches                & 17,966,181 & 15,634,284 & 29,049,644 & 32,272,266 & 1,217,049.77 & 1,276,968.46 & 19,289,352.21 & 19,304,926.29 \\ \hline
branch-misses           & 6,026,238 & 4,807,422 & 9,493,656 & 9,389,385 & 157,480.84 & 206,308.01 & 8,359,985.69 & 8,347,051.75 \\ \hline
cache-misses            & 1,905,018 & 1,935,723 & 3,294,973 & 3,393,875 & 89,387.24 & 95,823.33 & 2,272,277.01 & 2,259,892.11 \\ \hline
cache-references        & 56,549,194 & 52,146,008 & 90,716,025 & 110,217,462 & 3,823,844.07 & 4,010,831.51 & 60,564,424.71 & 60,527,581.94 \\ \hline
cpu-cycles              & 517,061,015 & 520,860,422 & 646,152,372 & 648,209,507 & 10,591,683.72 & 10,988,368.99 & 556,009,406.4 & 557,045,053.5 \\ \hline
instructions            & 177,321,599 & 168,377,524 & 245,462,058 & 350,625,667 & 5,074,403.29 & 5,984,552.81 & 191,179,338.1 & 191,641,573.7 \\ \hline
stalled-cycles-backend  & 376,762,721 & 372,718,288 & 524,937,791 & 511,850,597 & 7,629,991.32 & 7,885,738.00 & 425,839,237.6 & 426,475,802.2 \\ \hline
stalled-cycles-frontend & 87,446,592 & 88,138,223 & 141,962,111 & 143,933,966 & 3,456,586.16 & 4,544,204.32 & 115,741,544.9 & 115,176,038.7 \\ \hline
L1-dcache-load-misses   & 1,878,334 & 1,960,085 & 3,374,429 & 3,346,366 & 82,635.92 & 89,843.13 & 2,272,111.18 & 2,259,923.93 \\ \hline
L1-dcache-loads         & 56,739,767 & 55,283,572 & 89,000,662 & 115,208,522 & 3,588,934.18 & 3,828,156.85 & 60,570,155.28 & 60,546,280.35 \\ \hline
L1-dcache-store-misses  & 1,956,609 & 1,942,378 & 3,339,243 & 3,353,678 & 83,049.47 & 89,682.62 & 2,272,186.13 & 2,259,914.00 \\ \hline
L1-dcache-stores        & 56,972,961 & 54,419,554 & 88,811,829 & 140,522,484 & 3,832,611.75 & 4,057,428.67 & 60,571,193.08 & 60,544,835.71 \\ \hline
L1-icache-load-misses   & 13,827,766 & 11,988,798 & 24,108,450 & 23,845,377 & 370,845.55 & 542,301.83 & 21,692,797.93 & 21,747,695.11 \\ \hline
branch-load-misses      & 6,722,120 & 6,135,233 & 9,518,328 & 9,422,957 & 159,992.08 & 200,509.49 & 8,358,113.57 & 8,350,060.70 \\ \hline
branch-loads            & 30,159,233 & 28,740,801 & 42,451,703 & 51,885,196 & 1,212,427.67 & 1,342,774.59 & 31,495,177.82 & 31,465,806.96 \\ \hline
dTLB-load-misses        & 1,614,382 & 1,619,787 & 6,056,816 & 6,504,337 & 587,344.22 & 634,449.75 & 2,152,572.12 & 2,259,046.00 \\ \hline
dTLB-store-misses       & 1,547,597 & 1,624,507 & 5,964,164 & 6,174,406 & 587,151.27 & 632,784.82 & 2,151,872.26 & 2,258,214.79 \\ \hline
iTLB-load-misses        & 443,194 & 422,318 & 978,109 & 1,040,035 & 45,928.79 & 77,770.21 & 661,489.87 & 695,236.23 \\ \hline
    \end{tabular}
    \end{adjustbox}
    \caption{Comparison of Statistics between Router Normal and Router DoS}
    \label{static_comparison_router_normal_dos}
\end{table}

The analysis of the collected performance data reveals several critical observations. Table \ref{static_comparison_router_normal_dos} compares the router's performance metrics under normal conditions and during a DoS attack, highlighting several key changes. During the DoS attack, the mean CPU cycles increase slightly from 556 million to 557 million, with the standard deviation also rising from approximately 10.6 million to 10.9 million, indicating added computational strain and greater variability in processing. Cache performance shows a minor decline, with L1-dcache loads decreasing from 60.57 million to 60.55 million, and L1-dcache load misses dropping slightly from 2.27 million to 2.26 million, while the standard deviation for both metrics increases, suggesting less efficient and more erratic data retrieval. The mean number of instructions executed remains close, increasing from 191.18 million to 191.64 million, but the maximum number of instructions rises significantly from 245 million to 350 million during the attack, reflecting the additional overhead. Branch operations exhibit more variability, with the standard deviation of branch misses increasing from 157,480 to 206,308, indicating less predictable control flow. Memory access patterns, reflected in dTLB and iTLB load misses, show slight increases in both mean values and standard deviations, further contributing to overall performance instability. These changes underscore that while the DoS attack does not drastically alter mean performance metrics, it significantly increases the variability and unpredictability of the router's operations, affecting system stability.

\subsection{Router Data Box Plot Analysis}
\begin{figure}[htbp]
    \centering
    \begin{minipage}[b]{0.95\textwidth}
        \centering
        \includegraphics[width=\textwidth]{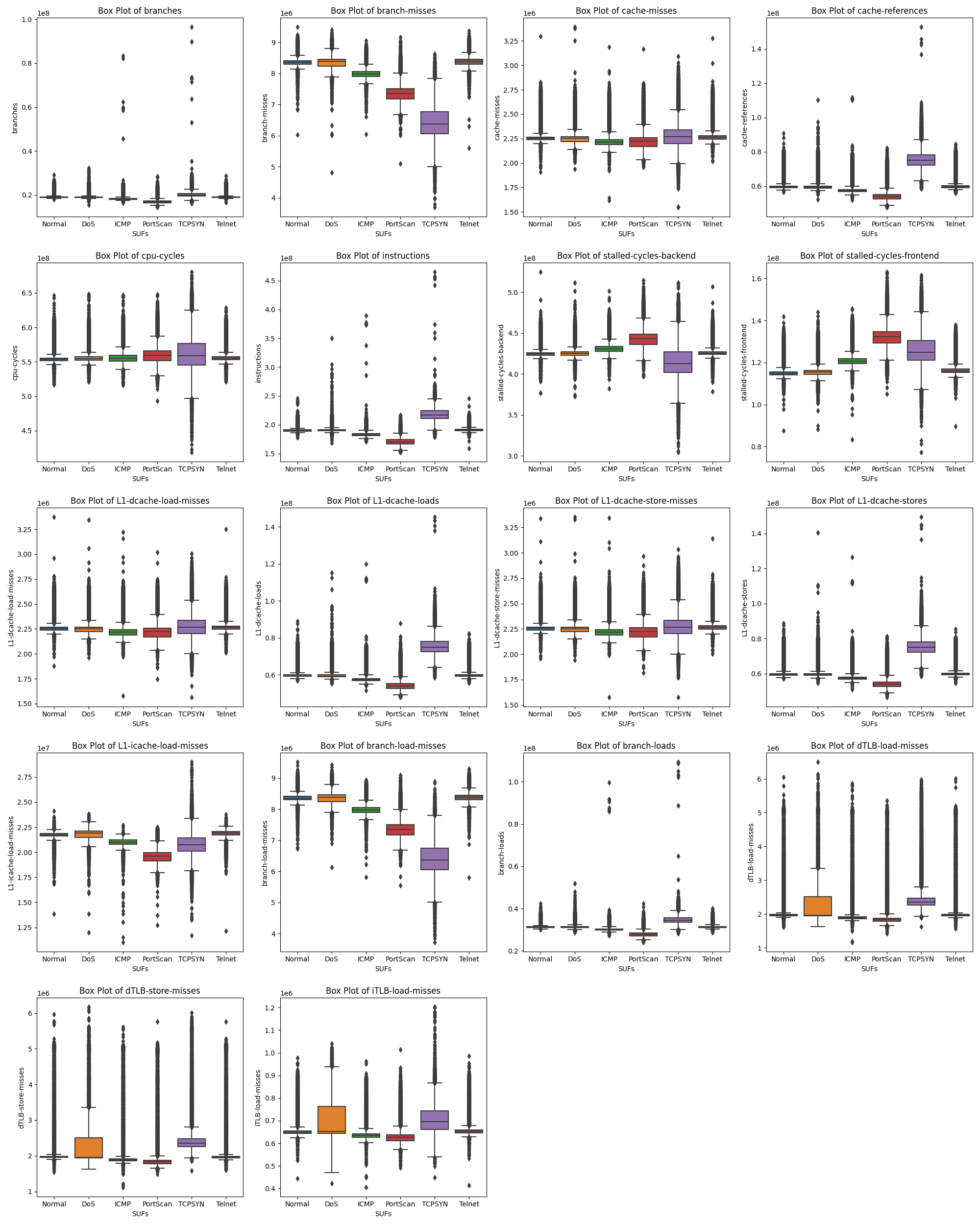}
        \caption{Box Plot of all features Router}
        \label{fig:allrouterfeatures}
    \end{minipage}
\end{figure}

A BoxPlot visualises data distribution and outliers for comparative analysis. Below is the observation for Figure \ref{fig:allrouterfeatures}

%explain more here ! 

\label{DataBoxPlotRouterAnalysis}

%I need to re write this since this data for the Pi not the router !!!!

\subsubsection{ICMP Flood}
\textbf{Significant Impact:} ICMP flood attacks exhibit a substantial increase across almost all performance metrics, notably in \texttt{Branch-Loads}, \texttt{Cache-References}, \texttt{CPU-Cycles}, and \texttt{Stalled-Cycles}. This suggests that ICMP floods are highly disruptive, placing a heavy load on both CPU and memory systems.

\subsubsection{PortScan}
\textbf{High Branch-Related Metrics:} PortScan attacks result in a high increase in \texttt{Branch-Load-Misses}, \texttt{Branch-Misses}, and \texttt{L1-dcache-Load-Misses}. This suggests that PortScan activities involve complex branching and data caching operations, leading to inefficiencies and higher CPU load.

\textbf{Moderate Impact:} PortScan also moderately increases metrics such as \texttt{Bus-Cycles}, \texttt{dTLB-Loads}, and \texttt{Stalled-Cycles}, reflecting its broader impact on system resources.

\subsubsection{DoS Attack}
\textbf{Minimal Impact:} DoS attacks show metrics similar to normal traffic for most performance indicators. This suggests that while DoS attacks can overwhelm systems through sheer volume, their per-request resource consumption is comparable to normal traffic.

\subsubsection{TCPSYN Flood}
\textbf{Moderate Increase in Specific Metrics:} TCPSYN floods show moderate increases in \texttt{dTLB-Loads}, \texttt{Branch-Loads}, and \texttt{Branch-Misses}, indicating that while the attack does not heavily tax the system across the board, it does create notable overhead in specific areas, especially related to memory and branching operations.

\subsubsection{Telnet Attack}
\textbf{Low to Moderate Impact:} Telnet attacks mainly show low to moderate rises across most of the metrics, with some of the notable increases in \textbf{dTLB-Load-Misses} and the \textbf{Branch-Load-Misses}. This indicates that while Telnet attacks do add some overhead, they are not as disruptive as ICMP floods or PortScans.

\subsubsection{General Observation}
\textbf{Branch Prediction and Caching:} Metrics related to branch prediction (e.g., \texttt{Branch-Load-Misses}, \texttt{Branch-Misses}) and caching (e.g., \texttt{Cache-Misses}, \texttt{L1-dcache-Load-Misses}) are notably impacted by attack traffic, especially by ICMP floods and PortScans. This highlights the inefficiencies introduced in the control flow and data retrieval operations due to malicious activities.

\textbf{Stalled Cycles:} Both \texttt{Stalled-Cycles-Backend} and \texttt{Stalled-Cycles-Frontend} show significant increases during ICMP floods and PortScan attacks, indicating that these attacks cause substantial pipeline stalls, delaying instruction execution.

\subsubsection{Efficient Attacks}
\textbf{DoS and TCPSYN:} DoS and TCPSYN floods appear to be more "efficient" in terms of resource usage per request, with many metrics remaining close to normal traffic levels. This suggests these attacks rely more on volume rather than per-request complexity to achieve their disruptive effects.

\input{chapters/AnalysisDeviceLowHighRouter}

\subsection{Camera Dataset Statistics} 

%Put the camera result statistics 
%The data looks 5ara so that we won't proceed with the experiments in AI. 

% After seeing the summary we will explain it.

%It will take a long time to debug such a problem so we decided to stop the experiments with the camera device. 

The analysis in Table \ref{static_comparison_camera_normal_dos} compares the Camera system’s performance metrics under normal conditions and during a DoS attack.

\begin{table}[!ht]
    \centering  % Center the table in landscape mode
    \begin{adjustbox}{max width=\textwidth}  % Adjust the box to fit the table within text width
    \begin{tabular}{|l|l|l|l|l|l|l|l|l|l|}
\hline
\textbf{Feature} & \textbf{Min (Normal)} & \textbf{Min (DoS)} & \textbf{Max (Normal)} & \textbf{Max (DoS)} & \textbf{Std (Normal)} & \textbf{Std (DoS)} & \textbf{Mean (Normal)} & \textbf{Mean (DoS)} \\ \hline
L1-dcache-load-misses  &     4,281,130 &    4,322,534 &     5,267,038 &    5,294,017 &  113,900 &  114,521 &   4,704,205 &  4,693,547 \\ \hline
L1-dcache-loads        &   117,124,080 &  117,291,625 &   140,714,585 &  141,185,913 &  2,754,811 &  2,791,226 &   126,937,900 &  126,626,200 \\ \hline
L1-dcache-store-misses &     4,297,568 &    4,311,756 &     5,215,186 &    5,266,076 &  114,203 &  116,002 &   4,704,670 &  4,694,242 \\ \hline
L1-dcache-stores       &   115,825,212 &  117,211,833 &   139,089,816 &  141,162,260 &  2,760,536 &  2,781,956 &   126,939,400 &  126,630,300 \\ \hline
branch-load-misses     &     3,603,866 &    3,653,834 &     4,324,037 &    4,309,900 &  81,757 &  82,127 &   3,938,898 &  3,931,060 \\ \hline
branch-loads           &    62,422,269 &   62,167,863 &    77,910,955 &   78,832,497 &  1,688,574 &  1,683,326 &   68,610,650 &  68,434,790 \\ \hline
branch-misses          &     3,571,262 &    3,563,203 &     4,344,720 &    4,438,531 &  89,116 &  88,561 &   3,939,422 &  3,929,731 \\ \hline
cache-misses           &     4,207,100 &    4,315,113 &     5,330,502 &    5,232,162 &  121,865 &  120,917 &   4,703,726 &  4,692,531 \\ \hline
cache-references       &   116,241,707 &  113,337,927 &   141,558,020 &  141,905,425 &  2,959,642 &  2,939,212 &   126,903,200 &  126,586,300 \\ \hline
cpu-cycles             &   636,781,334 &  630,483,070 &   781,004,435 &  758,318,664 &  15,344,260 &  15,280,020 &   694,333,900 &  692,280,000 \\ \hline
instructions           &   251,407,724 &  249,710,145 &   304,228,045 &  303,745,064 &  6,316,556 &  6,322,540 &   273,341,000 &  272,707,400 \\ \hline
\end{tabular}
    \end{adjustbox}
    \caption{Comparison of Statistics between Camera Normal and Camera DoS}
    \label{static_comparison_camera_normal_dos}
\end{table}

However, it is challenging to draw definitive conclusions due to issues encountered during data collection. The camera's architecture did not natively support the perf-tool, necessitating installing it from the source \cite{camera}. Unfortunately, this led to inaccuracies in how the metrics were captured by the tool, resulting in potential discrepancies in the data. As a result, the reliability of the collected data is questionable. Despite these challenges, the analysis shows only minor fluctuations in key metrics such as cache loads, branch operations, and CPU cycles between the normal and DoS states. The mean values for L1-dcache-loads and L1-dcache-load-misses, branch operations, and CPU cycles remained relatively stable, suggesting that the DoS attack might have had a limited impact on the camera's overall performance. However, the slight variations observed could be attributed to the data collection issues rather than actual changes in system behaviour, making it difficult to accurately assess the true impact of the DoS attack on the camera system. We conducted all planned experiments with the devices in person at TII over a one-month period. However, now that this period has ended, it was not possible to return to the company to redo the experiments as subsequent issues were identified. Based on the data gathered, we have excluded the camera from further analysis and removed it from future experiments.

%Given the complexities and time required to debug these issues, we decided to discontinue further experiments with the camera device, recognising that resolving the underlying problems would be lengthy and resource-intensive.

%\clearpage

\section{Model Implementation}
To fit the GRU model to our classification problem, we chose to go with a straightforward approach where we used the GRU model as a memory-based feature extractor. The role of the GRU is to learn which input sequences are relevant to a certain label. Given that the measurements could present a lot of variability even when belonging to the same label, we created sequences that gather a chain of almost instantaneous measurements that come from different counters. Based on the learned features, the model learns to output certain values (or embeddings) that are related to each label. The next layer is a linear fully connected layer that takes the output of the GRU on its input and learns to classify them according to the provided labels. We also chose to use an intermediate dropout layer to mitigate the dropout effect.

\input{chapters/CodeAI/GRU}

As shown in Listing \ref{code:GRU}, the GRU class has been defined with a first layer at the input that holds the embedding of the input sequence. The input sequence is transformed into an embedding vector that represents each input sequence and can be processed by the following layer. Next, the GRU layer takes the embedding vector at its input and outputs a high dimensional output based on the embedding vector. The final layer is a fully connected layer configured with an input dimension equal to the GRU output layer and output with the size of the provided labels.
The forward function defines how the different layers interact with each other, starting from an input sequence and getting a prediction by the end of the propagation by applying an argmax to the output probabilities provided by the GRU.

\section{Model Training}
Before training the model, we need to preprocess the data that have been loaded to normalise it and form the right sequences. We chose to use a MinMax scaler \cite{scikit-learn} for this purpose, which maps the measured values into a range between zero and one. The scaler formula is as follows:
%Put the right formula here 

\[
X_{\text{std}} = \frac{X_i - X_{\min(\text{axis}=0)}}{X_{\max(\text{axis}=0)} - X_{\min(\text{axis}=0)}}
\]

\[
X_{\text{scaled}} = X_{\text{std}} \cdot (\text{max} - \text{min}) + \text{min}
\]

We noticed that the HPCs' measurements are huge, which might interfere with the model's learning process. We opted to normalise the data using z-statistics. 

As shown in the following Listing \ref{code:Normalized} code snippet, after selecting only the relevant elements of the dataframe (The numerical values), we initialise our MinMaxScaler() by calling its fit function; this function goes through the whole dataframe and performs the computation of the scaling parameters for each feature on the tabular dataframe. The last lines output the scaled features on a copy of the original tabular data format, which will be used in the next step to generate the sequences and train the model. 
\input{chapters/CodeAI/normalizing}

The scaled tabular output is handled by the following Listing \ref{code:sequenceCreation} code snippet. The createsequences() function takes the labels and the scaled data separately; the main loop of the function iterates through the input and builds a sequence composed from the different feature values separated by a space. 
\input{chapters/CodeAI/Sequance}
%Dataset object padding of the input
Since we used the generic structure defined by the torch lightening framework, we implemented the following class, which represents and configures the sequences batches required to train the model. The Dataset class, as shown in Listing \ref{code:dataset}, takes the tabular frame as an input and then encodes the formed sequences using a very basic encoding that is formed from the input grammar. The object returns a tensor for each input sequence that could be consumed by the embedding layer of the model.
\input{chapters/CodeAI/HPCData}

The final step consists of implementing a LightningModule before being able to use the trainer object for the actual training of the model. As shown on the Listing \ref{code:modelTraining} code snippet, the defined class contains the necessary step to define the hyperparameters of the module during the instantiation of the object we chose to go with an input dimension equal to the maximum length of our input sequences and padding the shorter sequences. The truncation would also work for the model, but we opted for padding to avoid any information loss. The forward step operation is called from this class by just pointing to the default stepping function of the main model class. We also chose to use an Adam Optimiser because it fits our implemented model. After many alterations with other values, the learning rate has been fixed to 0.001. The learning rate that has been used generates fewer oscillations during the training in terms of validation loss. We also chose to use cross-entropy loss as a metric for the model, given that it provides better convergence during the training process.
%Model training
\input{chapters/CodeAI/HPCCLASIFIER2}

% Put the validation/training loss plots
% explain that the stop condition was validation loss; we based our training on selecting the best-performing epoch according to the validation loss.

\subsection{Model Training Evaluation}

%Here, I need to put a sample of the validation of one of the devices to see and say the reset could be found in the Appendix.  

We used W\&B tool \cite{wandb}, a popular tool used in machine learning for experiment tracking and visualisation. It was integrated into the code by initialising a WandB run and logging relevant data. The data logged to WandB is then accessible via the web dashboard, where we can monitor and analyse the experiments.

During the training process, we used validation loss, as shown in figure \ref{Val_Loss_Pi} and \ref{Val_Loss_Router}, as the stop condition to ensure optimal model performance and prevent overfitting. Specifically, we monitored the validation loss at the end of each epoch, as seen in Figure \ref{Epoch_Pi} and \ref{Epoch_Router}, and once it stopped decreasing, we halted the training process. This approach, known as early stopping, helps in selecting the model that generalises best to unseen data. Consequently, the model from the epoch with the lowest validation loss was chosen as the final model, as it represents the point where the model performed best on the validation set, striking an ideal balance between fitting the training data and maintaining generalisation to new data.

\input{chapters/wandbPi/wandbPI}

\input{chapters/wandbPi/wandbRouter}

%% file: chapters/Code/DoSCode.tex
\begin{lstlisting}[language=C, 
                   frame=lines, 
                   numbers=left, 
                   basicstyle=\footnotesize\ttfamily, 
                   keywordstyle=\bfseries, 
                   breaklines=true, 
                   showstringspaces=false,
                   caption={DoS Functions Code snippet},
                   label={code:DoSAttack}]
void httpFlood(char *address, int port){
    while (1){
        fd = tcpSockConnect(address, port);
        write(fd, "GET / HTTP/1.1\r\n", strlen("GET / HTTP/1.1\r\n"));
        bzero(buffer, BUFFER_SIZE);
    }
}
\end{lstlisting}

%% file: chapters/Code/HTTP.tex
\begin{lstlisting}[language=C, 
                   frame=lines, 
                   numbers=left, 
                   basicstyle=\footnotesize\ttfamily, 
                   keywordstyle=\bfseries, 
                   breaklines=true, 
                   showstringspaces=false,
                   caption={TCPSYN Functions Code snippet},
                   label={code:TCPSYNAttack}]
int SYNflood(char *source_ip, char *destination_ip, int port) {
    int s = socket(PF_INET, SOCK_RAW, IPPROTO_TCP);
    char datagram[4096];
    memset(datagram, 0, 4096);

    while (1) {
        sendto(s, datagram, iph->tot_len, 0, (struct sockaddr *) &sin, sizeof(sin));
    }

    return 0;
}
\end{lstlisting}

%% file: chapters/Code/port_scan.tex
\begin{lstlisting}[language=C,
                   frame=lines,
                   numbers=left,
                   basicstyle=\footnotesize\ttfamily,
                   keywordstyle=\bfseries,
                   breaklines=true,
                   showstringspaces=false,
                   caption={Port Scan Functions Code snippet},
                   label={code:PortAttack}]
int scanHost(char *ip_destination, int port){

    printf("scanning port no: %d", port);
    sa.sin_port = htons(port);
    // Create a socket of type internet
    sock = socket(AF_INET , SOCK_STREAM , 0);
    // Connect using that socket and sockaddr structure
    err = connect(sock , (struct sockaddr*)&sa , sizeof sa);
    if(sock < 0) 
    {
        printf("Couldn't open the socket !\n");
        perror("\nSocket");
        exit(1);
    }
    if(err < 0){
        printf("Port %d closed !\n", port);
        fflush(stdout);        
    }
    else
    {
        printf("%d open\n",  port);
        close(sock);
    }
    return port;
}
\end{lstlisting}

%% file: chapters/Code/telnet.tex
\begin{lstlisting}[language=C, 
                   frame=lines, 
                   numbers=left, 
                   basicstyle=\footnotesize\ttfamily,
                   keywordstyle=\bfseries, 
                   breaklines=true, 
                   showstringspaces=false,
                   caption={Telnet Functions Code snippet},
                   label={code:TelnetAttack}]
if (connect(sockfd, (struct sockaddr*)&serv_addr, sizeof(serv_addr)) < 0) {

    printf("Please enter the message: ");
    bzero(buffer, 256);
    fgets(buffer, 255, stdin);

    /* Send message to the server */
    n = write(sockfd, buffer, strlen(buffer));

    /* Now read server response */
    bzero(buffer, 256);
    n = read(sockfd, buffer, 255);

    printf("%s\n", buffer);
    return 0;
}
\end{lstlisting}

%% file: chapters/Code/ICMP.tex
\begin{lstlisting}[language=C, 
                   frame=lines, 
                   numbers=left, 
                   basicstyle=\footnotesize\ttfamily,
                   keywordstyle=\bfseries, 
                   breaklines=true, 
                   showstringspaces=false,
                   caption={ICMP Functions Code snippet},
                   label={code:ICMPAttack}]
while (1)
{
    memset(packet + sizeof(struct iphdr) + sizeof(struct icmphdr), rand() % 255, payload_size);
    icmp->checksum = 0;
    icmp->checksum = in_cksum((unsigned short *)icmp, sizeof(struct icmphdr) + payload_size);
     
    if ((sent_size = sendto(sockfd, packet, 
        packet_size, 0, (struct sockaddr*) &servaddr, sizeof(servaddr))) < 1) 
    {
        perror("send failed\n");
        break;
    }
    ++sent;
    printf("%d packets sent\r", sent);
    fflush(stdout);
     
    usleep(10000);  // microseconds
}
\end{lstlisting}

%% file: chapters/HPCeventsName.tex
\begin{table}[!ht]
\centering
\footnotesize
\begin{tabular}{|p{2cm}|p{4cm}|p{6.5cm}|}
\hline
\textbf{Category} & \textbf{Event} & \textbf{Description} \\ \hline
\textbf{CPU Events} & Instructions & Total instructions processed by the CPU \\ \cline{2-3}
 & CPU cycles & Total CPU cycles, indicating task duration \\ \cline{2-3}
 & Branches & Total branch instructions executed \\ \cline{2-3}
 & Branch misses & Incorrectly predicted branches, leading to performance penalties \\ \cline{2-3}
 & Branch loads & Load operations involving branch instructions \\ \cline{2-3}
 & Branch load misses & Load operations involving branch instructions that missed \\ \cline{2-3}
 & Stalled cycles backend & Cycles the backend of the pipeline is stalled \\ \cline{2-3}
 & Stalled cycles frontend & Cycles the frontend of the pipeline is stalled \\ \hline
\textbf{Cache Events} & Cache references & Total accesses to the cache \\ \cline{2-3}
 & Cache misses & Total cache accesses that miss \\ \cline{2-3}
 & L1 dcache loads & Load operations from the L1 data cache \\ \cline{2-3}
 & L1 dcache load misses & Load operations that miss in the L1 data cache \\ \cline{2-3}
 & L1 dcache stores & Store operations to the L1 data cache \\ \cline{2-3}
 & L1 dcache store misses & Store operations that miss in the L1 data cache \\ \cline{2-3}
 & L1 icache load misses & Instruction fetches that miss in the L1 instruction cache \\ \cline{2-3}
 & L1 icache loads & Instruction fetches from the L1 instruction cache \\ \cline{2-3}
 & LLC loads & Load operations from the last level cache \\ \cline{2-3}
 & LLC load misses & Load operations that miss in the last level cache \\ \hline
\textbf{Memory Events} & dTLB loads & Data accesses in the data TLB \\ \cline{2-3}
 & dTLB load misses & Failed data loads in the data TLB \\ \cline{2-3}
 & dTLB store misses & Failed data stores in the data TLB \\ \cline{2-3}
 & iTLB loads & Instruction accesses in the instruction TLB \\ \cline{2-3}
 & iTLB load misses & Failed instruction loads in the instruction TLB \\ \cline{2-3}
 & Bus cycles & Cycles involving bus activities \\ \hline
\textbf{Custom Events} & Vendor-specific events & Based on core implementation (e.g. ARM, Intel) \\ \hline
\end{tabular}
\caption{Perf Events}
\label{Table:PerfEvents}
\end{table}

%% file: tables/Design/HPCEventsdevices.tex
\begin{table}[!ht]
\begin{threeparttable}
    \centering
    \normalsize
    \large 
    \begin{tabular}{llccc}
        \toprule
        \textbf{Method} & \textbf{Description} &
        \textbf{Device A\tnote{1}} & \textbf{Device B\tnote{2}} & \textbf{Device C\tnote{3}} \\
        \midrule
        I   & Instructions &  \cmark  &  \cmark & \cmark \\
        B  & Branches & \xmark  &  \cmark & \xmark \\
        C\_C  & CPU cycles &  \cmark  &  \cmark & \cmark \\
        B\_M  & Branch misses &  \cmark  &  \cmark & \cmark \\
        B\_L  & Branch loads &  \cmark  &  \cmark & \cmark \\
        B\_L\_M & Branch load misses &  \cmark  &  \cmark & \cmark \\
        SCB & Stalled cycles backend &  \cmark  &  \cmark & \xmark \\
        SCF & Stalled cycles frontend &  \cmark  &  \cmark & \xmark \\
        CR  & Cache references &  \cmark  &  \cmark & \cmark \\
        CM  & Cache misses &  \cmark  &  \cmark & \cmark \\
        L1\_DL  & L1 dcache loads &  \cmark  &  \cmark & \cmark \\
        L1\_DLM & L1 dcache load misses  & \cmark  &  \cmark & \cmark \\
        L1\_DS  & L1 dcache stores & \xmark  &  \cmark & \cmark \\
        L1\_DSM & L1 dcache store misses & \xmark  &  \cmark & \cmark \\
        L1\_IL & L1 icache loads  & \cmark  &  \xmark & \xmark \\
        L1\_ILM & L1 icache load misses  & \cmark  &  \cmark & \xmark \\
        LLC\_L  & LLC loads & \cmark  &  \xmark & \xmark \\
        LLC\_LM  & LLC loads misses & \cmark  &  \xmark & \xmark \\
        DTLB\_L & dTLB loads & \cmark  &  \xmark & \xmark \\
        DTLB\_LM & dTLB load misses &  \cmark  &  \cmark & \xmark \\
        DTLB\_SM & dTLB store misses & \xmark  &  \cmark & \xmark \\
        ITLB\_L & iTLB loads & \cmark  &  \xmark & \xmark \\
        ITLB\_LM & iTLB load misses &  \cmark  &  \cmark & \xmark \\
        B\_C & Bus cycles & \cmark  &  \xmark & \xmark \\
        \bottomrule
    \end{tabular}
    \caption{HPC events used on a per-device basis.}
    \label{table:HPC-events}
    \begin{tablenotes}\footnotesize
\item[1] Pi5: Broadcom BCM2712 SoC 
\item[2] Turris Omnia: Marvell Armada 385 SoC 
\item[3] Camera : Tegra SoC 3 
\end{tablenotes}
\end{threeparttable}
\end{table}

\clearpage

%% file: chapters/CodeBash/Collection.tex
\begin{lstlisting}[language=bash, frame=lines, numbers=left,
                   basicstyle=\footnotesize\ttfamily,
                   caption={Collection.sh Code snippet},
                   label={code:collection}]
if [ -z "$5" ]; then
        while true; do
        ("./"$2 $3 $4 &)
        done
else
        ("./"$2 $3 $4 $5 &)
fi
(sudo perf stat -e <Events_Name> -a -I 200 2>"data"$2".txt" & )
\end{lstlisting}

%% file: chapters/CodeBash/StartCollection.tex
\begin{lstlisting}[language=bash, frame=lines, numbers=left,
                   basicstyle=\footnotesize\ttfamily,
                   caption={Startcollection.sh Code snippet},
                   label={code:startcollecting}]
echo "Collecting for Icmp"
sudo bash ./collection.sh $1 icmp_pi 192.168.1.133 192.168.1.235 10

echo "Collecting for Telnet"
sudo bash ./collection.sh $1 telnet_pi 192.168.1.235 0 0
    }
}
\end{lstlisting}

%% file: chapters/CodeData/Data_Parsing.tex
%data formating
\begin{lstlisting}[language=Python,
                   frame=lines,
                   numbers=left,
                   basicstyle=\footnotesize\ttfamily,
                   keywordstyle=\bfseries,
                   breaklines=true,
                   showstringspaces=false,
                   caption={Data Parsing Code snippet},
                   label={code:DataPreprocessingParsing}]
# Read the perf_output.txt file
file_name = 'data_telnet_arm.txt'
with open(file_name, 'r') as f:
    data = f.readlines()
 
#print(data)
# Clean and format the data
data_cleaned = []
for d in data[1:]:
    data_cleaned.append(d.split(" "))
 
data_formated = []
for d in data_cleaned:
    clean_elem = [e for e in d if e != '']
    data_formated.append(clean_elem)
 
#print(data_formated)
data_formated2 = []
 
for d in data_formated:
    if d[0] == "#" or d[1] == "#":
        pass
    else:
        data_formated2.append(d)
\end{lstlisting}

%% file: chapters/CodeData/Data_Prepocessing.tex
\begin{lstlisting}[language=Python,
                   frame=lines,
                   numbers=left,
                   basicstyle=\footnotesize\ttfamily,
                   keywordstyle=\bfseries,
                   breaklines=true,
                   showstringspaces=false,
                   caption={Data preprocessing Code snippet},
                   label={code:DataPreprocessing}]
data_exp = {}
for d in data_formated2:
    if '#' not in d[0] and '#' not in d[1]:
        tup = (d[2], d[1])
        mes = list()
        mes.append(tup)
        if d[0] not in data_exp.keys():
            data_exp[d[0]] = mes
        else:
            data_exp[d[0]].append(tup)
#print(data_exp[list(data_exp.keys())[0]])
 
with open(file_name+".csv", "+w") as f:
    first_line = "time," + ",".join([e[0] for e in data_exp[list(data_exp.keys())[0]]]) + "\n"
    #print(first_line)
    f.write(first_line)
    for k in list(data_exp.keys()):
        line_data = k + "," + ",".join([e[1] for e in data_exp[k]]) + "\n"
        #print(line_data)
        f.write(line_data)
\end{lstlisting}

%% file: chapters/AnalysisDeviceLOWHIGH.tex
\begin{center}
\scriptsize
\begin{longtable}{|l|c|c|c|c|c|c|}
    \hline
    \textbf{Feature} & \textbf{Normal} & \textbf{DoS} & \textbf{ICMP} & \textbf{PortScan} & \textbf{TCPSYN} & \textbf{Telnet} \\
    \hline
    \endfirsthead
    \hline
    \textbf{Feature} & \textbf{Normal} & \textbf{DoS} & \textbf{ICMP} & \textbf{PortScan} & \textbf{TCPSYN} & \textbf{Telnet} \\
    \hline
    \endhead
    \hline \multicolumn{7}{|r|}{{Continued on next page}} \\ \hline
    \endfoot
    \endlastfoot
    Branch-load-misses & Baseline & Slightly Low & High & High & Slightly Low & Slightly High \\
    \hline
    Branch-loads & Baseline & Slightly Low & Slightly Low & High & Slightly Low & Slightly High \\
    \hline
    Branch-misses & Baseline & Slightly Low & Slightly Low & High & Slightly Low & Slightly High \\
    \hline
    Bus-cycles & Baseline & Slightly Low & Low & High & Slightly Low & Slightly High \\
    \hline
    Cache-misses & Baseline & Slightly Low & Low & High & Slightly Low & Slightly High \\
    \hline
    Cache-references & Baseline & Slightly Low & Low & High & Slightly Low & Slightly High \\
    \hline
    CPU-cycles & Baseline & Slightly Low & Low & High & Slightly Low & Slightly High \\
    \hline
    dTLB-load-misses & Baseline & Slightly Low & Low & High & Slightly Low & Slightly High \\
    \hline
    dTLB-loads & Baseline & Slightly Low & Low & High & Slightly Low & Slightly High \\
    \hline
    Instructions & Baseline & Slightly Low & Low & High & Slightly Low & Slightly High \\
    \hline
    iTLB-load-misses & Baseline & Low & High & Slightly High & Low & Slightly High \\
    \hline
    iTLB-loads & Baseline & Low & High & Slightly High & Low & Slightly High \\
    \hline
    L1-dcache-load-misses & Baseline & Low & High & Higher & Slightly Low & Slightly High \\
    \hline
    L1-dcache-loads & Baseline & Low & High & Slightly High & Slightly Low & Slightly High \\
    \hline
    L1-icache-load-misses & Baseline & Low & High & Slightly High & Low & Slightly High \\
    \hline
    L1-icache-loads & Baseline & Low & High & Slightly High & Low & Slightly High \\
    \hline
    LLC-load-misses & Baseline & Low & High & Slightly High & Low & Slightly High \\
    \hline
    LLC-loads & Baseline & Low & High & Slightly High & Low & Slightly High \\
    \hline
    Stalled-cycles-backend & Baseline & Low & High & Slightly High & Low & Slightly High \\
    \hline
    Stalled-cycles-frontend & Baseline & Low & High & Slightly High & Low & Slightly High \\
    \hline
    \caption{BoxPlot Summary Pi}
    \label{Table:PiBoxPLOT}
\end{longtable}
\end{center}

%% file: chapters/AnalysisDeviceLowHighRouter.tex
\begin{center}
\scriptsize
\begin{longtable}{|l|c|c|c|c|c|c|}
\caption{BoxPlot Summary Router} \label{Table:RouterBoxPLOT} \\

\hline
\textbf{Feature} & \textbf{Normal} & \textbf{DoS} & \textbf{ICMP} & \textbf{PortScan} & \textbf{TCPSYN} & \textbf{Telnet} \\
\hline
\endfirsthead

\hline
\textbf{Feature} & \textbf{Normal} & \textbf{DoS} & \textbf{ICMP} & \textbf{PortScan} & \textbf{TCPSYN} & \textbf{Telnet} \\
\hline
\endhead

\hline \multicolumn{7}{|r|}{{Continued on next page}} \\ \hline
\endfoot

\endlastfoot

    Branches & Baseline & Slightly Low & Low & High & Slightly Low & Slightly High \\
    \hline
    Branch-Misses & Baseline & Slightly Low & Low & High & Slightly Low & High \\
    \hline
    Cache-Misses & Baseline & Slightly Low & Low & High & Slightly Low & High \\
    \hline
    Cache-References & Baseline & Slightly Low & Low & High & Slightly Low & High \\
    \hline
    CPU-Cycles & Baseline & Slightly Low & Low & High & Slightly Low & High \\
    \hline
    Instructions & Baseline & Slightly Low & Low & High & Slightly Low & High \\
    \hline
    Stalled-Cycles-Backend & Baseline & Low & High & Slightly High & Low & High \\
    \hline
    Stalled-Cycles-Frontend & Baseline & Low & High & Slightly High & Low & High \\
    \hline
    L1-dcache-Load-Misses & Baseline & Low & High & Slightly High & Low & High \\
    \hline
    L1-dcache-Loads & Baseline & Low & High & Slightly High & Low & High \\
    \hline
    L1-dcache-Store-Misses & Baseline & Low & High & Slightly High & Low & High \\
    \hline
    L1-dcache-Stores & Baseline & Low & High & Slightly High & Low & High \\
    \hline
    L1-icache-Load-Misses & Baseline & Low & High & Slightly High & Low & High \\
    \hline
    L1-icache-Loads & Baseline & Low & High & Slightly High & Low & High \\
    \hline
    LLC-Load-Misses & Baseline & Low & High & Slightly High & Low & High \\
    \hline
    LLC-Loads & Baseline & Low & High & Slightly High & Low & High \\
    \hline
    dTLB-Load-Misses & Baseline & Slightly Low & Low & High & Slightly Low & High \\
    \hline
    iTLB-Load-Misses & Baseline & Low & High & Slightly High & Low & High \\
    \hline

\end{longtable}
\end{center}

%% file: chapters/CodeAI/GRU.tex
\begin{lstlisting}[language=Python,
                   frame=lines,
                   numbers=left,
                   basicstyle=\footnotesize\ttfamily,
                   keywordstyle=\bfseries,
                   breaklines=true,
                   showstringspaces=false,
                   caption={GRU Code snippet},
                   label={code:GRU}]
class GRUModel(nn.Module):
    def __init__(self, input_dim, hidden_dim, output_dim, num_layers=1):
        super(GRUModel, self).__init__()
        self.embedding = nn.Embedding(input_dim, hidden_dim)
        self.gru = nn.GRU(hidden_dim, hidden_dim, num_layers, batch_first=True)
        self.fc = nn.Linear(hidden_dim, output_dim)
        
    def forward(self, x, lengths):
        x = self.embedding(x)
        lengths_cpu = lengths.cpu()  # Move lengths to CPU
        packed_input = pack_padded_sequence(x, lengths_cpu, batch_first=True, enforce_sorted=False)
        packed_output, _ = self.gru(packed_input)
        output, _ = pad_packed_sequence(packed_output, batch_first=True)
        out = self.fc(output[range(len(output)), lengths - 1])
        predictions = out.argmax(-1)
        #print(predictions)
        return out, predictions
\end{lstlisting}

%% file: chapters/CodeAI/normalizing.tex
%Normalization
\begin{lstlisting}[language=Python,
                   frame=lines,
                   numbers=left,
                   basicstyle=\footnotesize\ttfamily,
                   keywordstyle=\bfseries,
                   breaklines=true,
                   showstringspaces=false,
                   caption={Normalised Code snippet},
                   label={code:Normalized}]
from sklearn.preprocessing import MinMaxScaler

scaler = MinMaxScaler()

normalised_data = scaler.fit_transform(df)

normalised_df = pd.DataFrame(normalised_data, columns=df.columns)
\end{lstlisting}

%% file: chapters/CodeAI/Sequance.tex
\begin{lstlisting}[language=Python,
                   frame=lines,
                   numbers=left,
                   basicstyle=\footnotesize\ttfamily,
                   keywordstyle=\bfseries,
                   breaklines=true,
                   showstringspaces=false,
                   caption={Sequence creation code snippet},
                   label={code:sequenceCreation}]
import numpy as np
import pandas as pd
from sklearn.preprocessing import StandardScaler

def create_sequences(data, label):
    sequences = []
    labels = []
    for i in range(len(data)):
        sequences.append(" ".join(data.iloc[i].values.astype(str)))
        labels.append(label.iloc[i])
    return sequences, labels

X, Y = create_sequences(normalized_df, df["label"])
\end{lstlisting}

%% file: chapters/CodeAI/HPCData.tex
\begin{lstlisting}[language=Python,
                   frame=lines,
                   numbers=left,
                   basicstyle=\footnotesize\ttfamily,
                   keywordstyle=\bfseries,
                   breaklines=true,
                   showstringspaces=false,
                   caption={Dataset creation for the model trainer code snippet},
                   label={code:dataset}]
import torch
from torch.utils.data import Dataset, DataLoader

class HPCDataset(Dataset):
    def __init__(self, X, Y):
        #self.data = df
        self.hpc = X
        self.labels = Y
        
        self.char_to_idx = {char: idx for idx, char in enumerate(set(''.join(self.hpc)))}
        self.idx_to_char = {idx: char for char, idx in self.char_to_idx.items()}
        
    def __len__(self):
        return len(self.hpc)
    def __getitem__(self, idx):
        hpc = self.hpc[idx]
        label = self.labels[idx]
        hpc_encoded = [self.char_to_idx[char] for char in hpc]
        return torch.tensor(hpc_encoded, dtype=torch.long), torch.tensor(label, dtype=torch.long)
\end{lstlisting}

%% file: chapters/CodeAI/HPCCLASIFIER2.tex
%classifier
\begin{lstlisting}[language=Python,
                   frame=lines,
                   numbers=left,
                   basicstyle=\footnotesize\ttfamily,
                   keywordstyle=\bfseries,
                   breaklines=true,
                   showstringspaces=false,
                   caption={GRU Model Training},
                   label={code:modelTraining}]
class HPCClassifier2(pl.LightningModule):
    def __init__(self, input_dim, hidden_dim, output_dim, num_layers=1, learning_rate=0.001):
        super(HPCClassifier2, self).__init__()
        self.model = GRUModel(input_dim, hidden_dim, output_dim, num_layers)
        self.loss_fn = nn.CrossEntropyLoss()
        self.learning_rate = learning_rate
        self.validation_step_outputs = []
    
    def forward(self, x, lengths):
        return self.model(x, lengths)
    
    def training_step(self, batch, batch_idx):
        x, lengths, y = batch
        y_hat, pred = self(x, lengths)
        loss = self.loss_fn(y_hat, y)
        self.log('train_loss', loss)
        return loss
    
    def configure_optimizers(self):
        return optim.Adam(self.parameters(), lr=self.learning_rate)
        
input_dim = len(dataset.char_to_idx)
hidden_dim = 128
output_dim = len(set(dataset.labels))
model = HPCClassifier2(input_dim, hidden_dim, output_dim)
\end{lstlisting}

%% file: chapters/wandbPi/wandbPI.tex
\begin{figure}[!htb]
\minipage{0.3\textwidth}
\includegraphics[width=\linewidth]{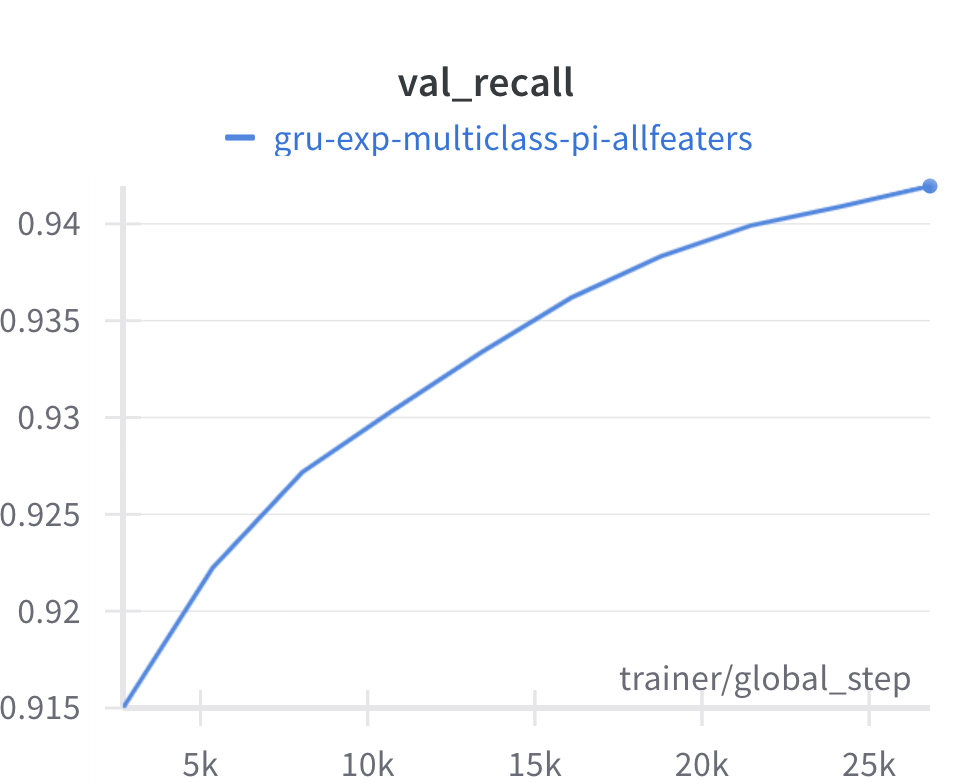}
\caption{Recall Pi}
\label{Recall_Pi}
\endminipage\hfill
\minipage{0.3\textwidth}
\includegraphics[width=\linewidth]{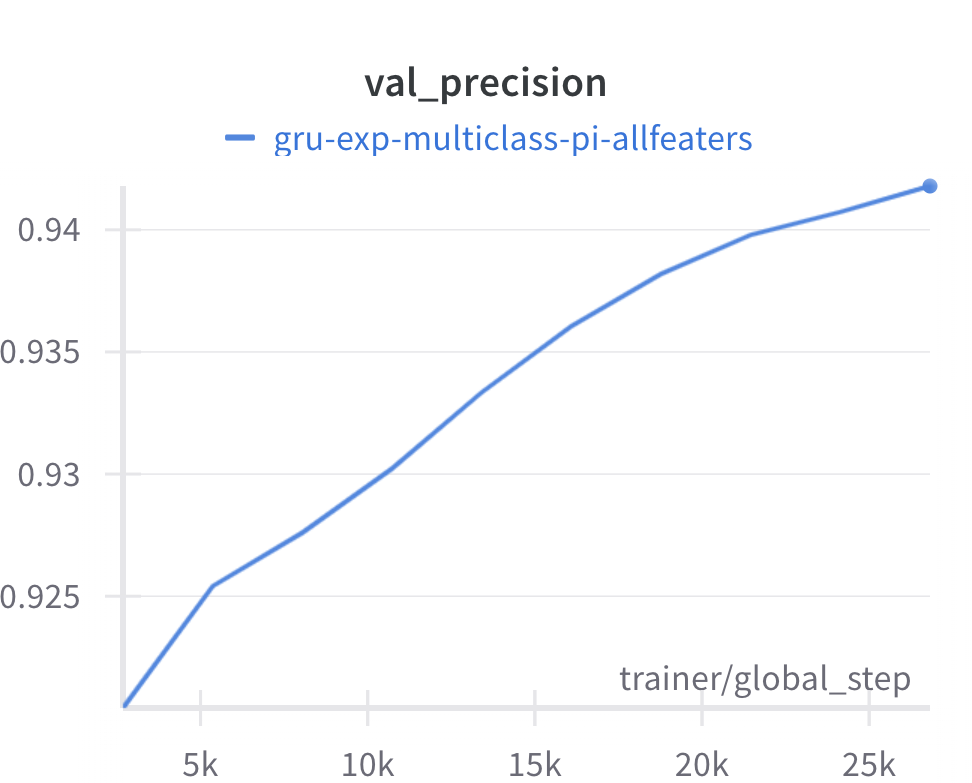}
\caption{Precision Pi}
\label{Precision_Pi}
\endminipage\hfill
\minipage{0.3\textwidth}
\includegraphics[width=\linewidth]{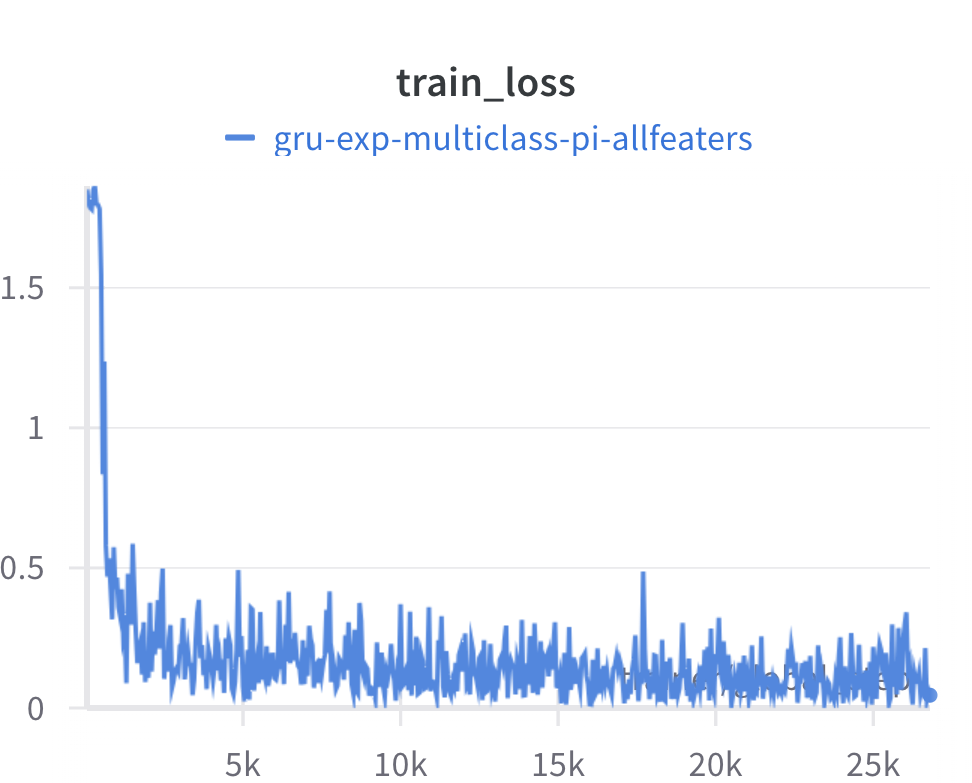}
\caption{Train loss Pi}
\label{Train_Loss_Pi}
\endminipage\hfill
\minipage{0.3\textwidth}
\includegraphics[width=\linewidth]{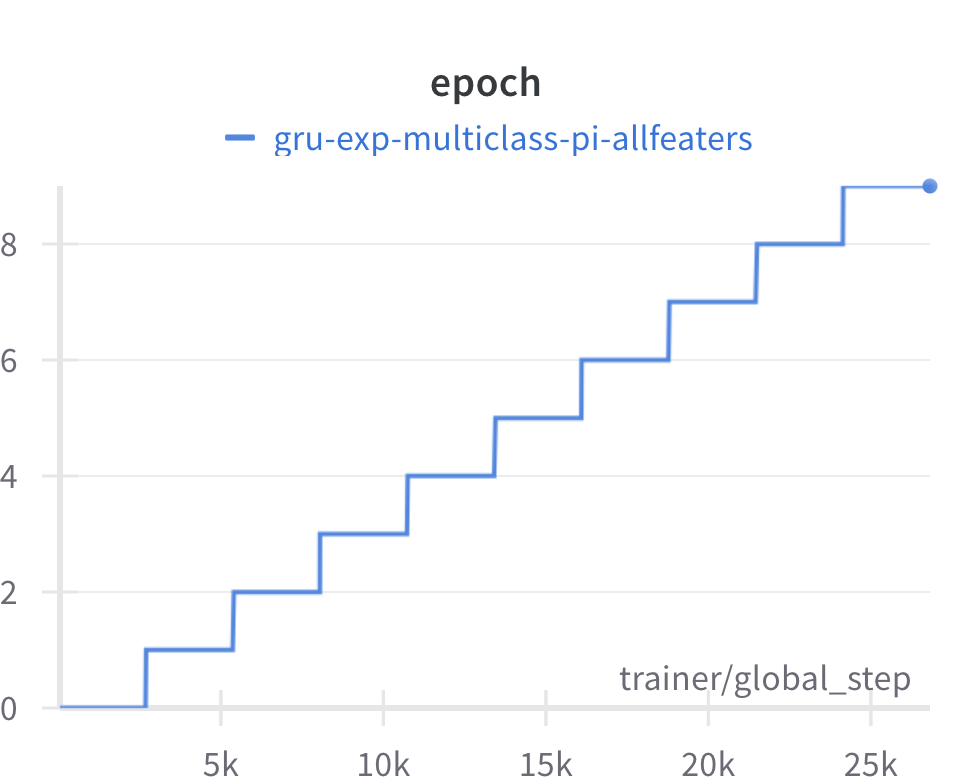}
\caption{Epoch Pi}
\label{Epoch_Pi}
\endminipage\hfill
\minipage{0.3\textwidth}
\includegraphics[width=\linewidth]{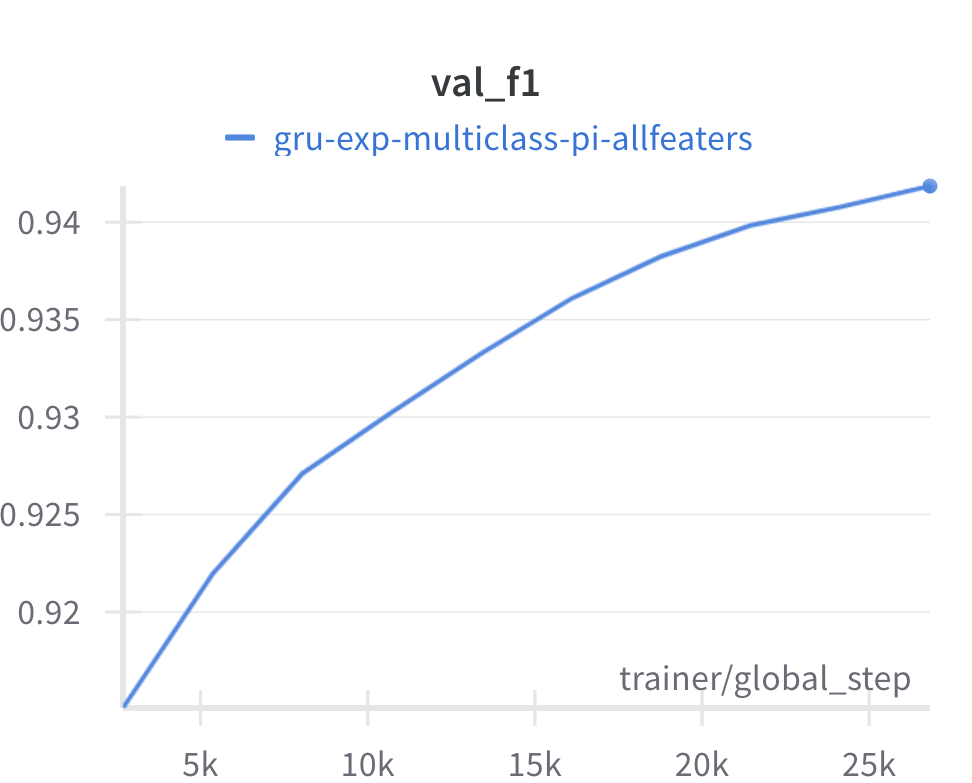}
\caption{F1 Pi}
\label{F1_Pi}
\endminipage\hfill
\minipage{0.3\textwidth}
\includegraphics[width=\linewidth]{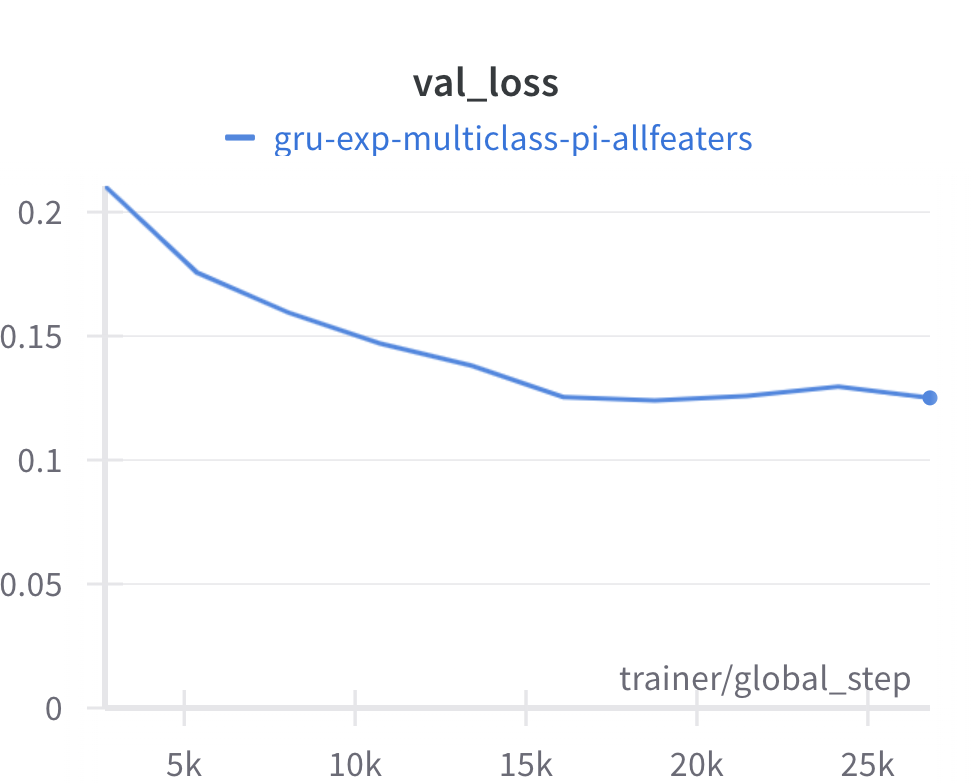}
\caption{Val\_loss Pi}
\label{Val_Loss_Pi}
\endminipage\hfill
\minipage{0.3\textwidth}
\includegraphics[width=\linewidth]{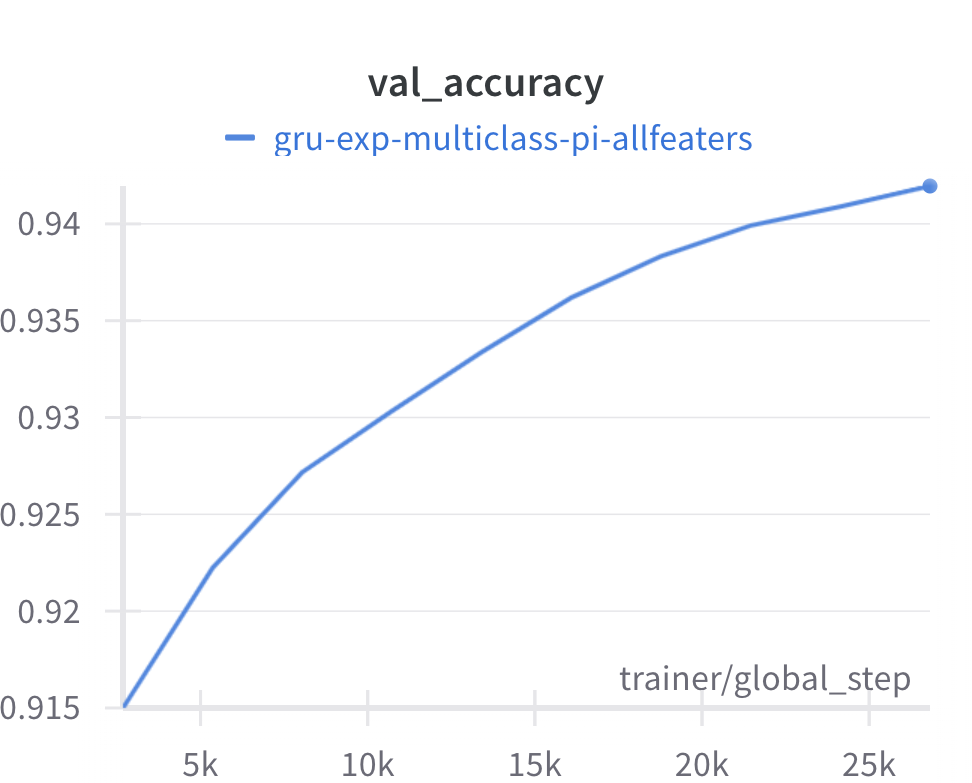}
\caption{Accuracy Pi}
\label{Accuracy_Pi}
\endminipage
\end{figure}

%% file: chapters/wandbPi/wandbRouter.tex
\begin{figure}[!htb]
\minipage{0.3\textwidth}
\includegraphics[width=\linewidth]{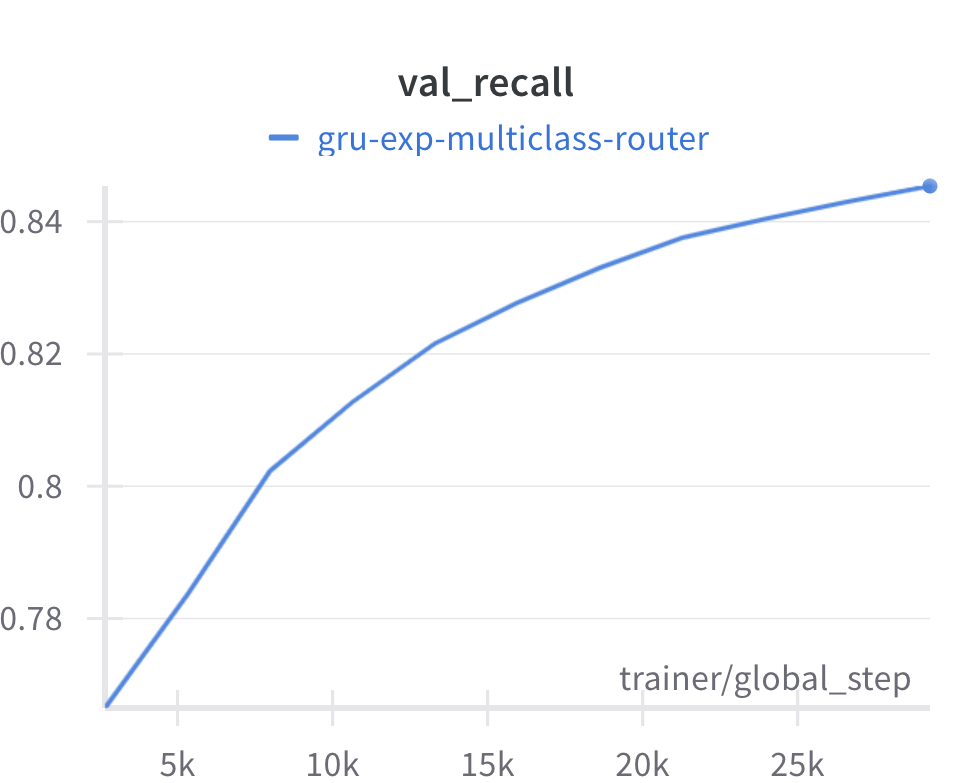}
\caption{Recall Router}
\label{Recall_Router}
\endminipage\hfill
\minipage{0.3\textwidth}
\includegraphics[width=\linewidth]{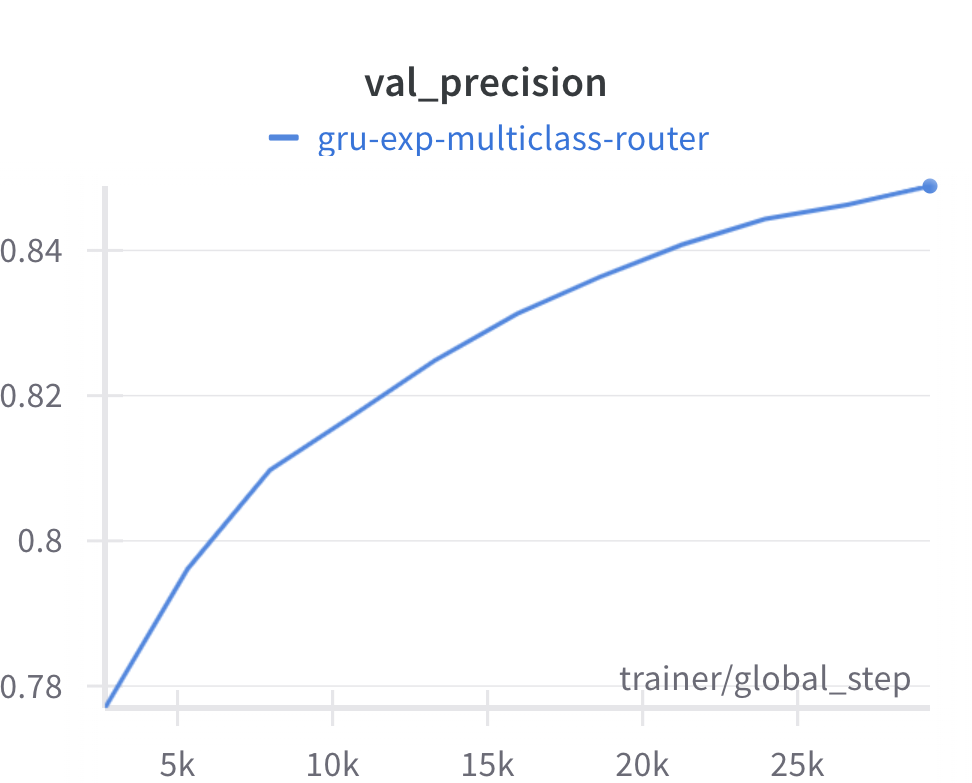}
\caption{Precision Router}
\label{Precision_Router}
\endminipage\hfill
\minipage{0.3\textwidth}
\includegraphics[width=\linewidth]{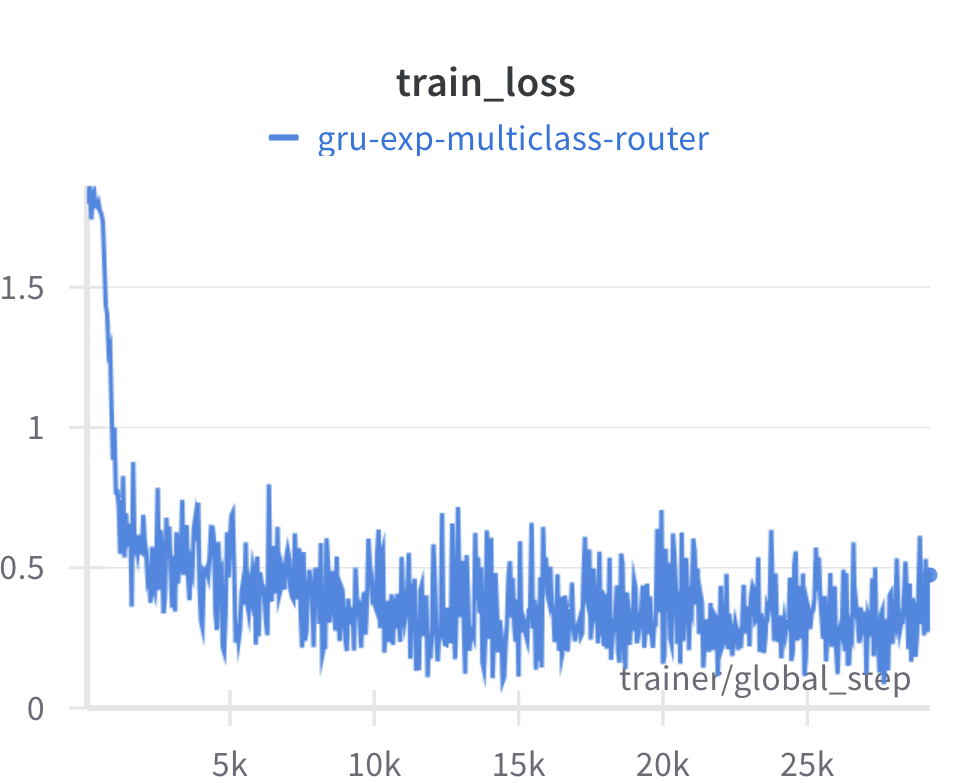}
\caption{Train loss Router}
\label{Train_Loss_Router}
\endminipage\hfill
\minipage{0.3\textwidth}
\includegraphics[width=\linewidth]{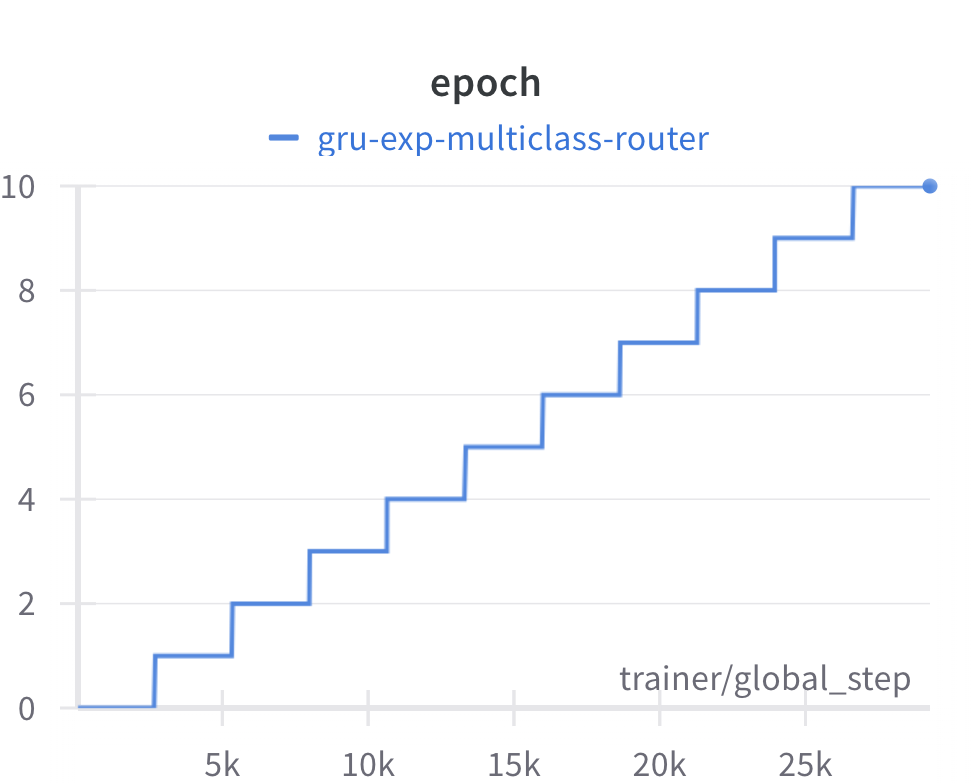}
\caption{Epoch Router}
\label{Epoch_Router}
\endminipage\hfill
\minipage{0.3\textwidth}
\includegraphics[width=\linewidth]{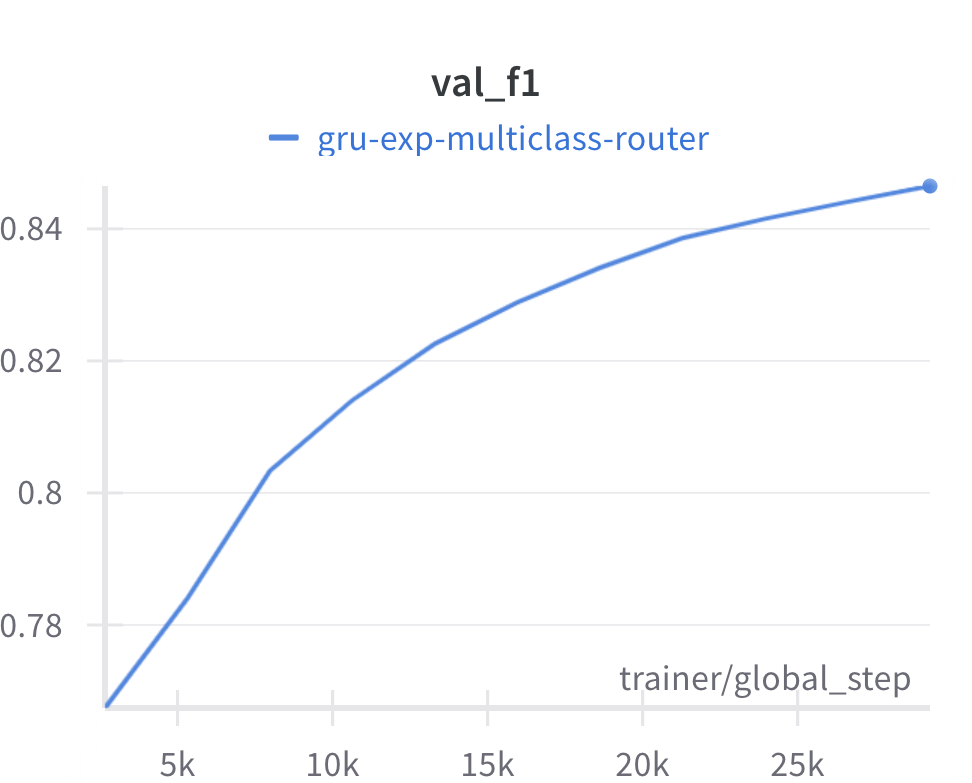}
\caption{F1 Router}
\label{F1_Router}
\endminipage\hfill
\minipage{0.3\textwidth}
\includegraphics[width=\linewidth]{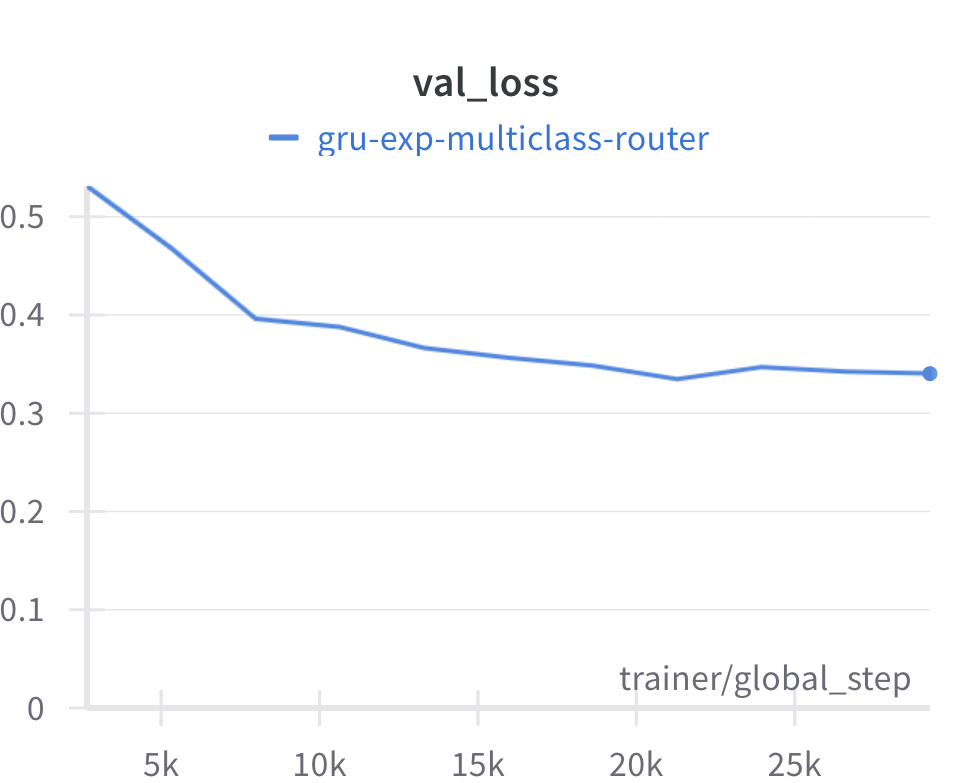}
\caption{Val\_loss Router}
\label{Val_Loss_Router}
\endminipage\hfill
\minipage{0.3\textwidth}
\includegraphics[width=\linewidth]{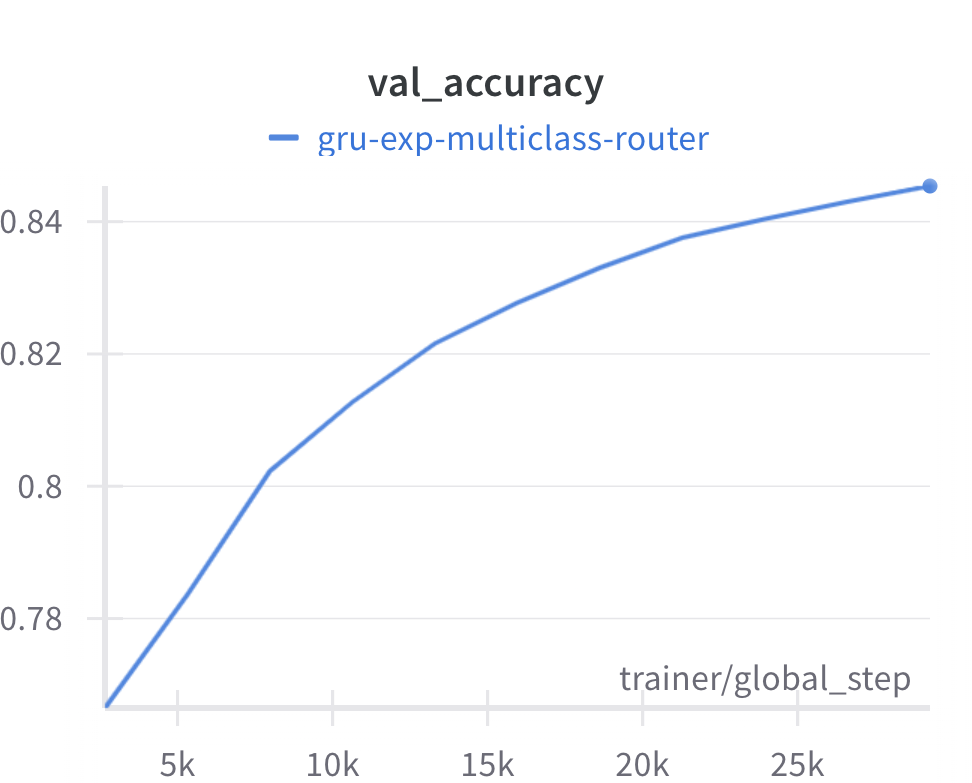}
\caption{Accuracy Router}
\label{Accuracy_Router}
\endminipage
\end{figure}

%% file: chapters/Experements.tex
\section{Metrics}

This section explains essential metrics for evaluating machine learning models: Validation Accuracy, Precision, Recall, F1 Score, and the Confusion Matrix. These metrics help us understand how well the model is performing. All the metrics are extracted from \cite{matrics}.

\subsection{Validation Accuracy (val\_accuracy)}

\textbf{Equation:}
\[
\text{Accuracy} = \frac{TP + TN}{TP + TN + FP + FN}
\]

Accuracy shows the percentage of correct predictions made by the model. It is calculated by adding the true positives (TP) and true negatives (TN) and dividing by the total number of predictions. Accuracy is useful but can be misleading if one class is much more common than others.

\subsection{Validation Precision (val\_precision)}

\textbf{Equation:}
\[
\text{Precision} = \frac{TP}{TP + FP}
\]

Precision tells us how many of the positive predictions are actually correct. To calculate precision, the true positives (TP) are divided by the total number of the positive directions, which is TP + FP. High precision shows fewer false positives, which is essential when the false alarms are expensive.

\subsection{Validation Recall (val\_recall)}

\textbf{Equation:}
\[
\text{Recall} = \frac{TP}{TP + FN}
\]

Recall is mainly the percentage of real positive cases that model successfully recognised. The true positives (TP) are divided by the total number of the actual positives (TP + FN) to get the desired result. When the missing positives could have major repercussions, the high recall model indicates that it is appropriate at identifying every positive case.

\subsection{Validation F1 Score (val\_f1)}

\textbf{Equation:}
\[
\text{F1 Score} = 2 \times \frac{\text{Precision} \times \text{Recall}}{\text{Precision} + \text{Recall}}
\]
The F1 Score is the balance between Precision and Recall. It integrates both into a single metric, making it highly important when we need to balance the false positives and the false negatives, especially in cases with imbalanced datasets.

\subsection{Confusion Matrix}

\textbf{Structure:}
\[
\begin{array}{c|c|c}
  & \text{Predicted Positive} & \text{Predicted Negative} \\
\hline
\text{Actual Positive} & TP & FN \\
\text{Actual Negative} & FP & TN \\
\end{array}
\]

Confusion Matrix table reflects the number of correct and incorrect predictions. It helps us to understand the different types of errors that the model is making, thus showing the counts of true negatives (TN), true positives (TP), false positives (FP), as well as false negatives (FN). \\ \\ \\ \\

After understanding the key metrics and their significance in evaluating the model performance, one can now proceed to the experimental phase in the section \ref{experiments} to analyse the given results. The analysis will allow us to apply the metrics in practice and also assess how well the model performs in actual scenarios. Before proceeding with the experiments, each function was assigned a number from 0 to 5, as shown in Table \ref{table:FunctionNum}.

\begin{table}[h!]
\centering
\begin{tabular}{|l|l|}
\hline
 Function & Num   \\ \hline
 Normal  & 0   \\ \hline
 DoS  & 1   \\ \hline
 ICMP & 2   \\ \hline
 Port\_Scan & 3   \\ \hline
 Telnet & 4   \\ \hline
 TCP\_SYN & 5   \\ \hline
 
\end{tabular}
\caption{Function Num}
\label{table:FunctionNum}
\end{table}

\section{Experiments}
This section will highlight the experiments conducted to evaluate a model capable of recognising different functions' signatures. We started with a binary classification that distinguished between two classes, benign and malicious. Then, we experimented with a multiclass model to identify the exact function being executed based on the HPC measures. Device A is the Pi5 Broadcom BCM2712 SoC and Device B is the Turris Omni Marvel Armada 385 SoC. 
\label{experiments}

\subsection{Device A Binary}
\label{BinaryPi} 

\begin{figure}[!t]
    \centering
    \includegraphics[scale=0.5]{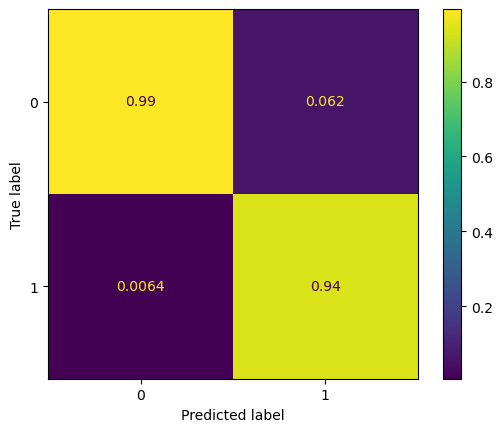}
    \caption{Device A Binary Confusion Matrix}
    \label{fig:DeviceAconfusionmatrixbinpi}
\end{figure}

As shown in Figure  \ref{fig:DeviceAconfusionmatrixbinpi}, a binary analysis for Device A between the DoS function and Normal behaviour. The model achieved a high accuracy of 94\%, indicating its strong ability to distinguish between the two functions, Normal and DoS. The high precision rate reflects a meagre rate of false positives, which is crucial in our case. Similarly, the recall suggests the model has a very low rate of false negatives. As illustrated by the confusion matrix, only 0.062 malicious functions are misclassified as benign, while just 0.0064 benign functions are incorrectly classified as malicious.

\subsection{Device A Multi-Class}

\begin{figure}[!h]
\minipage{0.5\textwidth}
\includegraphics[width=\linewidth]{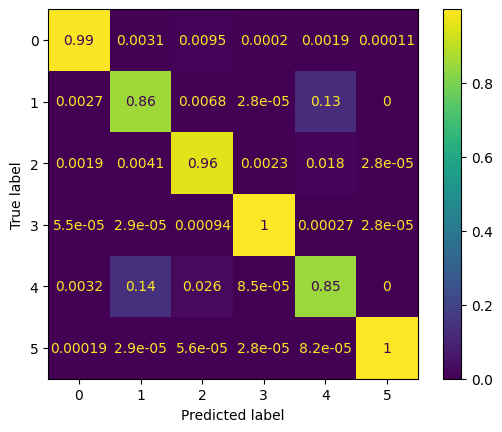}
\caption{Device A Multi-Class 1 Layer}
\label{fig:DeviceAconfusionmatrixmultipi}
\endminipage
\minipage{0.5\textwidth}
\includegraphics[width=\linewidth]{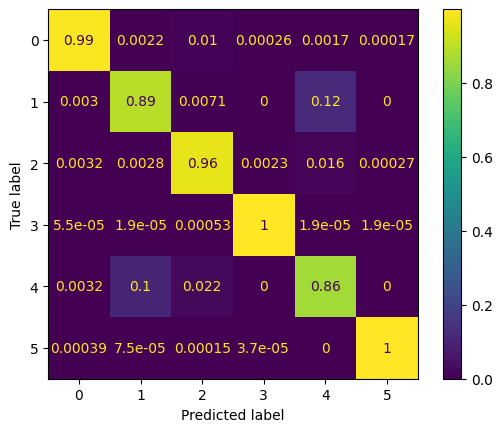}
\caption{Device A Multi-Class 2 Layer}
\label{fig:DeviceAconfusionmatrixmultipi1}
\endminipage
\end{figure}

Figure \ref{fig:DeviceAconfusionmatrixmultipi} highlights the model's performance across six functions as shown in Table \ref{MultiClassTable}. The model shows high accuracy, especially for Normal 99\%, ICMP 96\%, PortScan 100\%, and TCPSYN 100\%, indicating strong reliability in detecting these types of traffic. The 100\% accuracy for PortScan and TCPSYN is particularly important, as these are common methods used in network attacks. However, the slightly lower accuracy for DoS 86\% and Telnet 85\% suggests some difficulty in distinguishing these functions, particularly with 0.14 of Telnet instances misclassified as DoS. Despite this, the model maintains low false positive and false negative rates, making it reliable for most scenarios. Improving the model's ability to differentiate between DoS and Telnet could further enhance its effectiveness in network security applications.

The confusion matrix also reveals the model's classification accuracy in more detail. For Normal function, only 0.0031 instances were incorrectly classified as other types, demonstrating the model's high precision in identifying benign activities. The DoS had a slightly higher misclassification rate, with 0.14 Telnet function incorrectly classified as DoS, indicating some overlap between these functions. ICMP  was well-classified, with only minor confusion, and the PortScan and TCPSYN were perfectly classified, reflecting the model's robustness in detecting these attack types. Telnet, while generally well-classified, showed some confusion with DoS, as noted earlier. Overall, the matrix indicates a strong performance, with very low misclassification rates, particularly for critical attack types like PortScan and TCPSYN, essential for maintaining network security.

We tried to enhance our model by adding an extra layer to our GRU. Since it was one layer, we modified it to two layers. After enhancing our model by adding an extra layer to the GRU, we observed some changes in performance metrics as shown in Figure \ref{fig:DeviceAconfusionmatrixmultipi1}. Specifically, the accuracy for the DoS function increased from 86\% to 89\%, while the accuracy for Telnet slightly improved from 85\% to 86\%. These improvements indicate that the additional layer has enabled the model to better capture patterns related to these specific functions.
Moreover, the confusion between DoS and Telnet decreased, with the misclassification from DoS to Telnet reducing from 0.14 to 0.1, and from Telnet to DoS from 0.13 to 0.12. This reduction in confusion suggests that the model is now more effective at distinguishing between these two categories. Importantly, while these areas have seen improvement, the performance of other functions remained unchanged, indicating that the added complexity has not negatively impacted the model's overall stability. Therefore, the additional GRU layer has improved the model's performance, particularly in enhancing specificity and reducing errors between closely related classes.

%After adding the layer, it's essential to evaluate the model thoroughly on validation data to ensure that the changes lead to a meaningful improvement in performance. figure \ref{fig:DeviceAconfusionmatrixmultipi1} has slightly increase in some of the functions, DoS has increased from 86\% to 89\%. and Telnet slightly increased from 85\% to 86\% while the other remained unchanged. Also the confusion between DoS and Telnet has decreased from 1.4\% to 1\%. As well as for the Telnet to Dos from 1.3\% to 1.2\%. 

\subsection{Device B Binary}
\label{BinaryRouter}

\begin{figure}[htb!]
    \centering
    \includegraphics[scale=0.5]{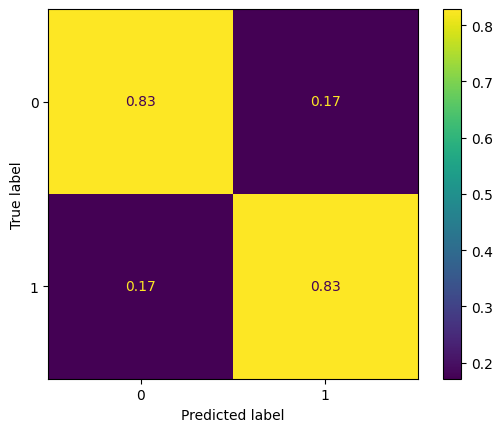}
    \caption{Device B Binary Confusion Matrix}
    \label{fig:DeviceBconfusionmatrixbinRouter}
\end{figure}

Figure \ref{fig:DeviceBconfusionmatrixbinRouter} presents a binary classification analysis, highlighting the model's performance distinguishing between Normal and DoS functions. The matrix shows that the model correctly classified 83\% for Normal and 83\% for DoS. However, 0.17 of the Normal instances were misclassified as DoS, and 0.17 of the DoS instances were misclassified as Normal. These misclassification rates indicate some challenges in distinguishing between the two functions, particularly with a slightly higher error rate for DoS. Despite this, the overall accuracy remains relatively high, reflecting the model's reasonable effectiveness in binary classification tasks.

\subsection{Device B Multi-Class}
\begin{figure}[!htb]
\minipage{0.5\textwidth}
\includegraphics[width=\linewidth]{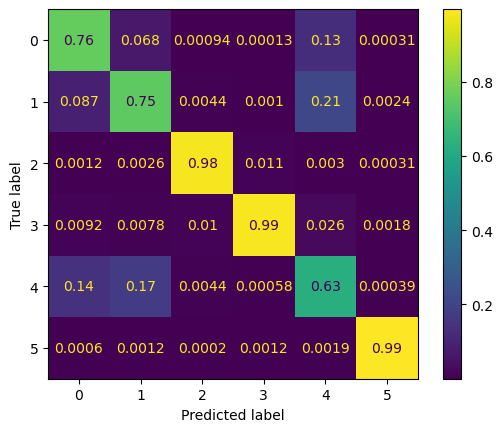}
\caption{Device B Multi-Class 1 Layer}
\label{fig:DeviceBconfusionmatrixmultiRouter}
\endminipage
\minipage{0.5\textwidth}
\includegraphics[width=\linewidth]{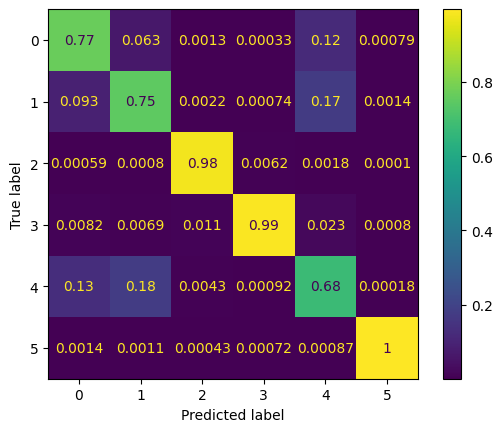}
\caption{Device B Multi-Class 2 Layer}
\label{fig:DeviceBconfusionmatrixmultiRouter1}
\endminipage
\end{figure}

Figure \ref{fig:DeviceAconfusionmatrixmultipi} illustrates the model's performance across six functions as shown in Table \ref{MultiClassTable}. The model shows varied accuracy across different functions, with the highest accuracy achieved in ICMP 98\%, PortScan 99\%, and TCPSYN 99\%. These high accuracy rates indicate the model's strong capability in correctly identifying these specific types of traffic.

For Normal, the model achieved 76\% accuracy, but 0.13 of the instances were misclassified as Telnet, and 0.068 were misclassified as DoS, indicating some difficulty in distinguishing between these functions. Similarly, DoS was correctly identified 75\% of the time but had notable confusion with Telnet 0.21 and Normal 0.087.

Telnet had a lower accuracy rate of 63\%, with significant confusion with DoS 0.17 and Normal 0.14 This suggests that the model struggles more with differentiating Telnet from other functions, which could be due to overlapping features or similarities in traffic patterns.

Overall, while the model performs very well for certain functions ICMP, PORTSCAN, TCPSYN, it faces challenges in accurately classifying others, particularly Telnet. This indicates that further refinement is needed, especially in feature selection or model adjustment, to improve the accuracy for those functions where the model currently struggles. Despite these challenges, the model's performance in correctly classifying a majority of the functions remains commendable, especially for critical functions with near-perfect accuracy.

After adding an additional layer to our GRU model, we compared the resulting confusion matrices to evaluate the impact on performance. In the confusion matrix Figure \ref{fig:DeviceBconfusionmatrixmultiRouter1}, we observed that the accuracy for Normal increased slightly from 76\% to 77\%, and the confusion between Normal and Telnet decreased from 0.13 to 0.12. Similarly, the accuracy for ICMP remained stable at 98\%, but the confusion from ICMP to PortScan decreased from 0.011 to 0.0062, indicating improved specificity. Moreover, the accuracy for Telnet improved from 63\% to 68\%, with a slight reduction in confusion from Telnet to TCPSYN, decreasing from 0.00039 to 0.00018. These changes suggest that the additional GRU layer has led to better discrimination between closely related classes, resulting in a modest but meaningful improvement in overall model performance.

\section{Results Summary}
\label{results:summary}
Table \ref{table:HPC-events} highlights that Devices A and B have varying capabilities in tracking certain HPC events, which likely contributes to the observed differences in model accuracy across these devices.
\begin{table}[h!]
\centering
\small
\begin{tabular}{|l|c|c|c|c|}
\hline
\textbf{Metric} & \textbf{Device A*} & \textbf{Device A Binary}  & \textbf{Device B*} & \textbf{Device B Binary} \\
\hline
Accuracy  & 0.9476 & 0.9647 & 0.8454 & 0.8377 \\
Precision & 0.9474 & 0.9663 & 0.8489 & 0.8388 \\
Recall    & 0.9476 & 0.9647 & 0.8454 & 0.8377  \\
F1-score  & 0.9475 & 0.9647 & 0.8465 & 0.8376\\
%Validation Loss & 0.1022 & 0.00003 & 0.3403 & 0.2663  
\hline
\end{tabular}

* Multiclass overall accuracy.
\caption{Comparison of Metrics between Device A and Device B}
\end{table}

\begin{table}[h!]
\centering
\small
\begin{tabular}{|l|c|c|c|c|}
\hline
\textbf{SUFs} & \textbf{Device A} & \textbf{Device B} & \textbf{Device A 2 Layers} & \textbf{Device B 2 Layers}\\
\hline
\textbf{Normal} & 99 & 76 & 99 & 77\\
\hline
\textbf{DoS} & 88 & 75 & 89 & 75\\
\hline
\textbf{ICMP}    & 96 & 98 & 96 &  98\\
\hline
\textbf{Port\_Scan}  & 100 & 99 & 100 &  99\\
\hline
\textbf{Telnet}  & 85 & 63 & 86 & 68\\
\hline
\textbf{TCP\_SYN\_Flood}  & 100 & 99 & 100 & 100\\
\hline
\end{tabular}
\caption{Multi-Class Device Accuracy}
\label{MultiClassTable}
\end{table}

The Binary Classification Device A in Section \ref{BinaryPi} demonstrated superior performance with an accuracy of 94.76\%. This high accuracy is likely due to Device A's ability to track a comprehensive set of HPC events that are unavailable on other devices. These additional features give the model richer data, enabling it to distinguish between normal and DoS functions more effectively. On the other hand, in Section \ref{BinaryRouter}, Device B achieved an accuracy of 83.77\%. It lacks some critical events, such as L1-icache-loads and LLC operations. Despite this, the Device B model still benefits from the HPC events, which shows reasonable accuracy. Both devices performed well in their environments, but they still differ because the device type influences the model, and each device has different HPC events. Overall, both of the devices performed well in binary classification.

In multi-class classification, Device A achieved an overall accuracy of 94.76\%., achieving high accuracy across most functions. The device identified PortScan and TCPSYN with a perfect accuracy of 100\%, indicating their robustness. Normal and ICMP also performed exceptionally well, with accuracy rates of 99\% and 96\%, respectively, indicating the model's strong ability to classify these functions. However, the accuracy for DoS and Telnet was slightly lower, suggesting some challenges in distinguishing these functions, although the overall performance remained strong. Device B, while also performing well, showed a slightly more varied performance, with PortScan and TCPSYn achieving a high accuracy of 99\%, but with lower accuracy observed for Normal and DoS. Telnet, in particular, struggles with lower accuracy, indicating the device's limitations in accurately classifying this function. Despite the challenges, ICMP had a high accuracy of 98\%, thus showing the model’s capability in various areas.

After seeing this limitation, we enhanced it by adding a second GRU layer; the multi-class classification performance saw notable improvements, particularly in areas where the initial accuracy was lower. On Device A, the second GRU layer increased the accuracy of DoS from 86\% to 89\% and Telnet from 85\% to 86\%, enhancing the model's ability to differentiate between these functions and making them more robust. The already high-performing functions, such as PortScan and TCPSYN, maintained their perfect accuracy, demonstrating the model’s stability even with the added complexity. Device B benefited significantly from the additional layer, especially for the Telnet function, where accuracy improved from 63\% to 68\%, addressing one of the device’s key weaknesses in function classification. Normal function accuracy also increased slightly, from 76\% to 77\%, while the accuracy for DoS, ICMP, PortScan, and TCPSYN remained stable, reinforcing the model's robustness in these areas. Adding the second GRU layer improved the model's performance across all devices, particularly for initially less robust functions. \\

% I could add limitation here ! as subsection 
\subsection{Limitation}
While we tried to improve the GRU by adding a second layer, several limitations still existed. Performance analysis is inherently tied to the specific set of HPC events tracked by each device. Devices with a limited set of features may not achieve the same level of accuracy as those with a broader range of events. 

The study hasn't explored other RISC-based architectures, such as RISC-V, which may show different performance traits because of variations in event tracking and processing capabilities. Thus, the findings may not be directly applicable across all hardware platforms, as each device supports different HPC events.

The study didn't investigate the effects of various compilation flags. Compilation flags could majorly impact software behaviour on certain hardware, potentially affecting the model’s performance.

This study also focused on the predefined set of functions; however, it didn't include all the possible malicious functions that could be encountered in actual scenarios. A more diverse set of wrong functions could reveal additional strengths or weaknesses in models.

The models were tested in controlled environments, which may infrequently fully capture the complexities of live deployments. Real-world factors such as real-time data processing, environmental noise, and varying network conditions could affect the model's performance.

%* not take other risc base arc such as  risc V(5)
%* Different compiling flags 
%* more malicious functions (we didn't take all of the malicious funcitons)
%* we didn't test the preformace of this model in live deployments. 

%% file: chapters/Disscusion.tex
%\section{Discussion}

%Explore the application of the model on other RISC-based architectures, such as RISC-V.

%Investigate the impact of different compilation flags on model performance and generalisability.

%Expand the dataset to include a broader range of malicious functions for more comprehensive testing.

%Conduct live deployment testing to evaluate model performance in real-time, dynamic environments.

%Develop strategies for cross-device generalization to ensure adaptability across devices with varying HPC event capabilities.

\section{Overview }

Through the creation and use of Hardware Performance Counters (HPCs) and Software Unclonable Functions (SUFs), this dissertation has efficiently presented an innovative strategy for improving the security of Internet of Things (IoT) devices. The urgent need for more robust security measures to shield IoT devices from increasingly potent cyber-attacks was the driving force behind the study. Further concentrated on utilising HPCs to generate distinct, device-specific signatures (SUFs) by utilising perf, hardware-level monitoring tools. These provide a robust method for identifying and distinguishing between legitimate and malicious behaviours in IoT devices. By creating an innovative, hardware-based strategy that combines the advantages of software and hardware security measures, it has majorly advanced the security of IoT devices. The suggested SUF framework is a useful response to the escalating IoT security issues because of its high accuracy, low influence on performance, and scalability. The research showed that these HPC-derived SUFs offer a dependable technique for detecting security vulnerabilities through in-depth testing and the deployment of deep learning models, mainly Recurrent Neural Networks (RNNs). The method is viable and scalable for use in real-world scenarios, as demonstrated by the results that demonstrated high accuracy in identifying malicious activity with no impact on device performance. This work thus creates new opportunities for extending similar approaches to other areas within the larger field of cybersecurity and furthering the growth of IoT security. The success of HPC-based SUFs in enhancing device security has been a significant advancement in developing more secure and resilient IoT ecosystems.

\subsection{Question Answering}

This subsection will answer the question that was asked at the beginning: \\

\textbf{Question 1: How effectively can HPC-derived SUF signatures distinguish between legitimate and compromised devices, enhancing the nuanced approach required for robust IoT security?} \\ 

\textbf{Answer:} The results of the experiments prove that the approach is highly effective in distinguishing between legitimate and compromised behaviours of an IoT device, achieving an accuracy ranging from 83.77\% to 96.47\% with high precision.\\

\textbf{Question 2: Which software-based techniques, integrated with HPC data, significantly fortify the deployment framework for SUFs, ensuring a resilient identification?} \\

\textbf{Answer:} Although we did not employ different compilation options, we experimented with different compilers to conduct the experiments on different devices. Under fixed compilation options, the binaries gave a stable signature that the proposed approach could easily identify. Using statistical tools to process the extracted data proved that software techniques such as scaling helped develop an approach with good performance.\\

\textbf{Question 3: In what ways can HPC-based SUFs be seamlessly incorporated into current IoT security paradigms without compromising operational efficiency? }\\

\textbf{Answer:} The proposed architecture has been developed to incorporate two architectural elements: a data collection pipeline and a classification pipeline that could function independently. The data collection pipeline has been designed to offer a lightweight integration. These design choices ensure a modular deployment that could be easily incorporated on any device with minimal impact.\\

\textbf{Main Question:} Can HPC patterns for specific functions serve as a reliable signature for identifying malicious threats that target IoT devices or firmware?  \\

\textbf{Answer:} Yes, HPC patterns for specific functions can serve as reliable signatures for identifying malicious functions that target IoT devices. These unique patterns generated by HPCs during the experiments were identified successfully using the proposed model. The model performed well, even when isolating each elementary function as a separate class. The multiclass model achieved comparable performance to the binary model with an accuracy ranging from 85\% to 95\% approximately.

%\textbf{Objective:} To explore avenues for integrating HPC-based SUFs into existing IoT security frameworks to strengthen firmware integrity checks and device authenticity verification while minimizing the impact on device performance.

\subsection{Research Significance}

This work is significant because its approach demonstrates the potential for using SUFs to improve the security of IoT devices through careful evaluation and application. Given the increasing use of IoT devices and the growing landscape of cyber threats against them, this research presents an approach to address those challenges. The key contributions are as follows:

\begin{itemize}
    \item Originality in Security Measures: By emphasising upon SUFs, this research presents a novel security system that makes use of the special qualities of the hardware to enable safe communication on Internet of Things devices. By being inherently tied to the physical device, the technique raises the amount of difficulty of cloning and alteration while addressing the security flaws in a way that complements the present software solutions.
    
    \item Comprehensive Evaluation Methodology: The research includes a thorough evaluation of both the reliability and performance of SUFs, investigating the feasibility of deploying such a solution in real environments. The reliability evaluation clarifies the conditions under which this approach remains functional, while the performance evaluation identifies the necessary efforts for deploying this approach into existing devices.
    
    \item Addressing a Critical Security Gap: As IoT devices become more complicated and widespread, current security frameworks frequently become inadequate, particularly when using regular security updates and patches is impractical. To close this gap, this study investigates SUFs as a hardware-level security solution that could provide a more reliable and long-lasting security framework for the devices.
    
    \item Practical Implications for IoT Security: The outcomes of this research have significant practical implications, providing valuable insights for IoT manufacturers, security professionals, and policymakers. By demonstrating the feasibility and effectiveness of SUFs, the study paves the way for their integration into future IoT devices, enhancing overall ecosystem security. Specific applications could include smart home devices, healthcare monitoring systems, industrial IoT platforms, and critical infrastructure, where enhanced security is critical.
    \item Interdisciplinary and Global Impact: The methodologies and insights developed through this research extend beyond cybersecurity, potentially influencing fields such as embedded systems design, hardware engineering, and AI-driven security solutions. Given the global deployment of IoT devices, the findings have worldwide relevance, contributing to international cybersecurity standards and protecting devices in diverse environments.
    \item Foundation for Future Research: This research contributes to the academic and practical understanding of IoT security, offering a solid foundation for future studies. The findings and methodologies developed through this study will inspire further exploration into hardware-based security solutions, driving innovation in the field. Also, the research opens opportunities for collaboration with industry partners along with other academic disciplines, thus fostering a multidisciplinary approach towards solving complex security problems.
\end{itemize}

\section{Future work}
While this research has made significant strides in enhancing IoT security using Software Unclonable Functions (SUFs) derived from Hardware Performance Counters (HPCs), there are several areas where future work could expand upon the current findings, particularly in addressing the limitations discussed in Section \ref{results:summary}. 

\begin{itemize}
    \item Application on other RISC-based architectures: Future research could explore the application of the current model on other RISC-based architectures, such as RISC-V. This would help determine the model's adaptability and effectiveness across different processor architectures, providing insights into its broader applicability in the IoT landscape.
    \item Expansion of the dataset: Expanding the dataset to include a broader range of malicious functions would allow for more comprehensive testing and validation of the model. This would enhance the model's ability to detect various threats, increasing its utility in diverse IoT environments.
    \item Cross-device generalisation: Developing strategies for cross-device generalisation is essential to ensure the model's adaptability across devices with varying HPC event capabilities. This would involve creating methods that allow the model to maintain high performance even when deployed on devices with different architectures or hardware configurations.
\end{itemize} 

\subsection{Potential Impact}
Through this dissertation, we successfully proved that it is possible to identify some functions through a behavioural analysis guided by hardware-based counters. Such a conclusion will drive more elaborate studies to apply this concept to various security applications related to the embedded systems sector. This project aims to build an innovative product that offers an advanced threat mitigation and identification technique to safeguard IoT devices at scale. The same concept could also be used to enhance the coverage of black-box fuzzing of complex binaries. Such an application would ease the security investigation of embedded target firmware by minimising the required reverse engineering effort. This same concept could be used to augment the application fuzzing coverage and identify the interesting software behaviours during the fuzzing campaign.